\documentclass[10pt,aps,prd,twocolumn,amsfonts,amsmath,floatfix,showpacs,superscriptaddress]{revtex4-2}
\usepackage{tikz} %
\usepackage{graphicx}%
\usepackage{bm}%
\usepackage{placeins} %
\usepackage{enumitem} %
\usepackage[caption=false]{subfig}
\usepackage[colorlinks=true, linkcolor=blue, citecolor=blue, urlcolor=blue]{hyperref}
\usepackage{makecell}
\usepackage{multirow}
\usepackage{fp}

\usepackage{acro}

\usepackage{booktabs}

\newcommand{\ccpi}{$\text{CC}1\pi^{\pm}$ } %
\newcommand{\mynumu}{\stackrel{\textbf{\fontsize{1pt}{0pt}\selectfont(---)}}{\nu}_{\mu}}

\DeclareAcronym{TPC}{
  short = TPC,
  long  = time projection chamber,
  short-plural = s,
  long-plural = s
}

\DeclareAcronym{LArTPC}{
  short = LArTPC,
  long  = liquid argon time projection chamber,
  short-plural = s,
  long-plural = s
}

\DeclareAcronym{LAr}{
  short = LAr,
  long  = liquid argon,
}

\DeclareAcronym{CCQE}{
  short = CCQE,
  long  = charged-current quasi-elastic,
}

\DeclareAcronym{DUNE}{
  short = DUNE,
  long  = Deep Underground Neutrino Experiment,
}

\DeclareAcronym{POT}{
  short = POT,
  long  = protons on target,
}

\DeclareAcronym{Fermilab}{
  short = Fermilab,
  long  = the Fermi National Accelerator Laboratory,
}

\DeclareAcronym{NuMI}{
  short = NuMI,
  long  = Neutrinos at the Main Injector,
}

\DeclareAcronym{ArgoNeuT}{
  short = ArgoNeuT,
  long  = Argon Neutrino Teststand,
}

\DeclareAcronym{BNB}{
  short = BNB,
  long  = Booster Neutrino Beam,
}

\DeclareAcronym{CV}{
  short = CV,
  long  = central value,
}

\DeclareAcronym{BDT}{
  short = BDT,
  long  = boosted decision tree,
}

\DeclareAcronym{MC}{
  short = MC,
  long  = Monte Carlo,
}

\DeclareAcronym{MIP}{
  short = MIP,
  long  = minimum ionizing particle,
}

\DeclareAcronym{CC}{
  short = CC,
  long  = charged-current,
}

\DeclareAcronym{NC}{
  short = NC,
  long  = neutral-current,
}

\DeclareAcronym{SIS}{
  short = SIS,
  long  = shallow inelastic scattering,
}

\DeclareAcronym{DIS}{
  short = DIS,
  long  = deep inelastic scattering,
}

\DeclareAcronym{RES}{
  short = RES,
  long  = resonance,
}

\DeclareAcronym{MEC}{
  short = MEC,
  long  = meson exchange current,
}

\DeclareAcronym{COH}{
  short = COH,
  long  = coherent pion production,
}

\DeclareAcronym{NCE}{
  short = NCE,
  long  = neutral-current elastic,
}

\DeclareAcronym{SBN}{
  short = SBN,
  long  = Short-Baseline Neutrino,
}

\DeclareAcronym{INC}{
  short = INC,
  long  = intra-nuclear cascade,
}

\DeclareAcronym{QE}{
  short = QE,
  long  = quasi-elastic,
}

\DeclareAcronym{RPA}{
  short = RPA,
  long  = Random Phase Approximation,
}

\DeclareAcronym{SCC}{
  short = SCC,
  long  = second-class current,
}

\DeclareAcronym{HARP}{
  short = HARP,
  long  = the Hadron Production Experiment,
}

\DeclareAcronym{FSI}{
  short = FSI,
  long  = final state interactions,
}

\DeclareAcronym{PMT}{
  short = PMT,
  long  = photomultiplier tube,
}

\DeclareAcronym{TMVA}{
  short = TMVA,
  long  = Toolkit for Multivariate Data Analysis,
}

\DeclareAcronym{SVM}{
  short = SVM,
  long  = support vector machine,
}

\DeclareAcronym{MCS}{
  short = MCS,
  long  = multiple Coulomb scattering,
}

\DeclareAcronym{BUU}{
  short = BUU,
  long  = Boltzmann-Uehling-Uhlenbeck,
} 

\DeclareAcronym{FSF}{
  short = FSF,
  long  = Free Spectral Function,
}  

\DeclareAcronym{NN}{
  short = NN,
  long  = nucleon-nucleon,
}

\begin{document}
\title{Measurement of single charged pion production in charged-current \texorpdfstring{$\nu_\mu$}{Muon Neutrino}-Ar interactions with the MicroBooNE detector}

\newcommand{\ANL}{Argonne National Laboratory (ANL), Lemont, IL, 60439, USA}
\newcommand{\Bern}{Universit{\"a}t Bern, Bern CH-3012, Switzerland}
\newcommand{\BNL}{Brookhaven National Laboratory (BNL), Upton, NY, 11973, USA}
\newcommand{\UCSB}{University of California, Santa Barbara, CA, 93106, USA}
\newcommand{\Cambridge}{University of Cambridge, Cambridge CB3 0HE, United Kingdom}
\newcommand{\CIEMAT}{Centro de Investigaciones Energ\'{e}ticas, Medioambientales y Tecnol\'{o}gicas (CIEMAT), Madrid E-28040, Spain}
\newcommand{\Chicago}{University of Chicago, Chicago, IL, 60637, USA}
\newcommand{\Cincinnati}{University of Cincinnati, Cincinnati, OH, 45221, USA}
\newcommand{\CSU}{Colorado State University, Fort Collins, CO, 80523, USA}
\newcommand{\Columbia}{Columbia University, New York, NY, 10027, USA}
\newcommand{\Edinburgh}{University of Edinburgh, Edinburgh EH9 3FD, United Kingdom}
\newcommand{\FNAL}{Fermi National Accelerator Laboratory (FNAL), Batavia, IL 60510, USA}
\newcommand{\Granada}{Universidad de Granada, Granada E-18071, Spain}
\newcommand{\IIT}{Illinois Institute of Technology (IIT), Chicago, IL 60616, USA}
\newcommand{\ICL}{Imperial College London, London SW7 2AZ, United Kingdom}
\newcommand{\Indiana}{Indiana University, Bloomington, IN 47405, USA}
\newcommand{\Kansas}{The University of Kansas, Lawrence, KS, 66045, USA}
\newcommand{\KSU}{Kansas State University (KSU), Manhattan, KS, 66506, USA}
\newcommand{\Lancaster}{Lancaster University, Lancaster LA1 4YW, United Kingdom}
\newcommand{\LANL}{Los Alamos National Laboratory (LANL), Los Alamos, NM, 87545, USA}
\newcommand{\Louisiana}{Louisiana State University, Baton Rouge, LA, 70803, USA}
\newcommand{\Manchester}{The University of Manchester, Manchester M13 9PL, United Kingdom}
\newcommand{\MIT}{Massachusetts Institute of Technology (MIT), Cambridge, MA, 02139, USA}
\newcommand{\Michigan}{University of Michigan, Ann Arbor, MI, 48109, USA}
\newcommand{\MSU}{Michigan State University, East Lansing, MI 48824, USA}
\newcommand{\Minnesota}{University of Minnesota, Minneapolis, MN, 55455, USA}
\newcommand{\Nankai}{Nankai University, Nankai District, Tianjin 300071, China}
\newcommand{\NMSU}{New Mexico State University (NMSU), Las Cruces, NM, 88003, USA}
\newcommand{\Oxford}{University of Oxford, Oxford OX1 3RH, United Kingdom}
\newcommand{\Pitt}{University of Pittsburgh, Pittsburgh, PA, 15260, USA}
\newcommand{\QMUL}{Queen Mary University of London, London E1 4NS, United Kingdom}
\newcommand{\Rutgers}{Rutgers University, Piscataway, NJ, 08854, USA}
\newcommand{\SLAC}{SLAC National Accelerator Laboratory, Menlo Park, CA, 94025, USA}
\newcommand{\SDSMT}{South Dakota School of Mines and Technology (SDSMT), Rapid City, SD, 57701, USA}
\newcommand{\Maine}{University of Southern Maine, Portland, ME, 04104, USA}
\newcommand{\TelAviv}{Tel Aviv University, Tel Aviv, Israel, 69978}
\newcommand{\UTA}{University of Texas, Arlington, TX, 76019, USA}
\newcommand{\Tufts}{Tufts University, Medford, MA, 02155, USA}
\newcommand{\VTech}{Center for Neutrino Physics, Virginia Tech, Blacksburg, VA, 24061, USA}
\newcommand{\Warwick}{University of Warwick, Coventry CV4 7AL, United Kingdom}

\affiliation{\ANL}
\affiliation{\Bern}
\affiliation{\BNL}
\affiliation{\UCSB}
\affiliation{\Cambridge}
\affiliation{\CIEMAT}
\affiliation{\Chicago}
\affiliation{\Cincinnati}
\affiliation{\CSU}
\affiliation{\Columbia}
\affiliation{\Edinburgh}
\affiliation{\FNAL}
\affiliation{\Granada}
\affiliation{\IIT}
\affiliation{\ICL}
\affiliation{\Indiana}
\affiliation{\Kansas}
\affiliation{\KSU}
\affiliation{\Lancaster}
\affiliation{\LANL}
\affiliation{\Louisiana}
\affiliation{\Manchester}
\affiliation{\MIT}
\affiliation{\Michigan}
\affiliation{\MSU}
\affiliation{\Minnesota}
\affiliation{\Nankai}
\affiliation{\NMSU}
\affiliation{\Oxford}
\affiliation{\Pitt}
\affiliation{\QMUL}
\affiliation{\Rutgers}
\affiliation{\SLAC}
\affiliation{\SDSMT}
\affiliation{\Maine}
\affiliation{\TelAviv}
\affiliation{\UTA}
\affiliation{\Tufts}
\affiliation{\VTech}
\affiliation{\Warwick}

\author{P.~Abratenko} \affiliation{\Tufts}
\author{D.~Andrade~Aldana} \affiliation{\IIT}
\author{L.~Arellano} \affiliation{\Manchester}
\author{J.~Asaadi} \affiliation{\UTA}
\author{A.~Ashkenazi}\affiliation{\TelAviv}
\author{S.~Balasubramanian}\affiliation{\FNAL}
\author{B.~Baller} \affiliation{\FNAL}
\author{A.~Barnard} \affiliation{\Oxford}
\author{G.~Barr} \affiliation{\Oxford}
\author{D.~Barrow} \affiliation{\Oxford}
% ...

\collaboration{The MicroBooNE Collaboration}
\thanks{microboone\_info@fnal.gov}\noaffiliation

\begin{abstract}
We present flux-integrated charged-current $\nu_\mu$ cross-section measurements on argon for final states containing exactly one $\pi^\pm$ and no other hadrons except nucleons. The analysis uses data from the MicroBooNE experiment in the Booster Neutrino Beam, corresponding to $1.11 \times 10^{21}$ \acl{POT}. Total and single-differential cross-section measurements are provided within a phase space restricted to muon momenta above 150~MeV, pion momenta above 100~MeV, and muon-pion opening angles smaller than 2.65~rad. Differential cross sections are reported with respect to the scattering angles of the muon and pion relative to the beam direction, their momenta, and their combined opening angle. The differential cross section with respect to muon momentum is based on a subset of selected events with the muon track fully contained in the detector, whereas the cross section with respect to pion momentum is based on a subset of selected events rich in pions that have not hadronically scattered on the argon before coming to rest. The latter has not been measured on argon before. The total cross section is measured as \( (3.75~\pm~0.07~\textrm{(stat.)}~\pm~0.80~\textrm{(syst.)}) \times 10^{-38} \, \text{cm}^2/\text{Ar} \) at a mean energy of approximately 0.8~GeV. Comparisons of the measured cross sections with predictions from multiple neutrino-nucleus interaction generators show good overall agreement, except at very forward muon angles. %
\end{abstract}

\maketitle

\section{\label{sec:introduction}Introduction}%
A good understanding of neutrino-nucleus interactions is crucial to interpret results from neutrino oscillation experiments, particularly to explore CP violation in the lepton sector \cite{T2K:2019bcf, NOvA:2021nfi} and to investigate physics beyond the Standard Model \cite{DUNE:2020fgq}. Simulating these interactions presents significant challenges, including modeling initial nucleon states, accounting for multi-nucleon effects, and handling final state interactions \cite{NuSTEC:2017hzk}. Cross-section measurements play a key role in improving these models and in reducing uncertainties for current and future neutrino experiments \cite{Ankowski:2016jdd}. Argon is an increasingly popular target material for large-scale neutrino oscillation experiments due to its use in \ac{TPC} detectors. Specifically, the long-baseline \ac{DUNE} \cite{DUNE:2020lwj} and the \ac{SBN} program \cite{Machado:2019oxb} at \ac{Fermilab} rely on these detector types for their precise particle tracking and calorimetry.

MicroBooNE has collected an extensive data set of neutrino interactions on argon. It operated in the \ac{BNB} with a mean neutrino energy of approximately 0.8~GeV and an energy spectrum that extends into the resonance production region and falls off with a long tail that goes beyond 2~GeV. At these energies, MicroBooNE is well suited to probe \(\Delta(1232)\) resonance production by measuring final state pions. 

This work presents results for \ac{CC} muon (anti\mbox{-})neutrino events with a single charged pion and any number of nucleons in the final state, referred to as \ccpi throughout this paper. Previously, MicroBooNE has published results for $\pi^0$ production in \ac{NC} and \ac{CC} neutrino scattering \cite{MicroBooNE:2018neo, MicroBooNE:2022zhr, MicroBooNE:2024sec, MicroBooNE:2024bnl} and \ac{CC} \(\nu_e,\bar{\nu}_e\) \(\pi^\pm\) production \cite{MicroBooNE:2025prw}. Improving \ac{CC} pion production models is important, as these interactions dominate at energies relevant to experiments such as NOvA \cite{NOvA:2021nfi} and \ac{DUNE}.

Muon (anti\mbox{-})neutrino \ccpi cross sections have been measured for several other target materials, including water \cite{T2K:2016cbz, MINERvA:2022djk}, carbon \cite{K2K:2008tus, MINERvA:2022djk}, hydrocarbons \cite{MiniBooNE:2010eis, T2K:2021naz, T2K:2019yqu, MINERvA:2014ogb, MINERvA:2016sfc, MINERvA:2019rhx, MINERvA:2022djk}, and some metals \cite{MINERvA:2022djk}. The MINERvA experiment \cite{MINERvA:2013zvz} has measured many of these and has shown challenges for existing models to describe data across a range of nuclear targets. Measurements on argon are needed to provide data for the tuning of event generators and to help understand how nuclear effects scale with the mass of the target nucleus.

Argon measurements have previously been made using the 0.24-ton active-mass \ac{ArgoNeuT} detector at \ac{Fermilab} \cite{Acciarri:2018ahy} using $1.25 \times 10^{20}$ \acf{POT} in the higher-mean-energy \acl{NuMI} beam \cite{Adamson:2015dkw}. That experiment measured \ccpi interactions for beams in neutrino and antineutrino mode separately, collecting 115 $\nu_\mu$ and 158 $\bar{\nu}_\mu$ events after background subtraction. This MicroBooNE analysis has selected 6816 events after background subtraction, offering a substantial increase in statistics.

At the vertex level, neutrino-nucleus scattering can lead to charged pion production through several interaction mechanisms, depending on the energy transferred to the nucleus. At lower energy transfers, \ac{COH} can occur, where a neutrino interacts with the entire nucleus, leaving it in its ground state while producing a pion. In the few-GeV neutrino energy range, the dominant mode of pion production is through the excitation of a nucleon \ac{RES}, primarily the $\Delta(1232)$, which decays into a nucleon and a pion. At higher neutrino energies, \ac{SIS} and \ac{DIS} become the prevailing neutrino interaction mechanisms. In \ac{DIS} pions are produced as part of the hadronization process following significant momentum transfer to a quark in the struck nucleon. \Ac{SIS} is the transition region between resonance production and \ac{DIS}. These inelastic interactions tend to produce multi-hadron, often multi-pion, final states. Events with more than one charged pion outside the nucleus are not part of the signal definition. Additionally, \ac{FSI} can add or remove pions from the set of observed particles for non-coherent interactions. This also allows for pion production in events where the primary interaction does not produce pions, including \ac{QE} scattering and \ac{MEC} interactions. These also count as signal events when they have exactly one charged pion exiting the nucleus, regardless of the initial number of pions produced in the neutrino interaction.

\section{\label{sec:MicroBooNE}The MicroBooNE Experiment}%
The MicroBooNE experiment uses a single-phase \ac{LArTPC} with an active mass of 85~tonnes \cite{MicroBooNE:2016pwy}. The detector is located 468.5~meters from the target of the \ac{BNB} at \ac{Fermilab}, which delivers a beam consisting primarily of sub-GeV muon neutrinos \cite{MiniBooNE:2008hfu}. MicroBooNE collected neutrino beam interactions between 2015 and 2020. This analysis uses $1.11 \times 10^{21}$ \ac{POT} from all five years of operation.

\subsection{The Booster Neutrino Beam}
\label{sec:BNB}
The \ac{BNB} generates a neutrino beam by accelerating protons to 8~GeV kinetic energy using the \ac{Fermilab} Booster synchrotron and directing them onto a beryllium target, producing secondary mesons \cite{Stancu:2001cpa}. These particles are then subjected to a toroidal magnetic field from the focusing horn, which operated in forward horn current mode for MicroBooNE's entire data taking period. This arrangement defocuses negatively charged mesons and focuses positive ones, which then decay primarily into neutrinos with a mean energy of 0.8~GeV. The predicted neutrino flux in MicroBooNE is dominated by muon-flavor neutrinos. The \(\nu_\mu\) component accounts for 93.66\%, while \(\bar{\nu}_\mu\) contributes 5.79\%. Electron-flavor neutrinos are much less common, with \(\nu_e\) and \(\bar{\nu}_e\) making up only 0.51\% and 0.05\%, respectively. The flux distribution of $\bar{\nu}_\mu$ also peaks at lower energies, resulting in fewer \ac{RES} interactions and fewer charged pions. Thus, the antineutrino component is expected to contribute less than $1\%$ of the cross sections presented. The combined $\nu_\mu + \bar{\nu}_\mu$ integrated neutrino flux prediction used for this analysis is \(8.73 \times 10^{11} \, \mathrm{cm}^{-2}\).

\subsection{The MicroBooNE Detector}
A comprehensive description of the MicroBooNE detector can be found in Ref.~\cite{MicroBooNE:2016pwy}. The core component of the detector, a liquid argon filled \ac{TPC}, serves as both the target for neutrino interactions and the detection medium for the resulting final state particles. Charged particles produced by neutrino interactions or background events, such as cosmic muons, ionize the argon as they traverse the detector volume. A uniform electric field of 273~V/cm drifts the free electrons toward three layered planes, each consisting of parallel wires spaced 3~mm apart, oriented at 60$^{\circ}$ angles relative to one another. The first two planes register induced signals from the drifting electrons, while the third plane collects the electrons directly. A photomultiplier-tube-based light collection system captures scintillation light, providing complementary timing information. Position reconstruction of the interaction is achieved along one dimension by measuring the charge distribution on the parallel wires, while the drift time of the electrons provides the orthogonal dimension. This configuration enables the detector to record projections of interactions from three different perspectives, allowing for 3D reconstruction of neutrino interactions.

\subsection{Simulation}
\label{sec:Simulation}
The flux simulation for MicroBooNE is adapted from the MiniBooNE experiment situated in the same beam line \cite{MiniBooNE:2008hfu}. It uses the G\textsc{eant}4 simulation toolkit \cite{GEANT4:2002zbu, Allison:2006ve, Allison:2016lfl} version \texttt{8.1} with additional constraints on $\pi^\pm$ and $K^+$ production from measurements taken by the HARP and SciBooNE experiments \cite{HARP:2007dqt, SciBooNE:2011sjq}. A tuned configuration of the neutrino \ac{MC} generator GENIE version \texttt{3.0.6} \cite{Andreopoulos:2009rq, GENIE:2021npt} is used to simulate neutrino interactions across all MicroBooNE analyses, according to this predicted flux. Based on the model configuration \texttt{G18\_10a\_02\_11a}, it constrains \ac{CC} \ac{QE} and \ac{CC} \ac{MEC} interactions with data from T2K \cite{MicroBooNE:2021ccs}. The effect of this tuning is minimal for the signal prediction of this analysis, but does affect the predicted background, which is rich in pionless \ac{CC} events. The configuration uses the hA 2018 intranuclear cascade model for \ac{FSI}~\cite{Dytman:2015taa}. A known shortcoming of the model is overprediction of the pion-nucleon charge exchange cross section below \(\sim400\)~MeV and underprediction above this. This affects the conversion of charged pions to neutral pions in the \ac{FSI} simulation. Using a modified simulation sample which aims to improve on this shortcoming, it was verified that this did not have a significant impact on the selection efficiency and that differences are covered by existing model uncertainties. Subsequent simulation steps are performed using the LArSoft software framework \cite{Snider:2017wjd}. G\textsc{eant}4 version \texttt{10.3.3} is used to simulate particle transport for the final state particles provided by GENIE. It produces groups of ionization electrons, which are smeared to account for electron attenuation, diffusion, and recombination with ions. Scintillation photons are also generated, with their probability of reaching a \acl{PMT} determined by their production position in the detector. Next, the electric field response of the electrons moving past the wires, as well as the response of the readout electronics, is simulated \cite{MicroBooNE:2018swd}. 

MicroBooNE records interactions for which the neutrino beam timing coincides with the detection of scintillation light. Since the MicroBooNE detector is surface-based, there are cosmogenic backgrounds that get included in the readout. This is modeled by combining the simulated waveforms with real recorded backgrounds taken when the beam was turned off.

Two additional types of samples are used to correctly model beam spills not producing a neutrino interaction in the detector. The first is pure beam-off data recordings without any simulated interactions, and the second is simulated neutrino interactions outside the active detector volume with overlaid beam-off background, referred to as dirt.

\subsection{Reconstruction}
The reconstruction process is identical for simulated and real data. First, signal processing and deconvolution are applied to the raw waveforms from the \ac{TPC} wires~\cite{MicroBooNE:2017qiu, MicroBooNE:2018swd, MicroBooNE:2018vro}, followed by the reconstruction of hits that convey calorimetric information associated with energy deposits on each wire. These hits are then provided as input to the Pandora pattern recognition software \cite{Marshall:2015rfa, Acciarri:2017hat}, which clusters them within each plane and matches these clusters between planes to identify groups of hits originating from the same particle.

Pandora first performs a cosmic ray muon reconstruction, identifying and removing clear cosmic ray muon tracks. The remaining hits are then used for event slicing, where the goal is to group together hits belonging to the same neutrino or cosmic ray interaction. Each slice is then reconstructed under both cosmic ray and neutrino interaction hypotheses. Pandora uses a \ac{SVM} to compute a topological score between 0 and 1 for each slice, assessing how cosmic-like or neutrino-like they are. Together with additional information from scintillation light, the most probable neutrino interaction slice is identified. Within this slice, hierarchies of reconstructed tracks and showers with 3D points and trajectories are produced. Finally, position and calorimetric calibration are applied to account for variations in the TPC wire response, charge loss due to electron attenuation and ion recombination, and distortions from space charge caused by the build-up of positive ions in the detector \cite{MicroBooNE:2019efx, MicroBooNE:2020kca}.

The reconstruction process has several performance limitations relevant to this analysis. One key issue is track fragmentation, where muons, whether originating from neutrinos or cosmic sources, can be incorrectly broken up into two separate tracks. This can mimic the signature of a back-to-back muon and pion pair, requiring the exclusion of this phase space from the analysis. Another limitation is Michel electron merging, where Michel electrons are occasionally incorrectly merged with muon and pion tracks. This merging introduces additional uncertainty in the estimation of the momentum.

\section{\label{sec:overview}Analysis Overview}%
\begin{figure}[htp]
    \centering
    \begin{tikzpicture}
        \clip [rounded corners=3mm] (0,0) rectangle (8.66,6.56); 
        \node[anchor=south west, inner sep=0pt] at (0,0)
            {\includegraphics[width=\columnwidth]{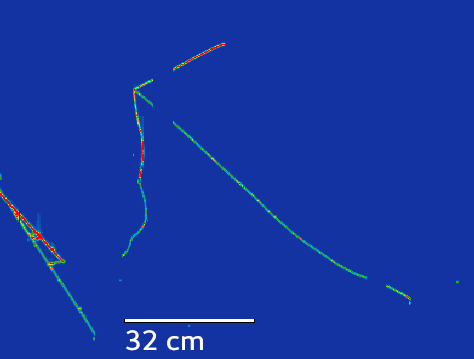}};
        \node[anchor=north east, inner sep=5pt] at (8.66,6.56)
            {\includegraphics[width=2cm]{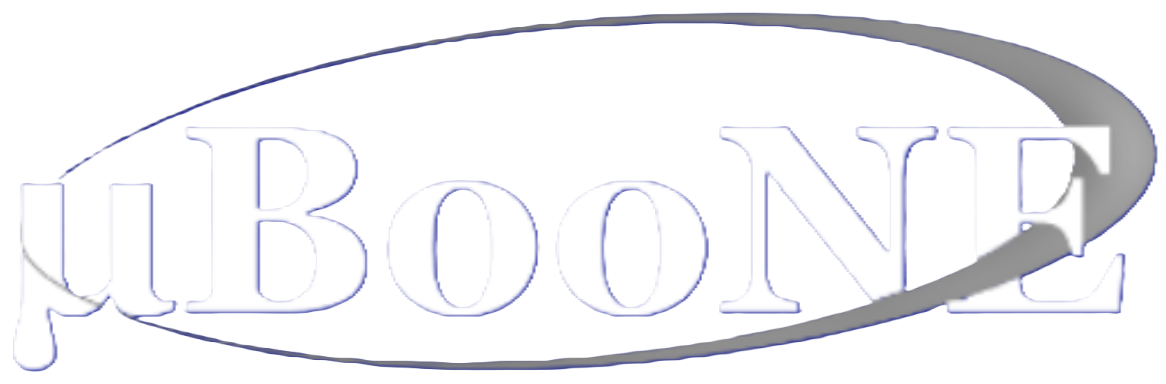}};
        \node[anchor=north west, text=white, font=\small] at (0.1,6.56) 
            {BNB Run 24727, Subrun 108, Event 5417};
        \draw [white, dashed, thick, ->] (1.5, 4.9) node[left, text=white, font=\small] {$\nu_\mu \, (\bar{\nu}_\mu)$} -- (2.2, 4.9);
        \node[anchor=east, text=white, font=\small, rotate=28] at (4.0, 6.0) {Candidate p};
        \node[anchor=east, text=white, font=\small, rotate=-42] at (5.6, 2.5) {Candidate $\mu$};
        \node[anchor=east, text=white, font=\small] at (2.5, 3.9) {Candidate $\pi^\pm$};
    \end{tikzpicture}
    \caption{Event display from the collection plane of the detector showing a selected event consisting of a long muon candidate, a shorter proton candidate, and a charged pion candidate appearing to decay to a Michel electron via a muon. Cosmic rays are visible in the bottom left corner, and sections of the tracks are missing due to unresponsive detector wires.}
    \label{fig:eventDisplay}
\end{figure}
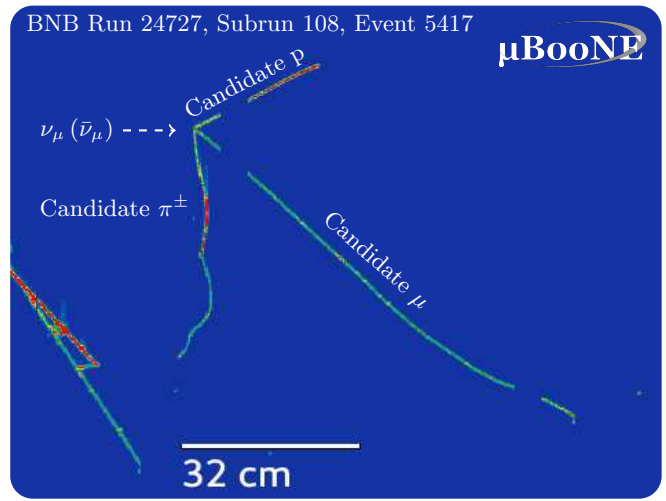

This analysis uses a topology-based signal definition imposed on the set of final state particles that exit the nucleus struck in the neutrino interaction. This includes events where a $\pi^{\pm}$ is produced after the initial interaction via \ac{FSI} inside the nucleus and excludes events where a $\pi^{\pm}$ is produced but captured before leaving the nucleus. The final states for this analysis consist of one (anti\mbox{-})muon, one charged pion, and any number of nucleons:

\begin{equation}
\mynumu + \text{Ar} \rightarrow 1 \mu^{\pm} + 1 \pi^{\pm} + \text{X}
\end{equation}
\noindent where $\text{X}$ is any set of nucleons and nuclei. A selected event matching this definition can be seen in Fig.~\ref{fig:eventDisplay}. 

Significant other meson production comes in the form of neutral pions, visible via their decay products, and charged kaons. Both are explicitly excluded from the signal definition. Without a magnetic field, MicroBooNE cannot distinguish between $\pi^+$ and $\pi^-$. Moreover, $\bar{\nu}_{\mu}$ interactions are included as part of the signal, although their contribution is minor (as described in Sec.~\ref{sec:BNB}). The neutrino interaction vertex is further required to be inside a fiducial volume, defined as the volume of the detector that is at least 10~cm away from all TPC borders and 50~cm away from the back face of the detector in the beam direction. The resulting volume is a rectangular cuboid of liquid argon with dimensions 236~cm $\times$ 213~cm $\times$ 977~cm.

The phase space restrictions for this analysis are: muon momentum $p_{\mu} >$~150~MeV, pion momentum $p_{\pi} >$~100~MeV, and muon-pion opening angle $\theta_{\mu\pi} <$~2.65~rad. The opening-angle phase space restriction excludes back-to-back muon-pion pairs, which are challenging to distinguish from long cosmic muon tracks and result in a phase space region with few signal events but high background contamination. %

In addition to the total cross section, single differential cross sections are presented for the following lab-frame variables: $p_{\mu}$ for the muon momentum, $\cos(\theta_{\mu})$ for the cosine of the muon angle with respect to the beam, $p_{\pi}$ for the pion momentum, $\cos(\theta_{\pi})$ for the cosine of the pion angle with respect to the beam, and $\theta_{\mu\pi}$ for the muon-pion opening angle. The angular variables are calculated from the directions of the reconstructed particle candidates, while the momentum measurements use separate subsets of events. For the muon momentum measurement, uncontained muon candidates are excluded from the selection to allow momentum estimation by range. Uncontained particles exit the detector and are defined for this analysis as tracks coming within 5~cm of any \ac{TPC} boundary. Pion momentum estimation also poses challenges due to the high rate of secondary interactions, such as inelastic scattering, which lack ionization signatures. This makes range- and calorimetry-based methods unreliable. The pion momentum is estimated using pions that do not undergo elastic or inelastic hadronic interactions on the argon and come to rest inside the detector, referred to as unscattered pions throughout this paper. For this, an unscattered-enhanced subset is selected, in which the composition of the selected signal events is much richer in these types of pions. This makes it possible to use the pion candidate's range to estimate the momentum. The charged pion momentum spectrum from neutrino interactions with argon has not been previously measured.

\section{\label{sec:selection}Event Selection}%
\subsection{Preselection}
The preselection inclusively selects \ac{CC} muon (anti\mbox{-})neutrino interactions in MicroBooNE and distinguishes them from NC and electron (anti\mbox{-})neutrino interactions. First, a series of particle-level cuts are applied to identify potential muon candidates. Pandora provides a score for each reconstructed particle that indicates how shower-like versus track-like it is on a scale from 0 to 1. This uses a set of metrics that include the total length, the distance from the interaction vertex, and changes along the length of the reconstructed particle (such as the transverse extent) \cite{Acciarri:2017hat}. Muon candidates must have unambiguous tracks with scores $\ge 0.85$. Potential muon tracks must start within 4~cm of the reconstructed neutrino vertex, which helps reduce cosmic rays being misidentified as muons originating from a neutrino interaction. Next, only primary tracks in Pandora's reconstruction hierarchy are considered. This removes tracks that are child particles of other reconstructed particles. The muon candidate must also have a track length of at least 20~cm. This corresponds to a momentum of 150~MeV, which aligns with the muon momentum phase space restriction that is enforced in the analysis. In the last steps, the energy loss curve $\mathrm{d}E/\mathrm{d}x$ for reconstructed calorimetric hits at the end of a track is compared to simulated distributions for muons and protons. The resulting agreement metrics (usually denoted \(\chi^2\), though not strictly following a \(\chi^2\) distribution) are used to determine how muon- or proton-like a given track is, with values of \(\chi^2_\mu < 30\) and \(\chi^2_p > 60\) required. The proton-to-muon \(\chi^2\) ratio cut \(\chi^2_p / \chi^2_\mu > 7\) is also applied to strongly favor muon agreement. Plots of the \(\chi^2\) distributions can be found in the supplemental material \cite{SupplementalMaterial}.  

Next, quality cuts are applied at the event level. The starting point of all child particles must be at least 10~cm from the TPC boundaries to reduce external backgrounds. Finally, the reconstructed neutrino interaction vertex must be in the defined fiducial volume.

\subsection{Boosted Decision Trees}
\label{sec:BDTs}
This work relies on three \acp{BDT} to identify particles in the final state of the interaction. These are the muon \ac{BDT}, the proton \ac{BDT} (used to separate protons and charged pions), and the unscattered pion \ac{BDT}. Each \ac{BDT} is trained to identify its target particle and reject the other types of tracks using the \ac{TMVA} framework \cite{TMVA:2007ngy} on contained reconstructed particles from signal events that pass the preselection. Scattering information for the simulated pions comes from the G\textsc{eant}4 transport simulation. To mitigate the impact of poorly reconstructed tracks, particles are weighted based on their completeness, defined as the fraction of calorimetric hits from a simulated particle that are assigned to the best-matching reconstructed particle. Overlaid cosmic rays have uniform weights. Training uses half of the simulated events, corresponding to 63,000 particles, while the remaining half is reserved for validation and cross-section extraction. Feature selection for the \acp{BDT} follows an iterative process. Initially, each \ac{BDT} is trained with a broad set of potentially suitable features. A series of retraining steps is then conducted, systematically removing one feature at a time to assess its impact on performance. The performance change resulting from a feature's removal serves as a measure of its importance. Features with little to no impact are permanently excluded. This process continues until each remaining feature significantly contributes to the final performance. Through this optimization, the muon and unscattered pion \acp{BDT} retain the six key features listed below. The proton \ac{BDT} uses only five features, as it does not depend on the number of descendant particles.

\begin{description}[leftmargin=0pt, labelwidth=0pt]
    \item[Calorimetric features]\hfill
        \begin{enumerate}[left=0pt]
            \item The ratio of the proton to \ac{MIP} energy loss likelihoods. \Ac{MIP} refers to particles that do not have a Bragg peak, which is a characteristic mass-dependent energy loss as particles slow down in matter. The likelihood is determined from the last 30~cm of a track by comparing the energy loss, $\mathrm{d}E/\mathrm{d}x$, along the track to simulated distributions. Specifically, for each hit in a reconstructed track, the likelihood of the energy loss at that point coming from a proton or a MIP is calculated. The ratio of these likelihoods is used as a metric for the likelihood of the particle being a proton.
            \item The ratio of the pion to \ac{MIP} energy loss likelihoods. The same as feature 1, but for charged pions.
            \item The mean change in deposited energy along the first third of a fitted track. The calculated mean excludes the first three calorimetric hits and also hits that are more than one standard deviation away from the mean.
        \end{enumerate}
\end{description}

The calorimetric features employ a multi-plane approach to mitigate issues when particles travel nearly parallel to wires \cite{MicroBooNE:2021ddy}. When possible, collection plane information is preferred due to its superior performance \cite{MicroBooNE:2018vro}. To maintain accurate $\mathrm{d}E/\mathrm{d}x$ estimates for particles that travel below an angular threshold relative to the collection plane wires, a weighted average of the induction planes is used. The weighting is based on the hit count on each plane.

\begin{description}[leftmargin=0pt, labelwidth=0pt]
    \item[Topological features]\hfill
        \begin{enumerate}[left=0pt]\addtocounter{enumi}{3}
            \item The number of all descendant particles in the reconstructed hierarchy associated with a primary particle. This provides information about the decay of primary particles.
            \item The standard deviation of the angular differences between the directions at adjacent points along the trajectory of the fitted track. More colloquially, this is a metric for the wiggliness of the track and provides information about reinteractions and decays.
            \item The particle classification score distinguishes between track-like and shower-like appearances. This score is computed by Pandora, as described in the preselection section.
        \end{enumerate}
\end{description}

\begin{figure}[htp]
  \centering
  \setlength{\tabcolsep}{0pt}      %
  \renewcommand{\arraystretch}{0.9} %

  \begin{tabular}{c}
    \includegraphics[trim=0.3cm 0 1.8cm 1.0cm, clip, width=\columnwidth]{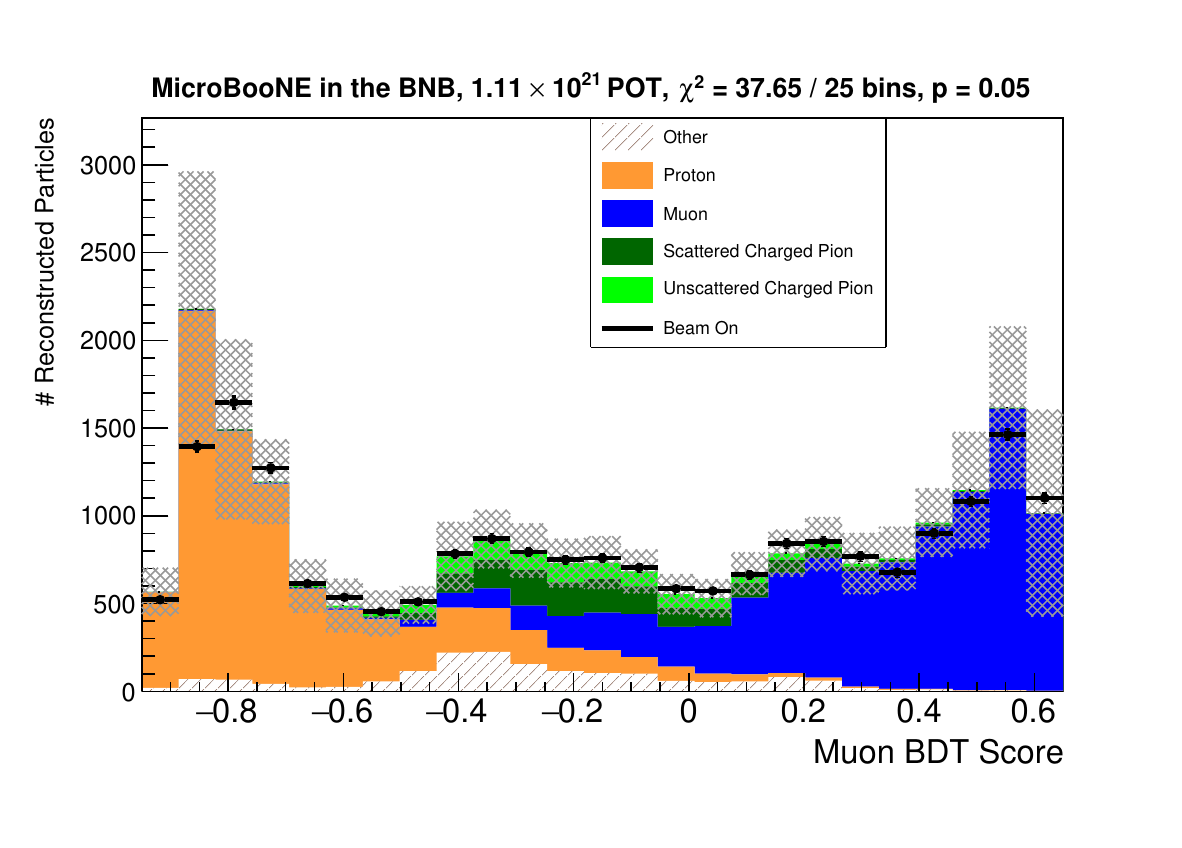} \\[-18pt]
    \makebox[0.49\textwidth][c]{\hspace*{7mm}(a) Muon BDT score }\\[6pt]
    \includegraphics[trim=0.3cm 0 1.8cm 1.0cm, clip, width=\columnwidth]{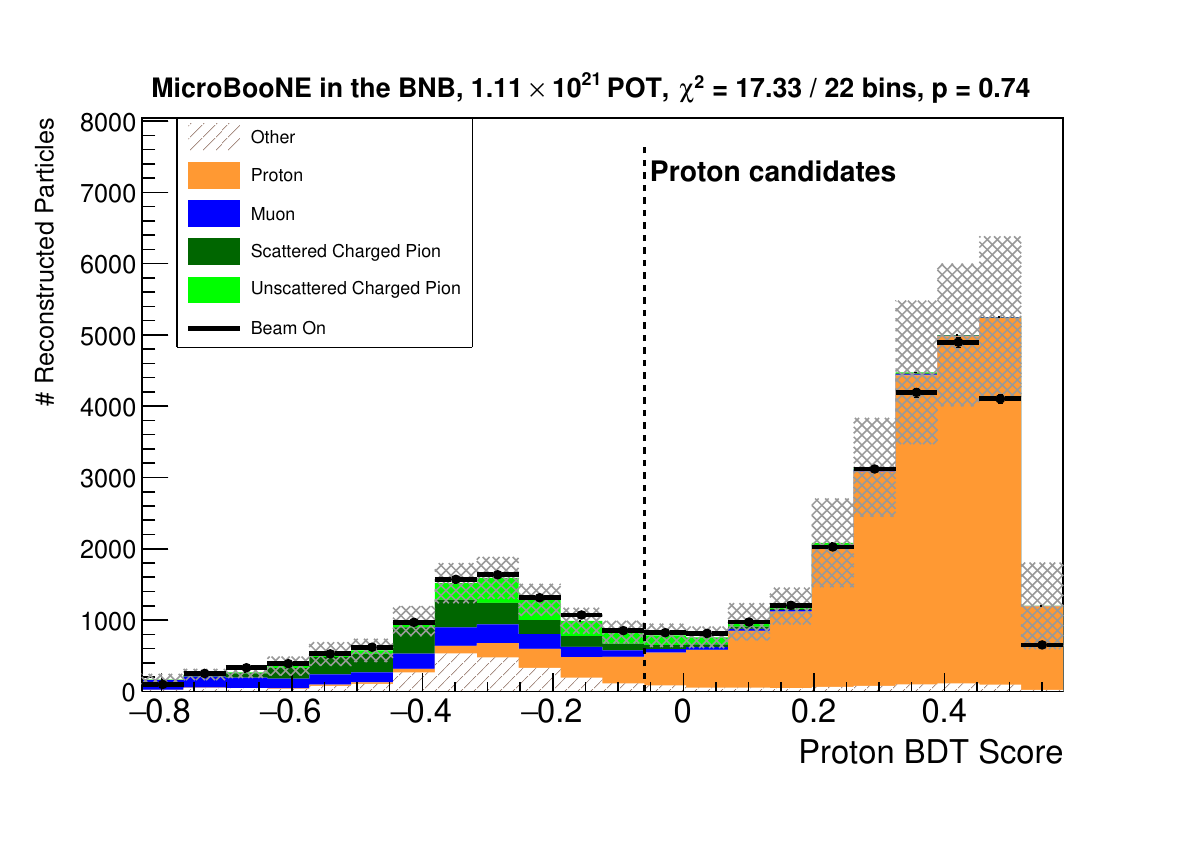} \\[-18pt]
    \makebox[0.49\textwidth][c]{\hspace*{7mm}(b) Proton BDT score}\\[6pt]
    \includegraphics[trim=0.3cm 0 1.8cm 1.0cm, clip, width=\columnwidth]{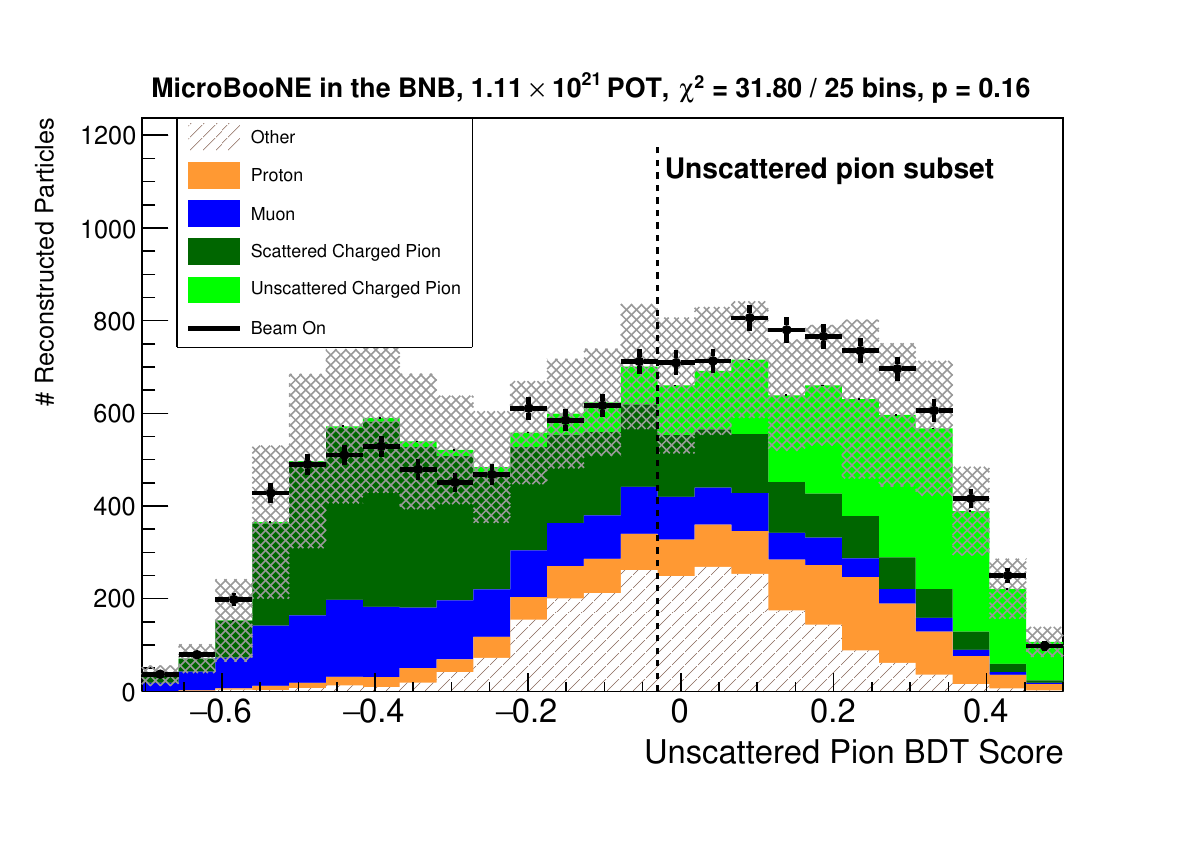} \\[-18pt]
    \makebox[0.49\textwidth][c]{\hspace*{7mm}(c) Unscattered pion BDT score}
  \end{tabular}

  \vspace{-4pt}
  \caption{Stacked histograms of muon, proton and unscattered pion \ac{BDT} scores of reconstructed particles. The plots show the distributions at the respective selection steps with all prior cuts applied as described in Sec.~\ref{sec:ccpiSelection} and \ref{sec:unsactteredPionSubset}. The particle with the highest muon \ac{BDT} score is used as the muon candidate in fully contained events. The proton \ac{BDT} plot then shows particles other than the selected muon candidates. Finally, the unscattered pion \ac{BDT} plot only shows pion candidates identified in the general selection.}
  \label{fig:bdts_combined}
\end{figure}

Figure~\ref{fig:bdts_combined} shows the \ac{BDT} score distributions for data and simulated events. Each distribution corresponds to the stage at which the respective \ac{BDT} is used in the selection, as described in the next section. Overall, good agreement is observed between data and simulation. Some shape differences can be seen in the protons at low muon and high proton \ac{BDT} scores. These differences are expected to arise from MicroBooNE's uncertainties in the modeling of calorimetric variables, primarily due to the treatment of recombination effects~\cite{Adams:2019ssg}, as well as from differences in proton multiplicity. While this leads to differences in the assigned \ac{BDT} scores at the extremes of the distributions, their impact on particle selection is minimal. The muon candidate is determined to be the particle with the highest muon \ac{BDT} score, making the selection insensitive to small local shifts at the low end. Similarly, discrepancies at high proton \ac{BDT} values do not influence the identification of pion candidates, which are selected from tracks with values below the cut value. %

\subsection{\texorpdfstring{\ccpi}{Single Pion Signal} Selection}
\label{sec:ccpiSelection}

The charged pion selection considers all events that pass the preselection and applies a series of selection cuts to identify events with a single charged pion, beginning with general quality cuts. First, a topological score computed by Pandora $>$ 0.67 is required to ensure that events clearly originate from a neutrino interaction. Next, the start points of all tracks must be within 9.5~cm of the reconstructed neutrino interaction vertex. This removes events where photons, particularly from \(\pi^0\) decays, and other backgrounds not directly attached to the neutrino vertex are misidentified as a pion candidate.

The first step of the topological selection is to identify one of the particles in the event as the muon candidate. Muons produced in \ac{CC} interactions tend to receive a significant fraction of the neutrino energy. Protons have a higher average ionization loss, \(\langle \mathrm{d}E/\mathrm{d}x \rangle\), compared to muons and charged pions. Hadrons also frequently scatter inelastically on the argon. As a result, muons typically have longer tracks and are much more likely to exit the detector. If there is one uncontained track, it is always identified as the muon candidate. If there are no uncontained tracks, the most muon-like track is chosen based on the highest score from the muon \ac{BDT}. Events with multiple uncontained tracks are rejected. Next, the proton \ac{BDT} is used to identify which of the remaining tracks are protons by requiring a high \ac{BDT} score. If exactly one particle remains that is not identified as either the muon or a proton, it is classified as the charged pion. Otherwise, the event is rejected.

Final quality cuts are applied to the identified particle candidates to improve the selection purity. The pion and muon candidates must have hits in all three wire planes to ensure they are not in an unresponsive region of the detector. A minimum threshold is enforced on the average \(\mathrm{d}E/\mathrm{d}x\) at the end of a pion candidate to suppress misidentified protons due to poor calorimetry. These protons originate mainly from reinteractions on argon or from traveling directly toward the wire planes.

\begin{figure*}[htbp]
    \centering

    \begin{minipage}[t]{0.49\textwidth} %
        \centering
        \includegraphics[width=\textwidth]{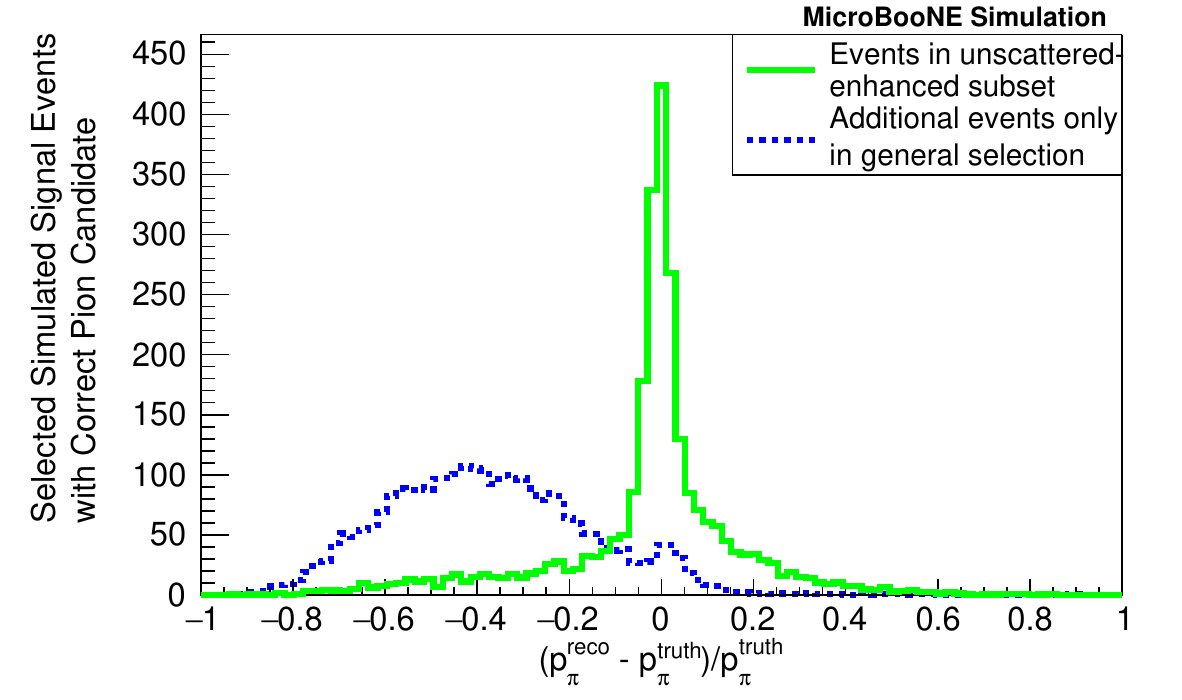}
        \caption{The relative difference between true and reconstructed pion momentum deviations as defined in Eq.~\ref{eq:resolution} for correctly identified pion candidates in selected signal events, with green representing events passing the unscattered pion BDT cut and blue representing all other events in the general selection. Together they show all events in the general selection.}
        \label{fig:momentumResolution_pi}
    \end{minipage}%
    \hfill
    \begin{minipage}[t]{0.49\textwidth} %
        \centering
        \includegraphics[width=\textwidth]{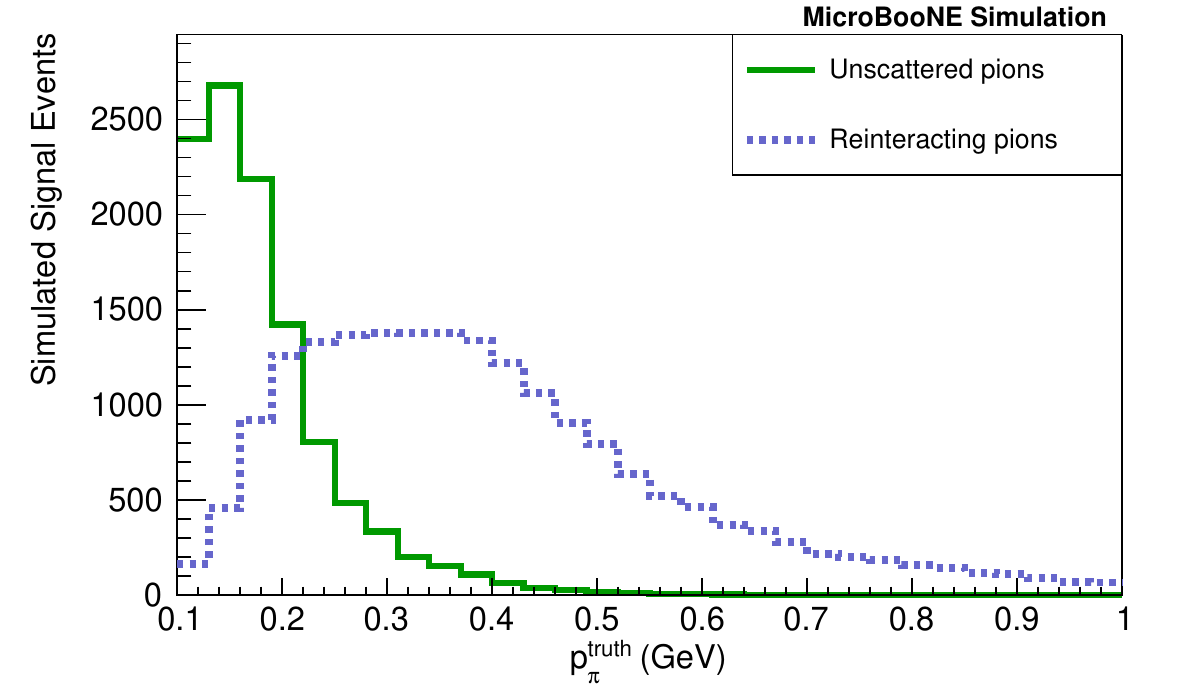}
        \caption{True pion momentum distribution of all signal events, showing contributions from true scattered and unscattered charged pions as predicted by GENIE + G\textsc{eant}4.}
        \label{fig:generic_vs_golden_truth_total}
    \end{minipage}%

    \vspace{1em}

    \begin{minipage}[t]{0.49\textwidth} %
        \centering
        \includegraphics[width=\textwidth]{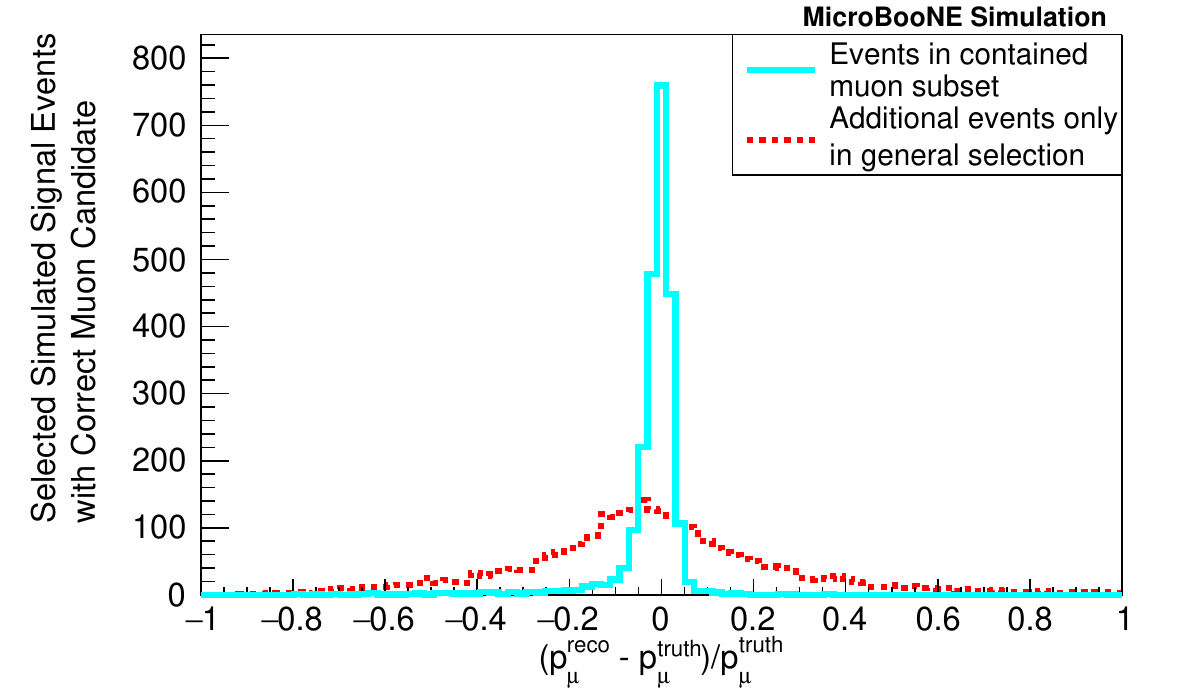}
        \caption{The relative difference between true and reconstructed muon momentum for correctly identified muon candidates in selected signal events, with cyan being contained and red being uncontained muons. Together they show all events in the general selection.}
        \label{fig:momentumResolution_mu}
    \end{minipage}
    \hfill
    \begin{minipage}[t]{0.49\textwidth} %
        \centering
        \includegraphics[width=\textwidth]{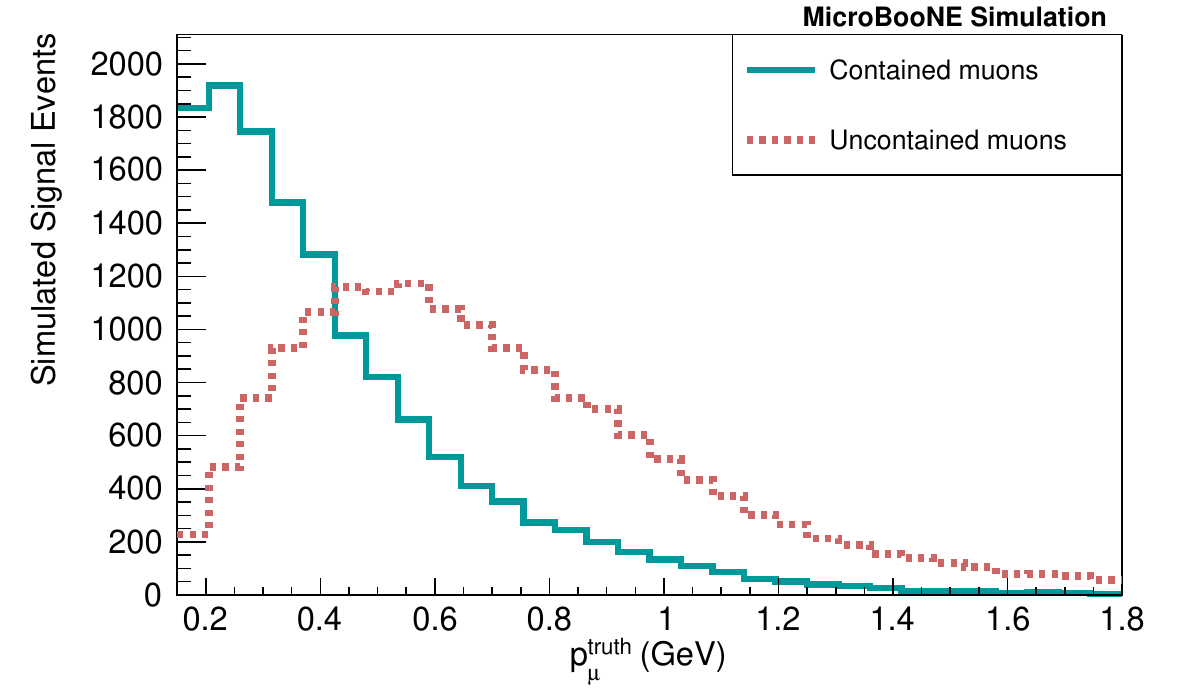}
        \caption{True muon momentum distribution of all signal events, distinguishing between true contained and uncontained muons as predicted by GENIE + G\textsc{eant}4.}
        \label{fig:contained_vs_uncontained_truth_total}
    \end{minipage}
\end{figure*}

\subsection{Unscattered Pion Subset}
\label{sec:unsactteredPionSubset}

The momentum of pions is measured using a subset of events rich in unscattered pions that have come to rest. These are identified by applying an additional cut on the unscattered pion \ac{BDT} score. This selection helps to exclude events where the pion is likely to have scattered on the argon. In simulation, the true scattering information is taken from the G\textsc{eant}4 transport model. The pion momentum for this subset is then determined from range. The resolution of the momentum estimation can be described as the standard deviation of the relative difference of the momentum $\delta_p$ for selected signal events:

\begin{equation}
 \delta_p = (p^\text{reco} - p^\text{truth})/p^\text{truth}
\label{eq:resolution}
\end{equation}
\noindent where $p^\text{reco}$ here is the reconstructed momentum for a correctly identified pion track and $p^\text{truth}$ is the simulated momentum. This yields a resolution of $\sigma = 0.20$ for the pion momentum estimator applied to the simulated subset.

Figure~\ref{fig:momentumResolution_pi} shows the differences between the reconstructed and true charged pion momentum for simulated events that pass or fail the unscattered pion \ac{BDT} cut. The same range-based momentum estimation is applied to all events, with the events not passing this cut exhibiting a biased and down-smeared distribution, which arises from momentum being underestimated for hadronically reinteracting charged pions. Excluding scattered pions is therefore necessary to accurately estimate the momentum from the track length.

The signal definition remains the same for the cross-section measurement performed as a function of pion momentum, but the selected unscattered pions are generally of lower energy. Figure~\ref{fig:generic_vs_golden_truth_total} presents the momentum distribution for scattered and unscattered $\pi^\pm$. The extractable pion momentum differential cross section using unscattered pions is thus limited to lower energies.

\subsection{Contained Muon Subset}
\label{sec:containedMuonSubset}

Some previous MicroBooNE analyses have used both range-based and \ac{MCS} muon momentum estimation techniques \cite{MicroBooNE:2019nio, MicroBooNE:2020fxd, MicroBooNE:2023cmw, MicroBooNE:2024bnl}. Figure~\ref{fig:momentumResolution_mu} compares the distributions for contained muons using range-based estimation and uncontained muons, for which only \ac{MCS} estimation is possible. The momentum estimation from range for contained muons provides a much higher resolution of $\sigma = 0.08$, compared to uncontained muons where \ac{MCS} is employed, yielding $\sigma = 0.29$. Because of this, the analysis relies solely on the more accurate range-based method \cite{Abratenko:2017nki} for the cross-section measurement as a function of muon momentum. However, this requires full containment, reducing the number of considered events for this differential cross section. As in Sec.~\ref{sec:unsactteredPionSubset}, the signal definition remains unchanged, but the muon momentum distribution is shifted lower for fully contained events, as shown in Fig.~\ref{fig:contained_vs_uncontained_truth_total}.

\subsection{Selection Performance}

The selection performance is assessed in terms of efficiency, purity, and the ability to correctly select unscattered pions. Efficiency is defined as the fraction of signal events that are selected. Purity is the fraction of selected events that are signal events. The unscattered charged pion fraction is the fraction of selected signal events in which the charged pion is unscattered in the simulation. Table~\ref{table:combinedSelection} shows these metrics and event rates for all three selections. The selections have similar purity between 49.5-56.1\%. The tradeoff for excluding uncontained muon tracks is a reduction in efficiency, since a large fraction of muons exit the detector. The same holds for the unscattered charged pion subset, which trades some efficiency for a higher fraction of unscattered pions.

\begin{table*}[htbp] %
    \centering
    \renewcommand{\arraystretch}{1.5} %
    \begin{tabular}{|c||c|c|c|c|c|}
    \hline
    \multirow{2}{*}{\textbf{Selections}} & \multicolumn{2}{c|}{\textbf{Selected Event Counts}} & \multirow{2}{*}{\textbf{Efficiency}} & \multirow{2}{*}{\textbf{Purity}} & \multirow{2}{*}{\makecell{\textbf{Unscattered}\\\textbf{$\boldsymbol{\pi^{\pm}}$ fraction}}} \\ \cline{2-3}
                      &               \textbf{Data} & \textbf{Prediction} & & & \\ \hline
    \textbf{General}                         & 12566 & \(11949~\pm~2426\)~(syst. + stat.) & \(20.3 \pm 1.1\)\% & \(52 \pm 9\)\% & \(36 \pm 5\)\% \\ \hline
    \makecell[l]{\textbf{Unscattered pion}}  & 6535  & \(5843~\pm~1120\)~(syst. + stat.) & \(9.5 \pm 1.0\)\% & \(50 \pm 11\)\% & \(68 \pm 7\)\% \\ \hline
    \makecell[l]{\textbf{Contained muon}}    & 4867  & \(4547~\pm~931\)~(syst. + stat.) & \(8.4 \pm 0.7\)\% & \(56 \pm 9\)\% & \(39 \pm 4\)\% \\ \hline
    \end{tabular}
    \caption{The number of data and predicted events selected by the general and the two subset selections. The prediction includes all statistical and systematic uncertainties, including uncertainties on the predicted signal. Also shown are the efficiency, purity, and the fraction of unscattered pion events for the prediction.}
    \label{table:combinedSelection}
\end{table*}

In addition to selecting events with the correct set of final state particles, the analysis must accurately distinguish between muon and pion candidates. Table~\ref{tab:track_length_comparison} highlights the strong dependence of this identification on the relative track lengths of the two particles. Specifically, when the pion track is longer than the muon track, the distinguishing power between them is limited. However, such events constitute a small fraction of the predicted signal, and the identification performance is good when the muon track length exceeds the pion track length. In these events, the identification accuracy ranges from 86\% - 99\% for the muon and charged pion subsets. Exiting particles tend to have higher energies, as their tracks are longer on average. As Figs.~\ref{fig:generic_vs_golden_truth_total} and~\ref{fig:contained_vs_uncontained_truth_total} show, higher energy tracks are largely muons. Thus, using exiting tracks as the muon candidate results in the high observed identification accuracy. Overall, this selection provides robust separation in the phase space regions where most events reside.

\begin{table}[htbp]
    \centering
    \renewcommand{\arraystretch}{1.5}
    \begin{tabular}{|c|c|c|}
        \cline{2-3}
        \multicolumn{1}{c|}{} & \multicolumn{2}{c|}{\boldsymbol{$R_{\mu} > R_{\pi}$}} \\
        \cline{2-3}
        \multicolumn{1}{c|}{} & \textbf{No} & \textbf{Yes} \\
        \hline
        \multicolumn{3}{|c|}{\textbf{Fraction of Correct Muon Candidates}}\\
        \hline
        \textbf{Contained Muons} & 53\% of 408 & 86\% of 2396 \\
        \hline
        \textbf{Uncontained Muons} & 95\% of 76 & 99\% of 3319 \\
        \hline
        \hline
        \multicolumn{3}{|c|}{\textbf{Fraction of Correct Pion Candidates}}\\
        \hline
        \textbf{Unscattered Pions} & 59\% of 398 & 88\% of 3554 \\
        \hline
        \textbf{Scattered Pions} & 65\% of 86 & 95\% of 2162 \\
        \hline
    \end{tabular}
    \caption{Fractions of correctly identified muon and pion candidates from selected GENIE signal events, categorized by whether the true track length $R$ of the muon exceeds that of the pion. Results are shown for contained and uncontained muons as well as scattered and unscattered charged pions as simulated by G\textsc{eant}4. The number of simulated events in each category is also shown.}
    \label{tab:track_length_comparison}
\end{table}

\begin{figure*}[htbp]
  \centering
  \setlength{\tabcolsep}{0pt} %
  \begin{tabular}{cc}
    \includegraphics[width=0.49\textwidth]{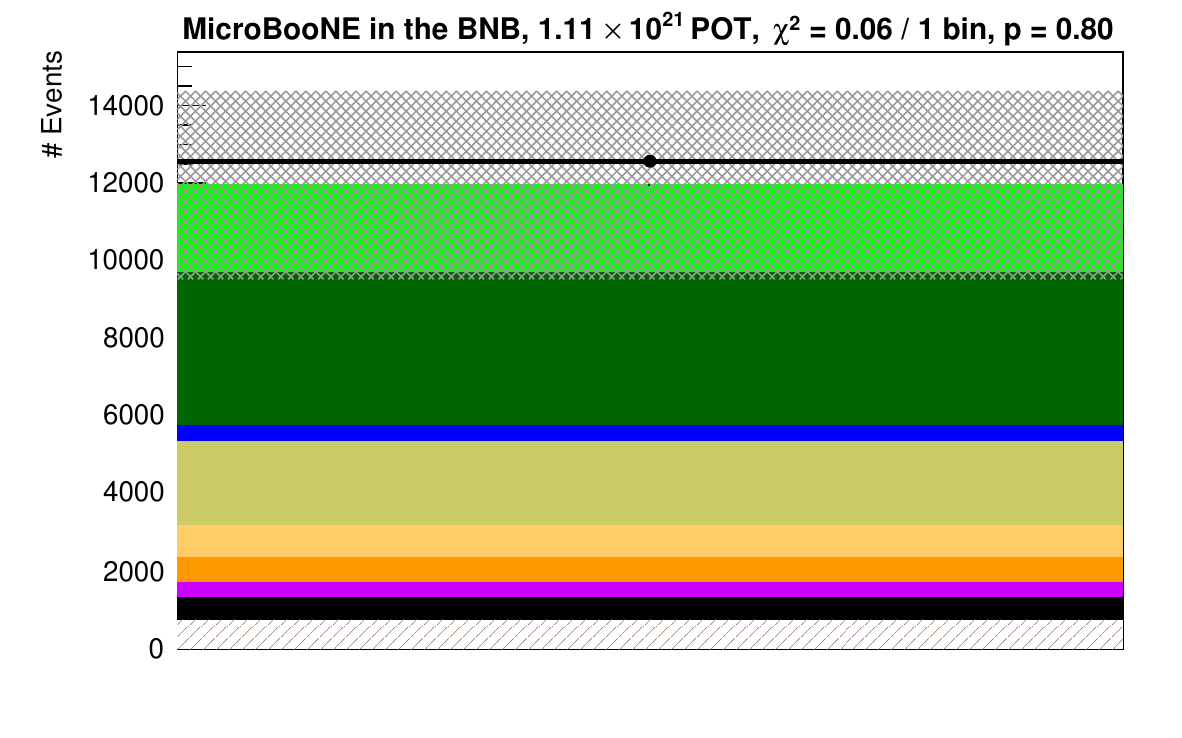} &
    \includegraphics[width=0.49\textwidth]{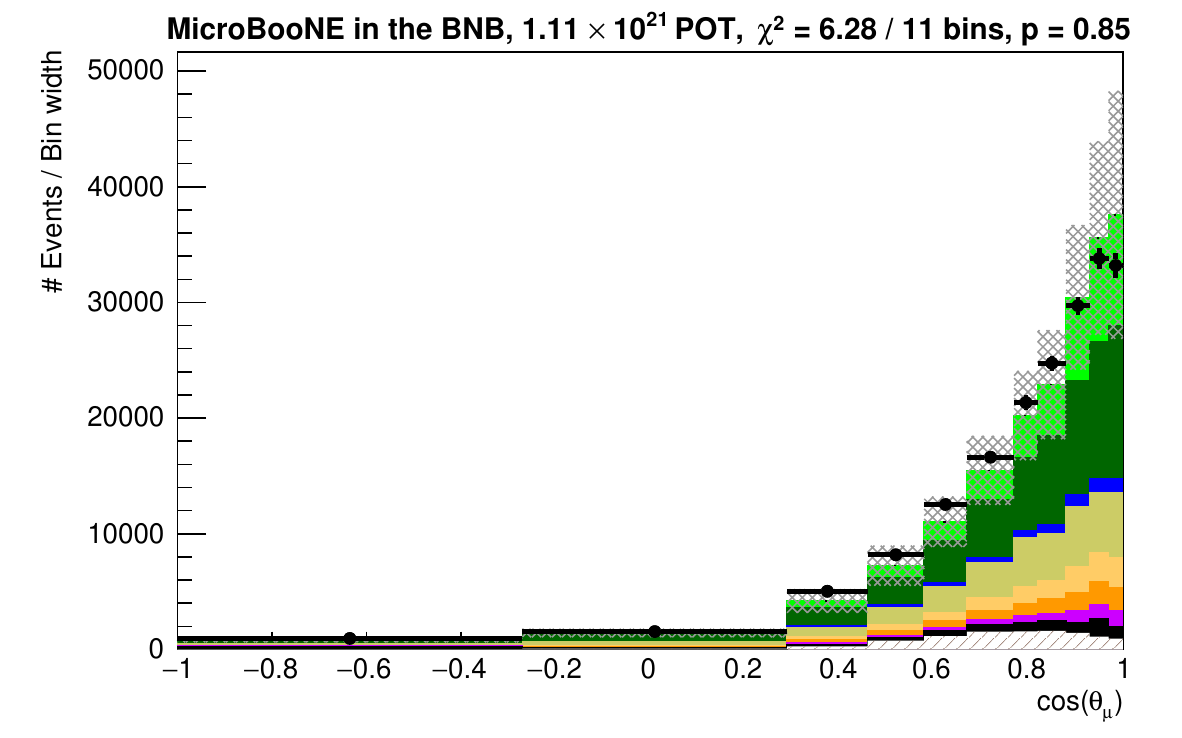} \\[-10pt]
    \makebox[0.49\textwidth][c]{\hspace*{5mm}(a) Total}  &
    \makebox[0.49\textwidth][c]{\hspace*{8mm}(b) Cosine of the muon angle} \\[6pt]
    \includegraphics[width=0.49\textwidth]{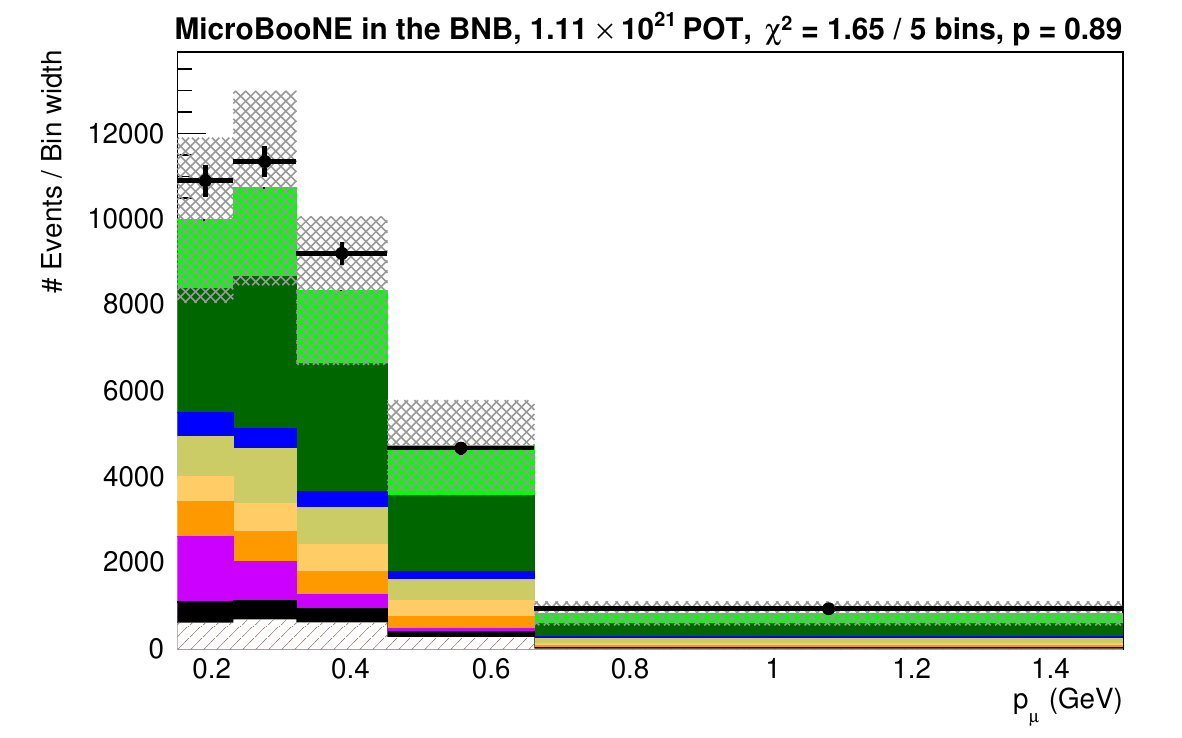} &
    \includegraphics[width=0.49\textwidth]{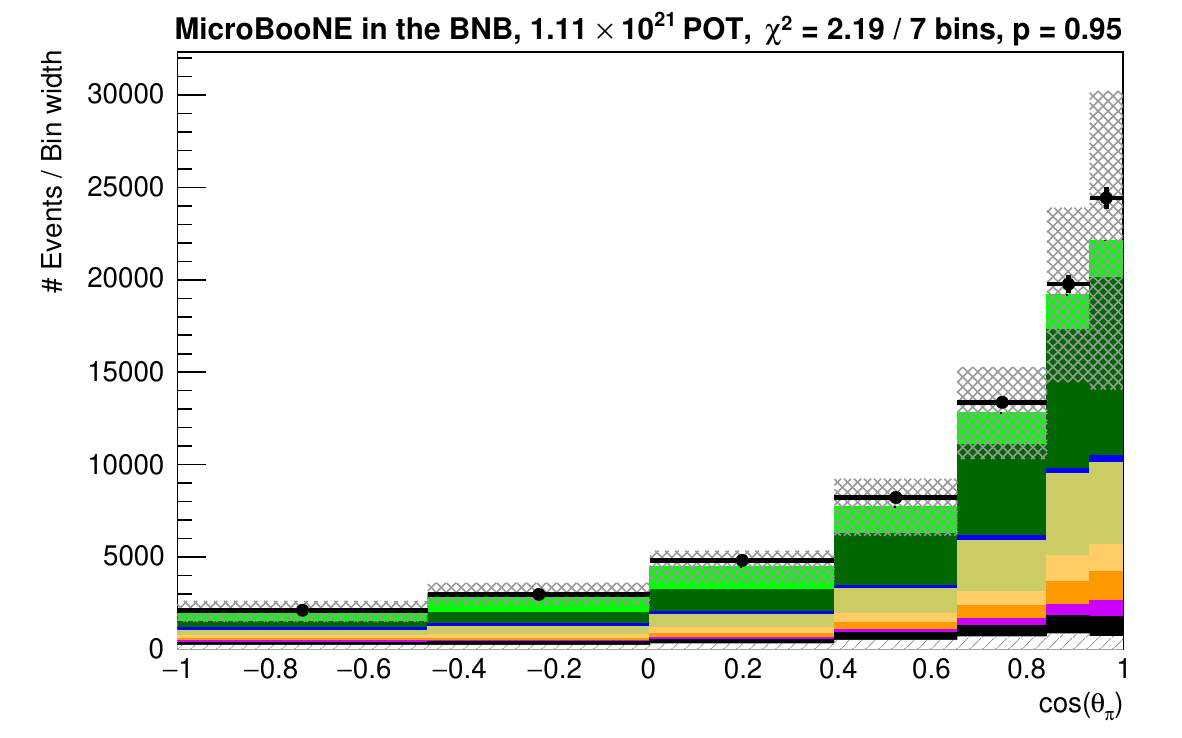} \\[-10pt]
    \makebox[0.49\textwidth][c]{\hspace*{5mm}(c) Muon momentum}  &
    \makebox[0.49\textwidth][c]{\hspace*{8mm}(d) Cosine of the pion angle} \\[6pt]
    \includegraphics[width=0.49\textwidth]{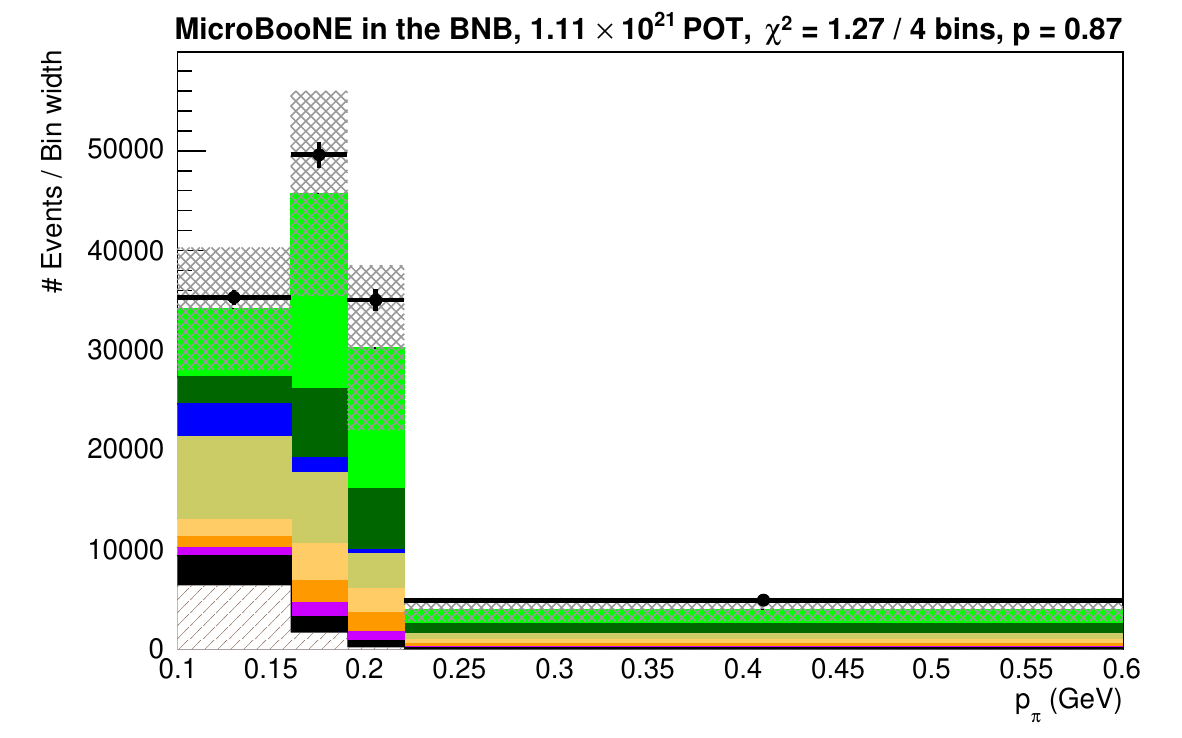} &
    \includegraphics[width=0.49\textwidth]{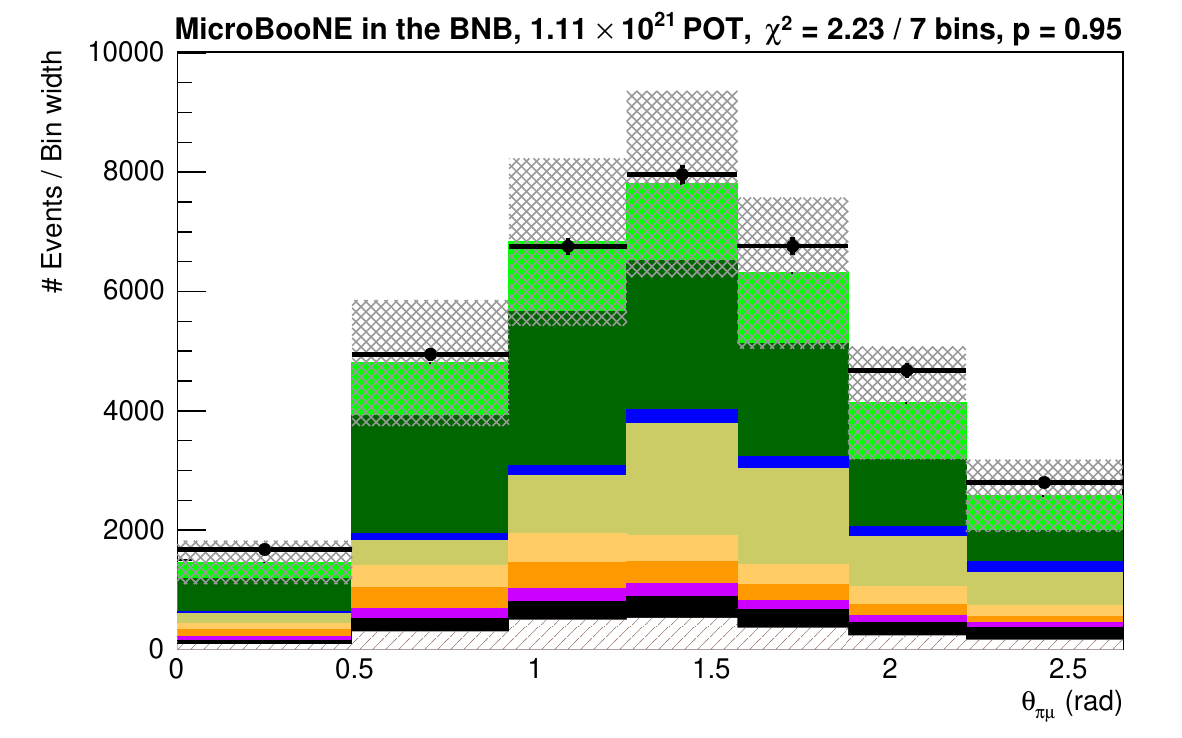} \\[-10pt]
    \makebox[0.49\textwidth][c]{\hspace*{5mm}(e) Pion momentum}  &
    \makebox[0.49\textwidth][c]{\hspace*{8mm}(f) Muon–pion opening angle}
  \end{tabular}

  \vspace{6pt}
  \fbox{\includegraphics[width=0.9\textwidth]{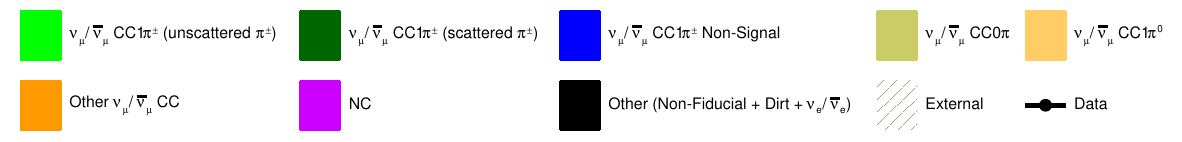}}
  \caption{The stacked histogram shows the predicted rate of events passing the selection. The grey grid is the total uncertainty on the signal plus background prediction. Black dots are measured event rates with statistical uncertainties. `External' are selected cosmic background events from the pure beam-off samples. The plots are: (a) the total event rate, (b) the scattering angle between the muon and the neutrino beam, (c) the muon momentum using the contained muon subset, (d) the scattering angle between the pion and the neutrino beam, (e) the pion momentum using the unscattered pion subset and (f) the opening angle between the muon and the pion. The last bins of the muon and pion momentum selections serve as overflow bins. The dark and light green stacks are scattered and unscattered charged pion signal events. The \textit{$\nu_{\mu}$ \ccpi Non‑Signal} events are selected events with the right topology but outside the defined phase space. The \textit{Other} category is mainly non-fiducial events, with minor contributions from dirt and electron (anti\mbox{-})neutrinos.}
  \label{fig:reco_combined}
\end{figure*}

Figure~\ref{fig:reco_combined} shows predicted and measured event counts across the five kinematic variables measured in the analysis, as well as for the total integrated event rate. The pion and muon momentum distributions are shown using their respective selection subsets. The largest background contribution comes from $\nu_\mu$ \ac{CC}0$\pi$Np with $N \ge 1$. In these pionless events, a proton is misidentified as a charged pion. The second largest background comes from cosmic ray interactions being misidentified as the neutrino interaction. This is followed by a mix of other \ac{CC} $\nu_\mu$ topologies, including $\pi^0$ and out-of-phase-space single charged pion events. The background compositions are slightly different for the additional selections, most prominently for the contained muon subset, where \ac{NC} events play a bigger role in the low momentum bins. Data-simulation agreement is good for both the total and differential observables.

\subsection{Background Modeling}
\label{sec:sideband}

\begin{figure}[htbp]
    \centering
    \includegraphics[width=\columnwidth]{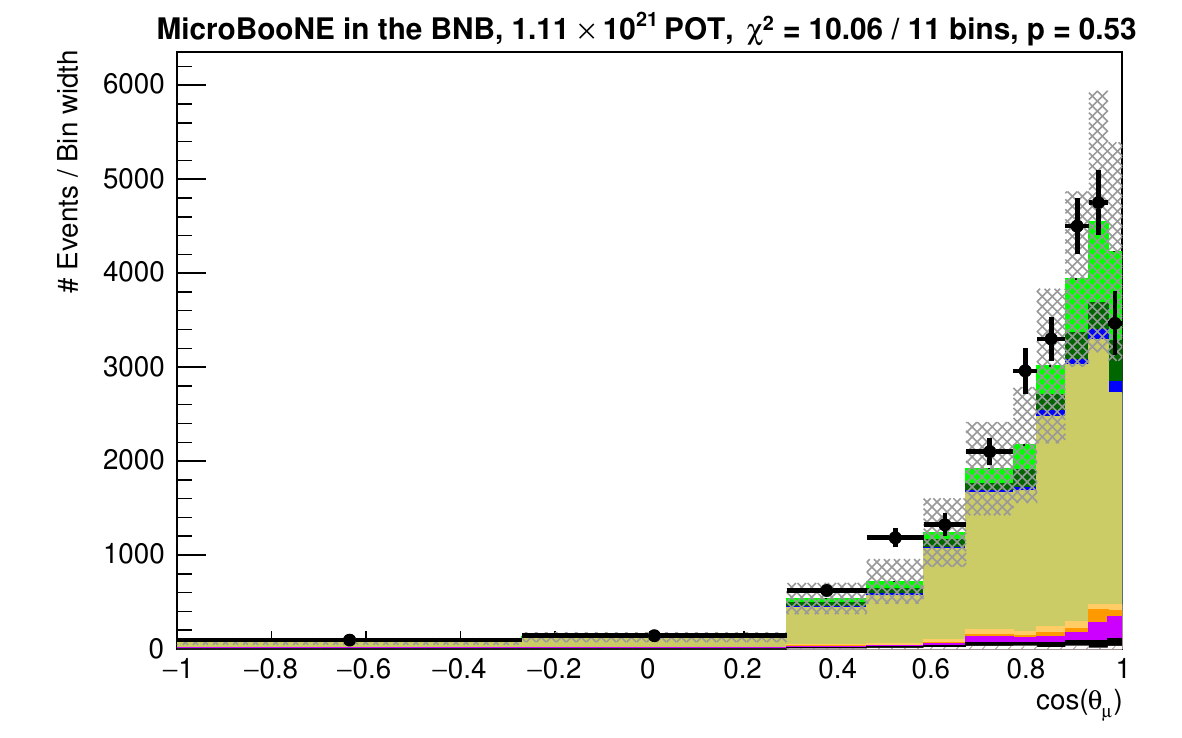}
    \caption{Stacked histogram plot of events passing the sideband selection using the same binning as the main selection for the muon angle. The legend is the same as shown in Fig.~\ref{fig:reco_combined}.}
    \label{fig:sidebandMuAngle}
\end{figure}

The dominant background of the analysis is muon neutrino \ac{CC} interactions with zero pions arising primarily from proton tracks misidentified as charged pions. To verify that the background is adequately modeled, an orthogonal sideband selection is employed by modifying the cuts used in the main selection. Instead of requiring one muon, any number of identified protons, and one other particle (the charged pion candidate), the sideband selection targets events with one muon, one or more protons, and no other particles. To better match the kinematics of protons misidentified as pions in selected background events, additional cuts are applied to select events with more \ac{MIP}-like tracks. Specifically, proton candidates are required to have a low proton \ac{BDT} score close to the selection cut value and to pass a minimum muon \ac{BDT} score requirement. Other cuts, including the one used to define the unscattered pion subset, are applied using the proton candidate, treating it as the pion candidate in the main selection. In cases with multiple protons, the one with the longest track is used. This track is also used to compute the pion kinematic variables. The sideband selection uses the same binning and uncertainty treatment as the main selection and all comparisons can be found in the supplemental material~\cite{SupplementalMaterial}. Of the simulated events selected by the sideband for the generic selection, 19\% are signal events with a single pion. Fig.~\ref{fig:sidebandMuAngle} shows the muon angle comparison, highlighting known GENIE modelling differences with MicroBooNE data at forward angles. These discrepancies are dominated by low-\(Q^2\) \ac{CC} events with protons and no pions, as reported in Ref.~\cite{MicroBooNE:2020akw}. However, the sideband data-simulation comparison shows very good agreement within uncertainties across all kinematic variables, indicating that differences are well covered by the included uncertainties.

\section{\label{sec:Uncertainties}Uncertainties}%
The main sources of systematic uncertainty in this analysis arise from the modeling of the neutrino flux, neutrino interactions on argon, particle propagation in the detector, and detector response.

A multiple-universe approach is used to estimate the systematic uncertainties by comparing alternative predictions (universes) against the central value and using this to compute a covariance matrix: 
\begin{equation}
    V_{a b} = \frac{1}{N_{\text{univ}}} \sum_{k=1}^{N_{\text{univ}}} \left( n_a^k - n_a^{CV} \right) \left( n_b^k - n_b^{CV} \right),
\end{equation}
\noindent where $n_a^k$ and $n_b^k$ are the number of events in bins $a$ and $b$ of universe $k$, $CV$ denotes the \acl{CV} universe, and $N_{\text{univ}}$ is the number of universes. Uncertainties on the prediction also arise from the finite statistics of the simulated events and the measured beam-off data, which are combined into a diagonal covariance matrix, \(V_{a b}^\text{Stat.}\). The total covariance matrix is then the sum of the individual systematic and statistical uncertainties:
\begin{equation}
    V_{a b}^\text{Total} = V_{a b}^\text{Stat.} + \sum^\text{Syst.}_i V_{a b}^i,
\end{equation}
\noindent with the following sets of systematic uncertainties considered:

\textbf{Flux:} Uncertainties on hadron production from proton interactions with the beam target, as well as beam simulation uncertainties related to the focusing horn and cross-section uncertainties on secondary interactions, are all propagated together by reweighting events \cite{MicroBooNE:2018efi}. 

\textbf{Interaction model:} Interaction model uncertainties are assessed for most of the considered model parameters by randomly and simultaneously varying them within their Gaussian uncertainties and reweighting events based on this. Some parameters are individually varied across one or two alternative predictions, as discussed in Ref.~\cite{MicroBooNE:2021ccs}. 

\textbf{Reinteraction:} The Geant4Reweight framework \cite{Calcutt:2021zck} is used to reweight events to account for systematic uncertainties on the modeling of charged pion and proton transport in argon.

\textbf{Detector:} Separate samples are used to vary components of the detector simulation, including the light collection system, the wire response simulation, space charge effects, and electron-ion recombination of the drift electrons \cite{MicroBooNE:2021roa}. The last one in particular is a significant source of uncertainty in this analysis, as the particle identification relies heavily on \(\mathrm{d}E/\mathrm{d}x\).

\textbf{POT:} A flat 2\% normalization uncertainty is used to account for the uncertainty on the protons delivered to the proton-beam target. This uncertainty stems from the observed disagreement between the two toroids that monitor the beam \cite{MiniBooNE:2008hfu}.

\textbf{Target:} A 1\% normalization uncertainty is assumed on the number of argon nuclei in the fiducial volume of the detector.

\begin{table}[htp]
  \centering
  \renewcommand{\arraystretch}{1.1}
  \begin{tabular}{|lr|}
    \hline
    \textbf{Category}      & \textbf{Uncertainty (\%)} \\ 
    \hline
    Total                  & 21.4                    \\ 
    \hspace{0.8em}Systematic      & 21.3                    \\ 
    \hspace{1.6em}Flux            & 14.1                    \\ 
    \hspace{1.6em}Interaction model & 12.6                    \\ 
    \hspace{1.6em}Detector        &  9.1                    \\ 
    \hspace{1.6em}POT             &  3.3                    \\ 
    \hspace{1.6em}Target          &  1.6                    \\ 
    \hspace{1.6em}Reinteraction   &  1.5                    \\ 
    \hspace{0.8em}Statistical     &  2.0                    \\ 
    \hspace{1.6em}Data            &  1.6                    \\ 
    \hspace{1.6em}Background      &  1.1                    \\ 
    \hline
  \end{tabular}
  \caption{Fractional uncertainties for the extracted total cross section in Fig.~\hyperref[fig:unfolded_combined]{\ref*{fig:unfolded_combined}a}. The background statistical uncertainty contains the simulated neutrino prediction and the beam-off sample.}
  \label{tab:fractionalUncertainties}
\end{table}

The dominant sources of uncertainty are flux, followed by interaction model and detector uncertainties. Variations exist across the different differential variables, with interaction model uncertainties, for instance, playing a larger role at backward muon angles and low pion momenta. Uncertainty distribution plots can be found in the supplemental material~\cite{SupplementalMaterial}. Table~\ref{tab:fractionalUncertainties} shows the breakdown of the fractional uncertainties for the extracted total cross section.

\section{\label{sec:extraction}Cross-Section Extraction}%
The cross sections presented in this work are given in a regularized truth space, obtained through an unfolding procedure applied to the measured event rates. This allows for direct comparisons between different experiments and facilitates interpretation in the context of theoretical models. The cross section in the regularized truth-space bin $a$ is defined as:

\begin{equation}
    \left\langle \frac{d\sigma}{dx} \right\rangle_a = \frac{\sum_\beta U_{a\beta} (D_\beta - B_\beta)}{N_{\text{Ar}} \times \Phi \times \Delta x_a}
\end{equation}
\noindent where $D_\beta$ and $B_\beta$ represent the selected data events and the background prediction, including cosmic backgrounds, in the measured bin $\beta$; $N_{\text{Ar}}$ is the number of argon nuclei within the fiducial volume; $\Phi$ denotes the integrated flux; $\Delta x_a$ is the width of bin $a$ in the kinematic variable $x$. The term $U_{a\beta}$ is the unfolding matrix and corrects for both bin-migration effects and selection efficiencies. When applied to the background-subtracted data, the unfolding matrix provides an estimate of the true signal event counts for each bin.

Determining the unfolding matrix is a non-trivial task. A straightforward inversion of the response matrix, which maps predicted signal events from truth space bins to measured bins, can amplify noise caused by statistical fluctuations and systematic uncertainties. This noise amplification leads to oscillations in the unfolded spectrum. To address this, regularization techniques are applied to all differential cross sections, which enforce constraints to stabilize the unfolding. These constraints trade some variance for a controlled bias toward a prediction from simulation. In this work, the Wiener-SVD technique with a first derivative regularization term is used to obtain an unfolding matrix \cite{Tang:2017rob}. The regularization effects are encapsulated in a regularization matrix $A_C$ and the unfolded data results for the kinematic variables live in the regularized truth space described by that matrix. To compare truth predictions from generators to the unfolded kinematic variable distributions, $A_C$ needs to be applied first. This has been done for all differential cross section comparisons shown in this analysis. The regularization matrix is provided in the supplemental material \cite{SupplementalMaterial}.

This analysis takes a blockwise approach to unfolding and reports a full covariance matrix between all measured bins across all observables in the supplemental material \cite{SupplementalMaterial}. Details of the method are described in Ref.~\cite{Gardiner:2024gdy}.

The robustness of the unfolding and regularization is evaluated using a fake data study, in which events produced with the interaction generator NuWro \cite{Golan:2012rfa} are treated as real data. Since this NuWro sample is produced with the same pipeline as the GENIE simulation, the unfolding is performed with model and statistical uncertainties only. The extracted fake data cross sections show good $\chi^2$ agreement with the true NuWro distributions and are also provided in the supplemental material \cite{SupplementalMaterial}.

\section{\label{sec:results}Results}%

\begin{figure*}[p]
  \centering
  \setlength{\tabcolsep}{0pt} %
    
  \begin{tabular}{cc}
    \includegraphics[width=0.49\textwidth]{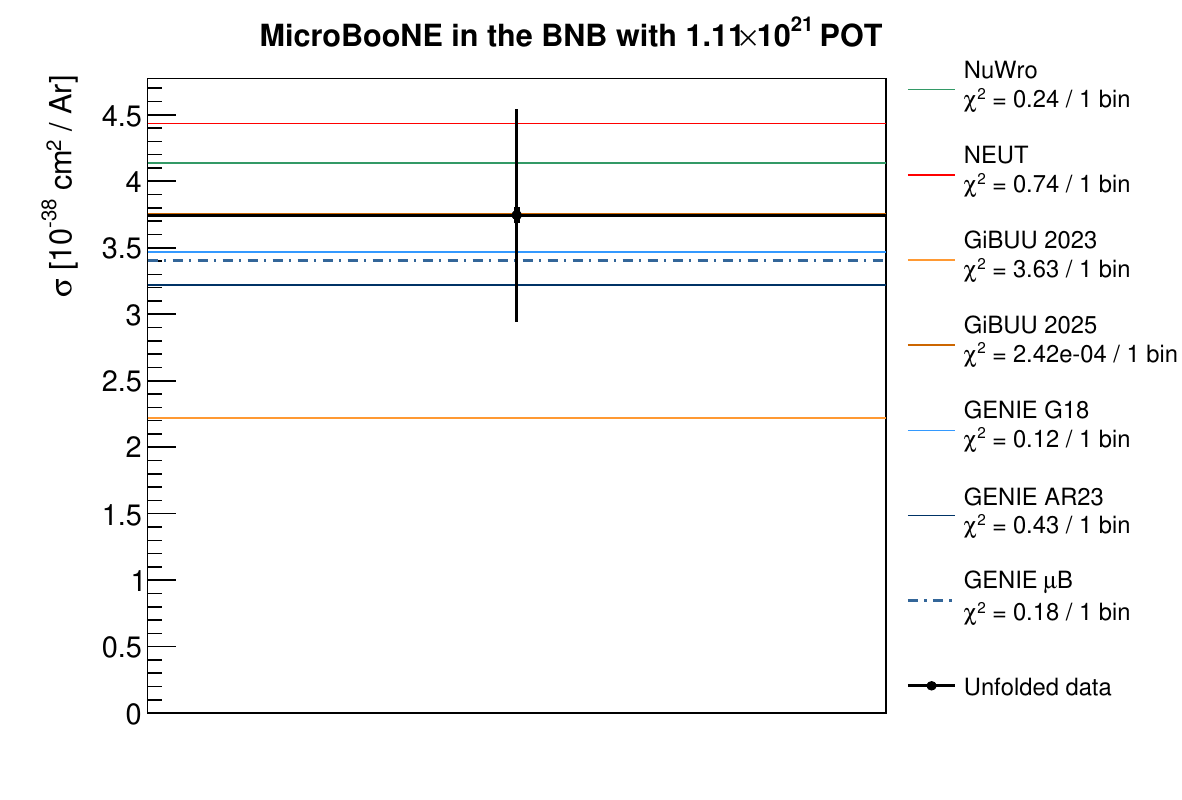} &
    \includegraphics[width=0.49\textwidth]{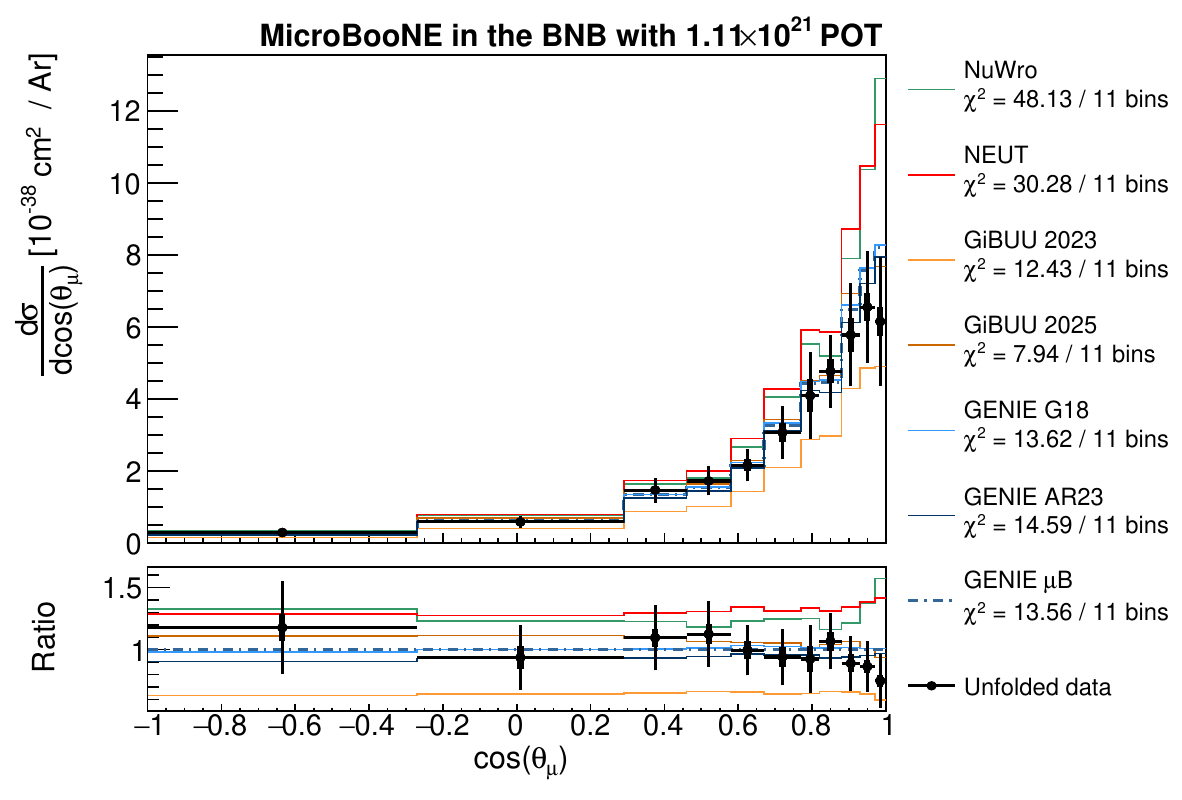} \\[-3pt]
    \makebox[0.49\textwidth][c]{\hspace*{-15mm}(a) Total}  &
    \makebox[0.49\textwidth][c]{\hspace*{-10mm}(b) Cosine of the muon angle} \\[6pt]
    \includegraphics[width=0.49\textwidth]{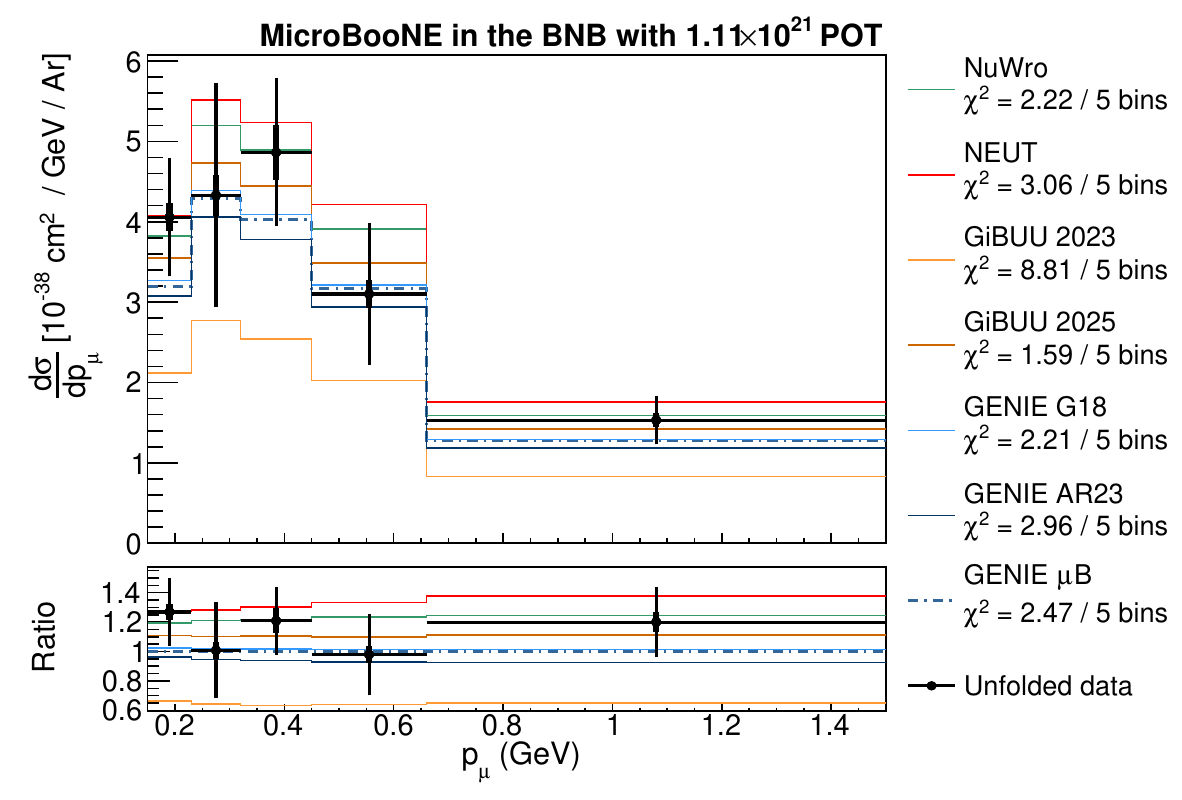} &
    \includegraphics[width=0.49\textwidth]{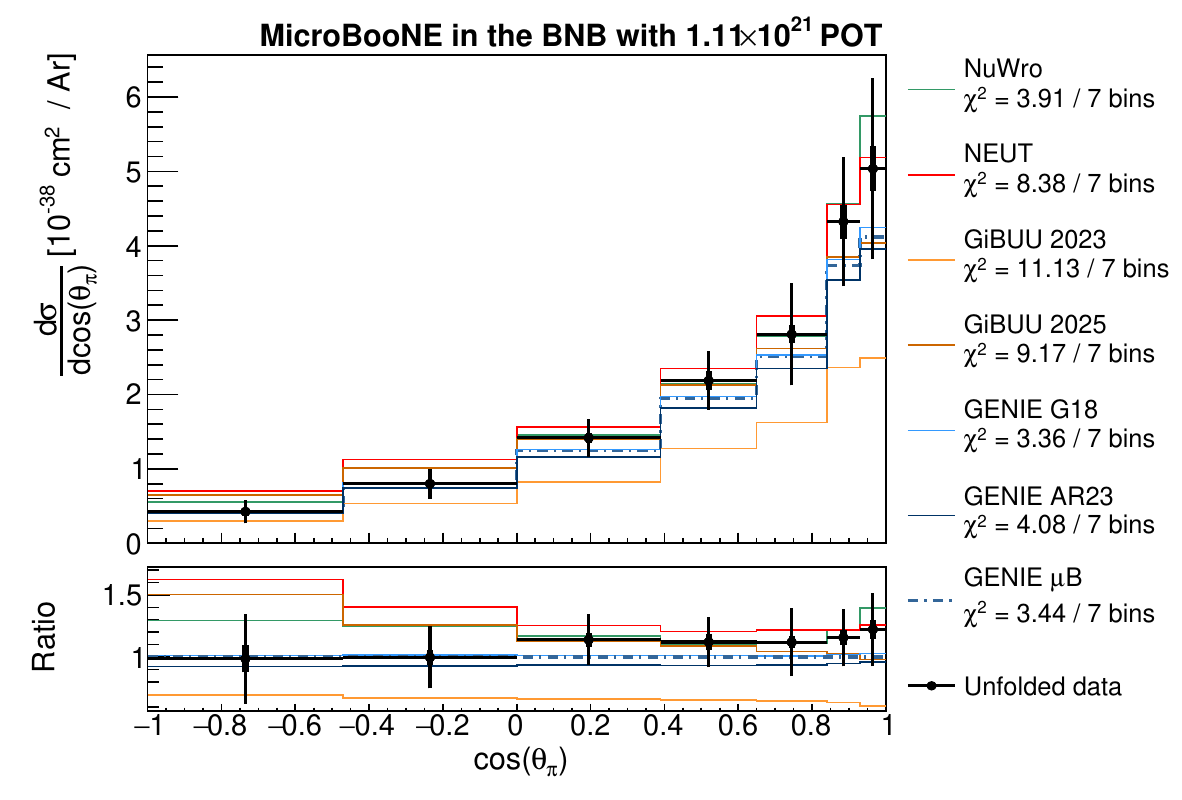} \\[-3pt]
    \makebox[0.49\textwidth][c]{\hspace*{-15mm}(c) Muon momentum}  &
    \makebox[0.49\textwidth][c]{\hspace*{-10mm}(d) Cosine of the pion angle} \\[6pt]
    \includegraphics[width=0.49\textwidth]{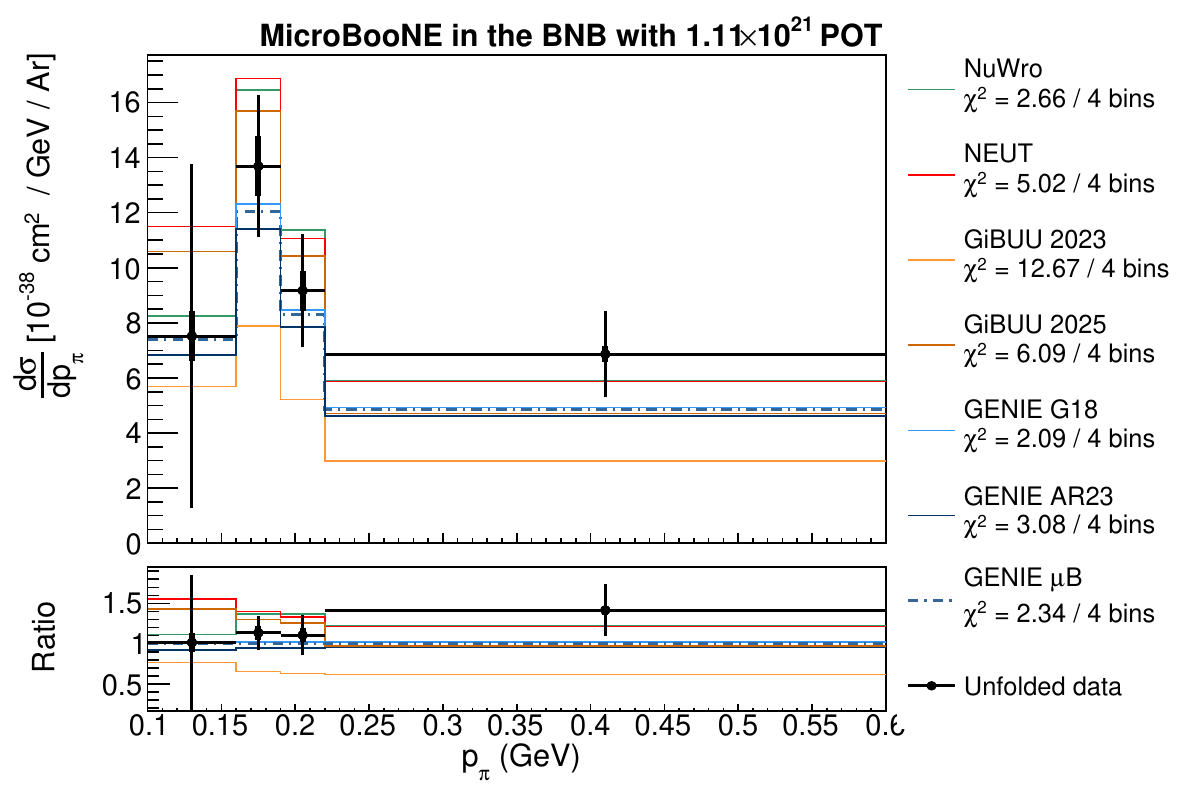} &
    \includegraphics[width=0.49\textwidth]{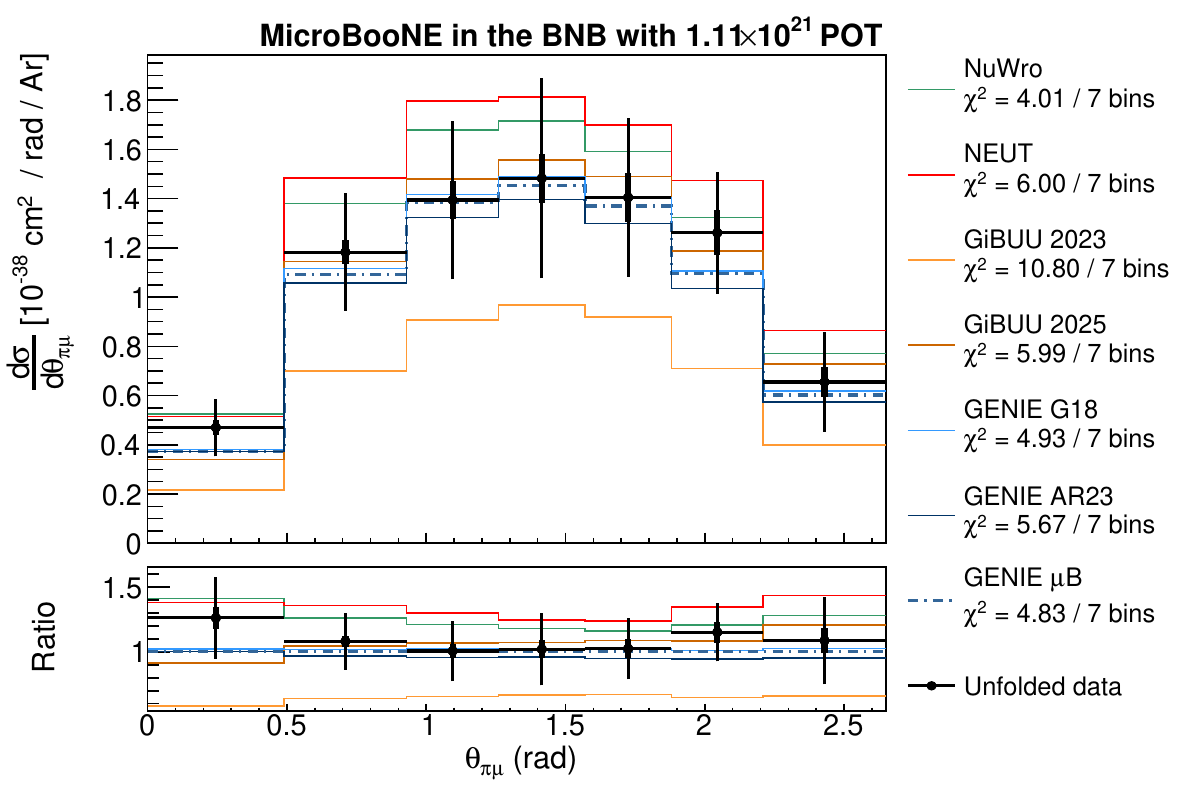} \\[-3pt]
    \makebox[0.49\textwidth][c]{\hspace*{-15mm}(e) Pion momentum}  &
    \makebox[0.49\textwidth][c]{\hspace*{-10mm}(f) Muon–pion opening angle}
  \end{tabular}

  \caption{Unfolded cross-section results. The five differential cross sections are in regularized truth space with the regularization matrix \(A_c\) applied to the predictions. The single bin total cross-section measurement is not regularized. For the particle momentum kinematic variables, the dedicated selection subsets are used. The black dots represent the unfolded selection with the associated total uncertainty. The thicker inner error bars are the statistical uncertainty only. The dashed blue line is the GENIE \ac{CV} truth prediction, and the ratio plots are shown with respect to it. The other lines are predictions from additional generators and tunes. The line of the GiBUU 2025 prediction is under the data for the total cross section.}
  \label{fig:unfolded_combined}
\end{figure*}

\begin{figure}
    \centering
    \includegraphics[width=\columnwidth]{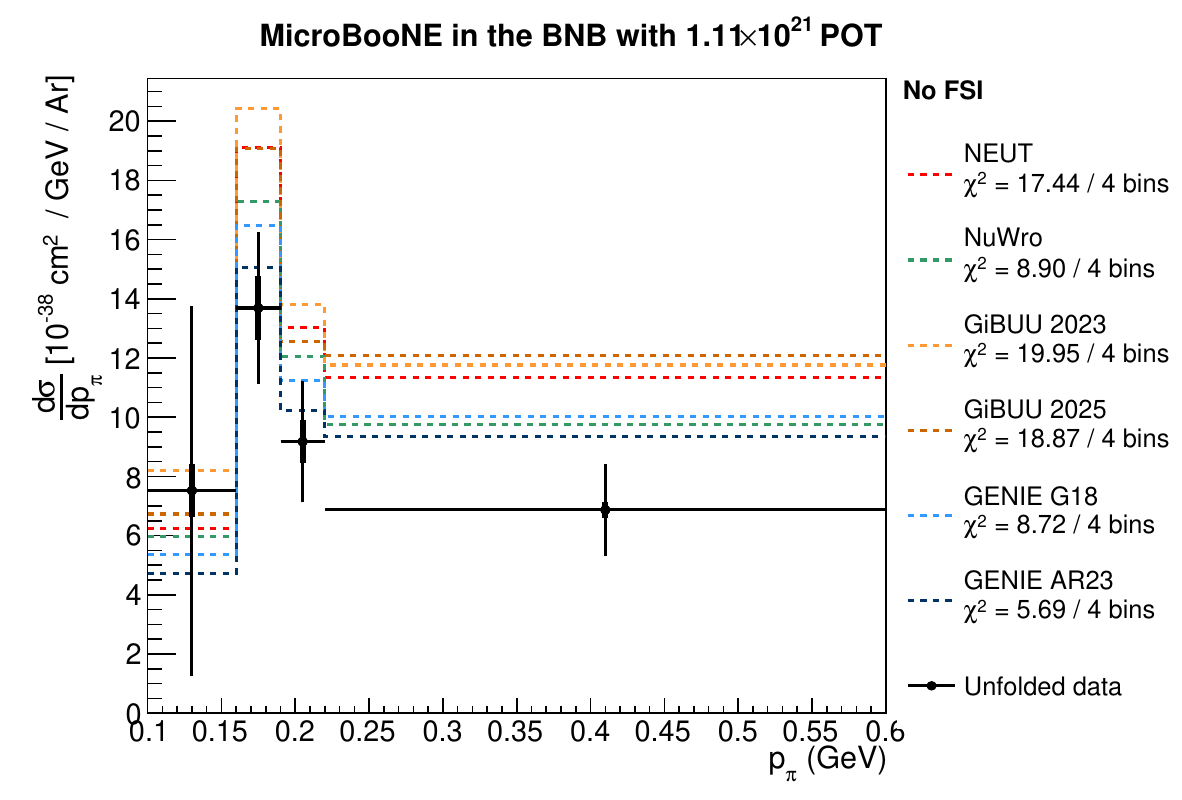}
    \caption{Comparison between the unfolded pion momentum cross section and generator predictions with \ac{FSI} disabled.}
    \label{fig:unfolded_slice_05_nofsi}
\end{figure}

The flux-integrated cross sections are presented in regularized truth space in Fig.~\ref{fig:unfolded_combined}. The effects of unfolding, in the form of the regularization matrix $A_c$, are applied to predictions of the differential cross sections to preserve the data-generator $\chi^2$ agreement through unfolding. Figure~\ref{fig:reco_combined} has smaller \(\chi^2\) values than the GENIE \(\mu\)B comparisons to unfolded data in Fig.~\ref{fig:unfolded_combined}, as the latter does not include uncertainties on the signal. No regularization is applied to the single bin measurement of the total cross section. Table~\ref{tab:generatorPValue} summarizes the \textit{p}-values quantifying data-generator agreement.

\NewDocumentCommand{\rot}{O{40} O{-0.5em} m}{\makebox[#2][r]{\rotatebox{#1}{\makebox[0em][l]{#3\hspace{1mm}}}}}%
\begin{table}
\centering
\begin{tabular}{|l|cccccc|}
\cline{1-1}
\textbf{Generators}    & \multicolumn{1}{c}{\rot{Total}} & \multicolumn{1}{c}{\rot{$\cos(\theta_{\mu})$}} & \multicolumn{1}{c}{\rot{$p_{\mu}$}} & \multicolumn{1}{c}{\rot{$\cos(\theta_{\pi})$}} & \multicolumn{1}{c}{\rot{$p_{\pi}$}} & \multicolumn{1}{c}{\rot{$\theta_{\pi \mu}$}} \\
\hline
GENIE $\mu$B   & 0.85 & 0.26 & 0.78 & 0.84 & 0.67 & 0.68 \\
GENIE G18      & 0.91 & 0.25 & 0.82 & \textbf{0.85} & \textbf{0.72} & 0.67 \\
GENIE AR23     & 0.67 & 0.20 & 0.71 & 0.77 & 0.54 & 0.58 \\
NuWro                   & 0.44 & 0.00 & 0.82 & 0.79 & 0.62 & \textbf{0.78} \\
GiBUU 2023              & 0.08 & 0.33 & 0.12 & 0.13 & 0.01 & 0.15 \\
GiBUU 2025              & \textbf{0.99} & \textbf{0.72} & \textbf{0.90} & 0.24 & 0.19 & 0.54 \\
NEUT                    & 0.24 & 0.00 & 0.69 & 0.30 & 0.29 & 0.54 \\
\hline
\end{tabular}
\caption{\textit{P}-values of the different generator-data comparisons. Highlighted in bold is the best fit for each cross section.}
\label{tab:generatorPValue}
\end{table}

The total unfolded cross section is measured to be 
$$ \sigma = (3.75~\pm~0.07~\textrm{(stat.)}~\pm~0.80~\textrm{(syst.)}) \times 10^{-38} \, \text{cm}^2/\text{Ar} $$
at a mean neutrino beam energy of approximately 0.8 GeV. All extracted cross section values in numerical format, along with the regularization matrices and covariances, are reported in the supplemental material \cite{SupplementalMaterial}.

The results are compared with predictions from various event generators using the NUISANCE framework~\cite{Stowell:2016jfr}. All generators employ a local Fermi gas model for the initial nucleus.

\textbf{GENIE $\mu$B:} The first comparison is with the GENIE MicroBooNE tune \ac{CV} sample produced using version \texttt{3.0.6}, as described in Sec.~\ref{sec:Simulation}. It incorporates the full València model \cite{Nieves:2004wx, Nieves:2011yp, Gran:2013kda} for \ac{CC} \ac{QE} and \ac{MEC} interactions, the Kuzmin-Lyubushkin-Naumov Berger-Sehgal model \cite{Kuzmin:2003ji, Graczyk:2007bc, Berger:2007rq, Nowak:2009se} for \ac{RES}, and Berger-Sehgal \cite{Berger:2008xs} for \ac{COH}. For \ac{DIS}, it relies on the Bodek-Yang model \cite{Bodek:2002ps} using PYTHIA \cite{Sjostrand:2006za}, while the hA 2018 model \cite{Dytman:2015taa} is used for \ac{FSI}.

\textbf{GENIE G18:} This is the untuned version of GENIE $\mu$B, using the same model set \texttt{G18\_10a\_02\_11a}. The comparison uses the latest version of GENIE \texttt{3.04.02}.

\textbf{GENIE AR23:} The DUNE/SBN configuration \texttt{AR23\_20i\_00\_000} of GENIE is produced with version \texttt{3.04.02} and uses the SuSAv2 model \cite{Amaro:2019zos} instead of the València model.

\textbf{NuWro:} Version \texttt{21.09.2} of NuWro is used, which employs the Adler-Rarita-Schwinger formalism for delta resonance calculations \cite{Graczyk:2009qm} and uses only the \ac{MEC} component of the València model. For \ac{QE}, it relies on the Llewellyn Smith model \cite{LlewellynSmith:1971uhs} with \ac{RPA} corrections. Berger-Sehgal is used for \ac{COH}, Bodek-Yang for \ac{DIS}, and an \ac{INC} model for \ac{FSI}.

\textbf{NEUT:} Version \texttt{5.4.0.1} of NEUT \cite{Hayato:2002sd} uses the full València model for \ac{CC} \ac{QE} and \ac{MEC}, Kuzmin-Lyubushkin-Naumov Berger-Sehgal for \ac{RES}, Berger-Sehgal for \ac{COH}, and the Bodek-Yang model for \ac{DIS}. For \ac{FSI}, it employs an \ac{INC} model.

\textbf{GiBUU 2023:} The 2023 (Patch 3) version of GiBUU \cite{Buss:2011mx} has models similar to those of the other generators for \ac{QE}, \ac{MEC}, and \ac{RES} that are implemented for use with the \ac{BUU} equations. Using the supplied configuration, resonance production is modeled following the MAID analysis \cite{Mosel:2019vhx}. The non-resonant background single pion production term uses the Bosted-Christy parameterization \cite{Bosted:2007xd}. GiBUU does not simulate \ac{COH} interactions. For the angular differential cross section, \ac{COH} interactions are concentrated in the forward pointing bins close to $\cos(\theta) = 1$, but overall they constitute only a small fraction of the signal events (\(\sim 1\%\) in GENIE G18). \ac{FSI} is handled exclusively via the \ac{BUU} equations. Unlike the other generators, GiBUU does not rely on empirical tuning to neutrino cross-section data.

\textbf{GiBUU 2025:} The recently released 2025 (Patch 1) version of GiBUU is also shown. The supplied configuration switches to MAID-like calculations of the single pion background terms. Publications on MicroBooNE's $\pi^0$ cross sections have found high sensitivity to in-medium effects for the GiBUU generator \cite{Bogart:2024gmb, MicroBooNE:2024bnl}. This includes a slight preference for the absence of collisional broadening of the \(\Delta\) resonance with the 2023 configuration and is now turned off in this version.

The total flux-integrated cross section using all selected \ccpi events in Fig.~\hyperref[fig:unfolded_combined]{\ref*{fig:unfolded_combined}a} closely matches the GiBUU 2025 prediction and also shows very good agreement with all other generators apart from GiBUU 2023. NuWro and NEUT predict higher, while GENIE predicts lower cross sections than the unfolded data. GiBUU 2023 tends to systematically underpredict pion production cross sections \cite{MicroBooNE:2024bnl}.

The differential cross section with respect to the cosine of the muon angle, shown in Fig.~\hyperref[fig:unfolded_combined]{\ref*{fig:unfolded_combined}b}, reveals significant divergence among generator predictions at very forward angles, where momentum transfer to the nucleus is minimal. GiBUU 2025 performs best, followed by the GENIE predictions. NuWro and NEUT overpredict in this region and do not agree well with data, as the \textit{p}-values in Table~\ref{tab:generatorPValue} show. Models overpredicting at low momentum transfer, \(Q^2\), for single pion final states has been well documented by several experiments, including MINOS, MiniBooNE, and MINERvA \cite{MINOS:2014axb, MiniBooNE:2010eis, MINERvA:2014ogb}. It also matches MicroBooNE's result for the \ac{CC} $\nu_{\mu}$ $\pi^0$ cross section \cite{MicroBooNE:2024bnl}. MINOS and MINERvA have improved their data-simulation agreement by introducing ad hoc \(Q^2\)-dependent suppression functions for resonance production \cite{MINERvA:2014ogb, MINERvA:2019kfr}.

The muon momentum differential cross section shown in Fig.~\hyperref[fig:unfolded_combined]{\ref*{fig:unfolded_combined}c}, using the subset of completely contained events described in Sec.~\ref{sec:containedMuonSubset}, demonstrates excellent agreement with all models except GiBUU 2023, which systematically underpredicts.

The pion angle differential cross section in Fig.~\hyperref[fig:unfolded_combined]{\ref*{fig:unfolded_combined}d} shows strong agreement with GENIE models in the backward direction, while NuWro, NEUT, and GiBUU 2025 overpredict there. In the forward direction, measured cross sections consistently exceed GENIE and GiBUU 2025 predictions, aligning better with NuWro and NEUT due to their higher total cross-section predictions. Overall, data-simulation agreement is best for the GENIE configurations and NuWro, with NEUT and GiBUU performing worse.

Figure~\hyperref[fig:unfolded_combined]{\ref*{fig:unfolded_combined}e} shows the differential cross section with respect to pion momentum, using the unscattered-enhanced subset of events described in Sec.~\ref{sec:unsactteredPionSubset}. The large uncertainties in the first bin are driven by uncertainties on the background prediction arising from the neutrino interaction model. GENIE and NuWro predictions sit around $7 \times 10^{-38}$~cm$^2$/GeV/Ar, but NEUT and GiBUU 2025 significantly deviate, predicting much higher cross sections at low pion momentum. Comparison with Fig.~\ref{fig:unfolded_slice_05_nofsi}, which has \ac{FSI} disabled for the generators, shows that these higher values are driven by their final state simulations. At higher energies, \ac{FSI} simulation suppresses the predictions, contributing to the underprediction in the last bin. Overall, data-simulation agreement is best with GENIE and NuWro, followed by NEUT and GiBUU 2025. GiBUU 2023 shows poor agreement with the data.

Finally, Fig.~\hyperref[fig:unfolded_combined]{\ref*{fig:unfolded_combined}f} shows the differential cross section with respect to the opening angle between muon and pion using all selected events. Data-simulation agreement is generally good, with NuWro performing best, followed by GENIE, NEUT, and GiBUU. 

In summary, GiBUU 2025 performs best for the total and muon kinematic variables, but, apart from GiBUU 2023, shows the worst agreement with data for the pion variables. GENIE and NuWro give the closest predictions for those differential cross sections. Predictions of the muon angular distribution generally exceed the data, and are especially high for NuWro and NEUT, signaling deficiencies in low-$Q^2$ modeling. At backward pion angles, NEUT, GiBUU 2025, and, to a lesser extent, NuWro have higher predictions than both GENIE and the data. Finally, uncertainties and differences between generators are both large at low pion momentum, with the latter driven by differences in \ac{FSI} modeling.

\section{\label{sec:Conclusion}Conclusion}
Flux-integrated measurements of charged-current muon (anti\mbox{-})neutrino cross sections on argon for events with a charged pion in the final state are reported using MicroBooNE data from all five experimental runs. The analysis uses a \ac{BDT}-based selection and defines subsets of events for muon and pion momentum estimation. The pion momentum differential cross section represents the first such measurement on argon. Total and angular cross sections are also extracted, constituting the highest-statistics measurements of this interaction to date.

The total cross section and most pion kinematic distributions are reproduced within uncertainties by a suite of modern neutrino event generators. Disagreement arises in the muon angle relative to the beam, where most generators overestimate the cross section at very forward angles. This results in lower data-simulation agreement for almost all generators compared to the other differential cross sections, with NuWro and NEUT not agreeing with the data. This finding echoes previous results from several experiments, including recently published MicroBooNE \(\pi^0\) measurements, and suggests that the generator models have room for improvement at low momentum transfer \(Q^2\).

The measurements here are limited by systematic uncertainties in the detector simulation, the background estimation, and the neutrino flux modeling. Future work to reduce these uncertainties, especially at low pion momentum, along with exploration of double differential cross sections, could help further refine resonant pion production in empirical models. Additionally, new methods for pion momentum estimation, such as Ref.~\cite{DUNE:2024tdb}, are essential for extending differential cross sections to the full range of pion energies.

\section*{Acknowledgments}%
This document was prepared by the MicroBooNE collaboration using the resources of the Fermi National Accelerator Laboratory (Fermilab), a U.S. Department of Energy, Office of Science, Office of High Energy Physics HEP User Facility. Fermilab is managed by Fermi Forward Discovery Group, LLC, acting under Contract No. 89243024CSC000002. MicroBooNE is supported by the following: the U.S. Department of Energy, Office of Science, Offices of High Energy Physics and Nuclear Physics; the U.S. National Science Foundation; the Swiss National Science Foundation; the Science and Technology Facilities Council (STFC), part of the United Kingdom Research and Innovation; the Royal Society (United Kingdom); the UK Research and Innovation (UKRI) Future Leaders Fellowship; and the NSF AI Institute for Artificial Intelligence and Fundamental Interactions. Additional support for the laser calibration system and cosmic ray tagger was provided by the Albert Einstein Center for Fundamental Physics, Bern, Switzerland. We also acknowledge the contributions of technical and scientific staff to the design, construction, and operation of the MicroBooNE detector as well as the contributions of past collaborators to the development of MicroBooNE analyses, without whom this work would not have been possible. For the purpose of open access, the authors have applied a Creative Commons Attribution (CC BY) public copyright license to any Author Accepted Manuscript version arising from this submission.

\FloatBarrier
\bibliography{bib_clean}

\end{document}

\maketitle
\section{Introduction}
This document provides numerical results and supplementary information for the publication. Section~\ref{sec:cov} presents covariances and correlations between all bins. Section~\ref{sec:xsec} is organized by cross-section category and lists the numerical values for all bins. In addition, the block-diagonal elements of the regularization matrix are shown for the differential cross sections. Section~\ref{sec:sideband} shows a sideband study of the main background. Results of a fake data test are shown in Section~\ref{sec:fake_data}. Section~\ref{sec:chi2_score_dist} presents calorimetric track-fitting particle identification (PID) scores used in the preselection for reconstructed GENIE events under proton and muon hypotheses. The impact of phase space restrictions on the analysis is shown in Section~\ref{sec:phase_space}. Fractional uncertainties for the individual unfolded cross-section bins are shown in Section~\ref{sec:fracUncert}. Finally, Section~\ref{sec:noFSI} reports generator comparisons performed with final state interactions (FSI) disabled.

\section{Covariance and Correlation Matrix}
\label{sec:cov}

\begin{figure}[htb]
    \centering
    \begin{overpic}[width=0.79\textwidth]{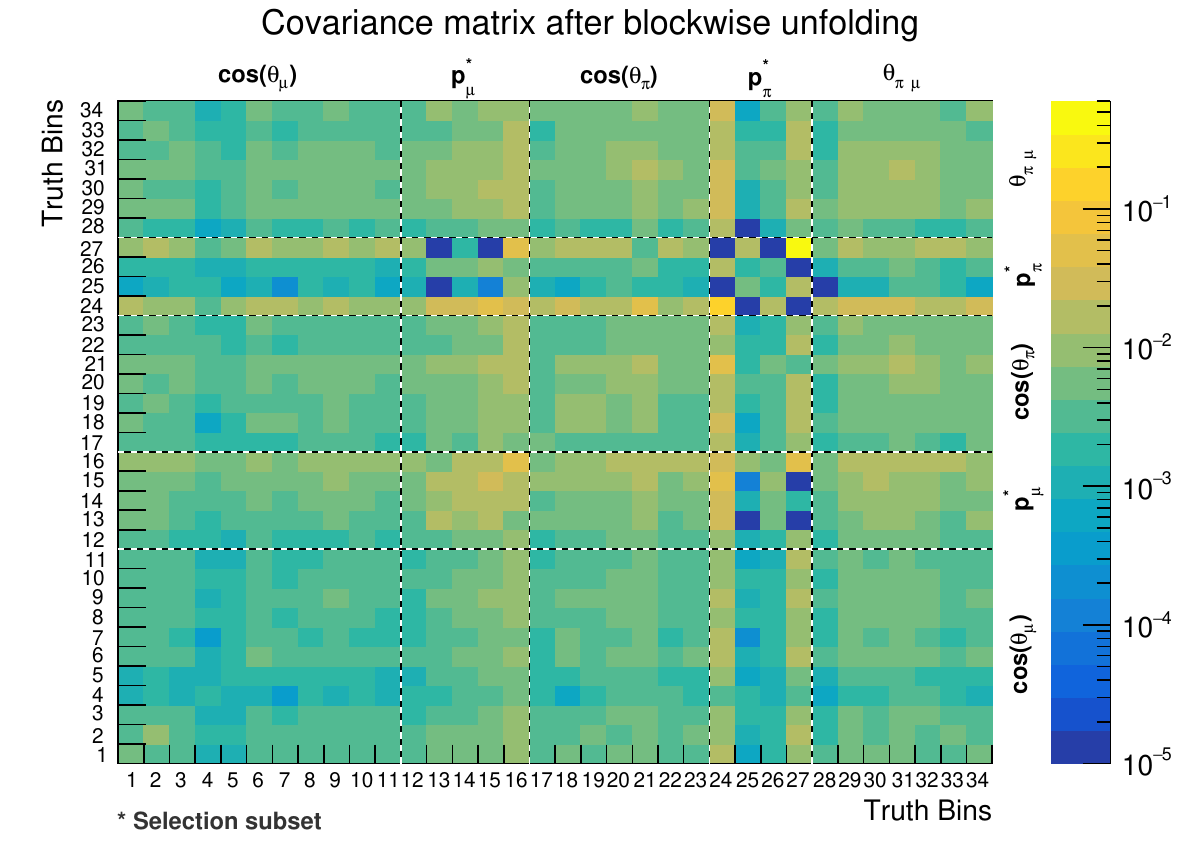}
        \put(3.5,7){\rotatebox{90}{\textbf{MicroBooNE Simulation}}}
    \end{overpic}
    \caption{Covariance matrix of the unfolded bins without bin width scaling applied, so all entries here have units of \((10^{-38}~\text{cm}^2/\text{Ar})^2\). The dashed lines separate bins from different differential cross sections.}
    \label{fig:cov_all}
\end{figure}

Figure~\ref{fig:cov_all} shows the covariance matrix of the unfolded cross sections for all differential bins. Along both axes, the matrix elements are grouped as follows, with the corresponding bin edges defined in Section~\ref{sec:xsec}:
\begin{itemize}
    \item Bins 1--11: Muon angle
    \item Bins 12--16: Muon momentum
    \item Bins 17--23: Pion angle
    \item Bins 24--27: Pion momentum
    \item Bins 28--34: Muon–pion opening angle
\end{itemize}

\begin{figure}[!htb]
    \centering
    \begin{overpic}[width=0.79\textwidth]{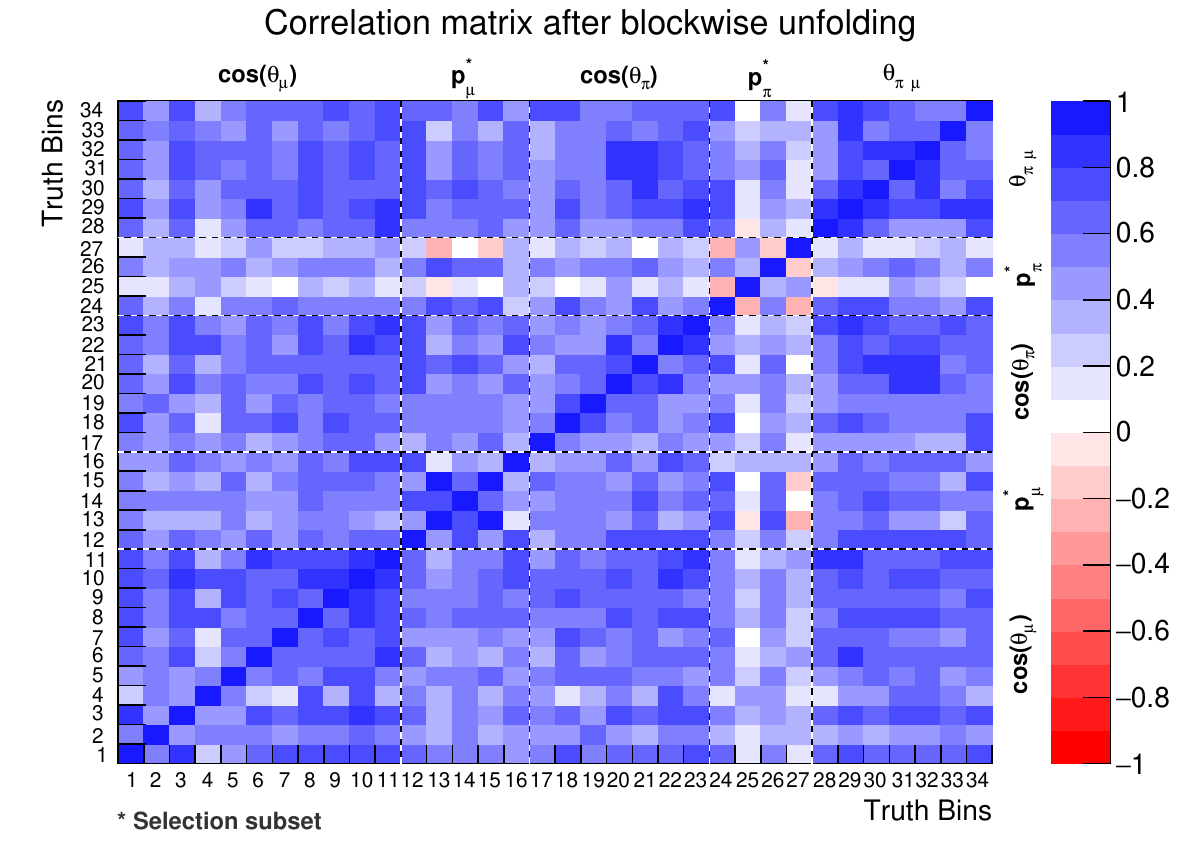}
            \put(3.5,7){\rotatebox{90}{\textbf{MicroBooNE Simulation}}}
    \end{overpic}
    \caption{Correlation matrix of the unfolded bins. The dashed lines separate bins from different differential cross sections.}
    \label{fig:corr_all}
\end{figure}

The correlation matrix shown in Fig.~\ref{fig:corr_all} is derived from the covariance matrix of the unfolded bins by normalizing the covariances with the standard deviations of the variables. Given the covariance matrix elements \(\sigma_{ij}\), the correlation matrix elements are defined as:

\begin{equation}
    r_{ij} = \frac{\sigma_{ij}}{\sqrt{\sigma_{ii} \sigma_{jj}}},
\end{equation}
\noindent where \(\sigma_{ii}\) and \(\sigma_{jj}\) are the variances of variables \(i\) and \(j\), respectively.

\clearpage
\section{Cross sections}
\label{sec:xsec}
\begin{figure}[!htb]
    \centering
     \begin{overpic}[width=0.6\textwidth]{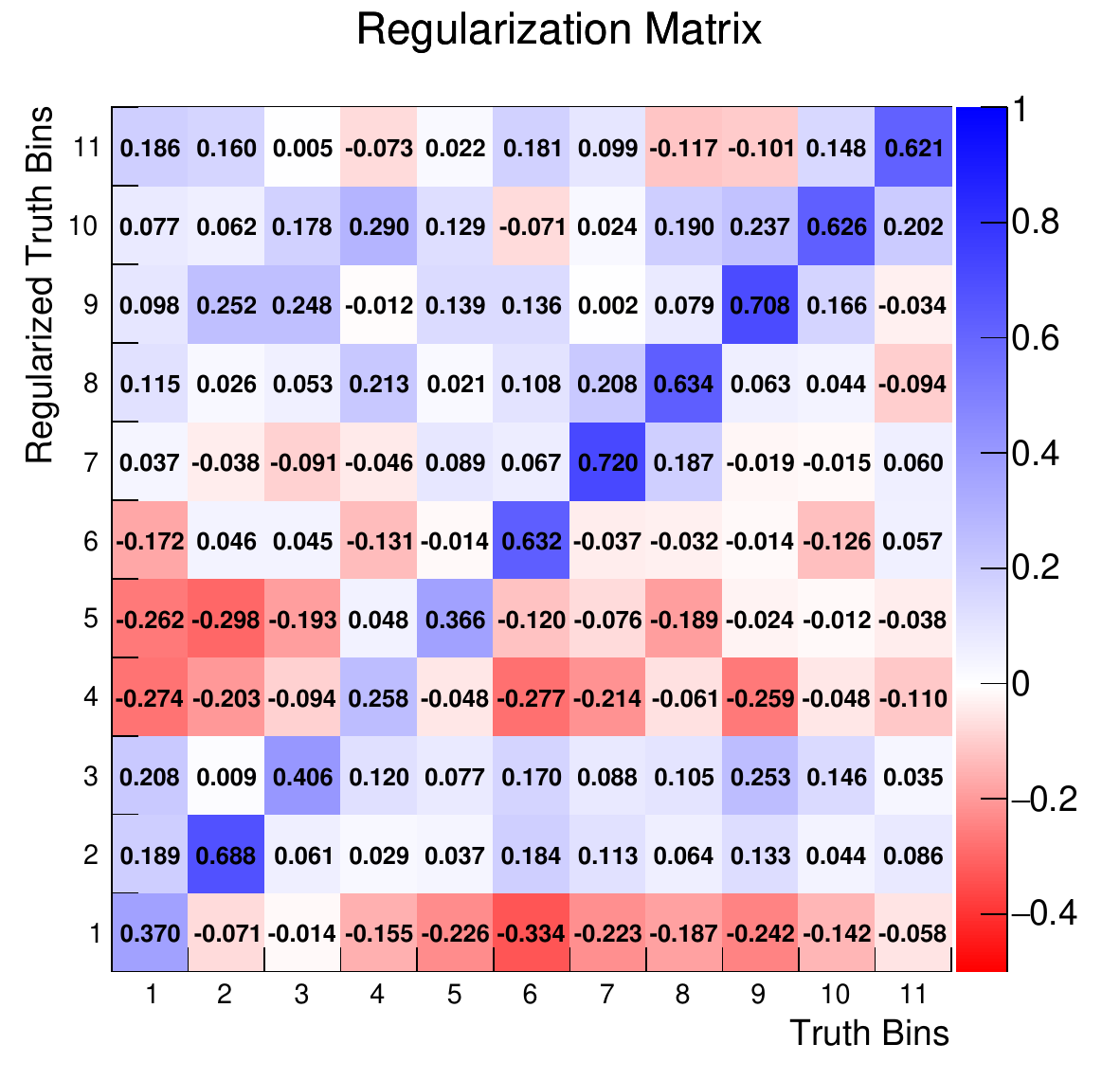}
         \put(2.5,10){\rotatebox{90}{\textbf{MicroBooNE}}}
    \end{overpic}     
    \caption{Regularization matrix for the $\cos(\theta_{\mu})$ cross section. This needs to be applied to truth predictions when comparing them to the unfolded data results.}
    \label{fig:reg_mu_ang}
\end{figure}

\begin{figure}[!htb]
    \centering
     \begin{overpic}[width=0.6\textwidth]{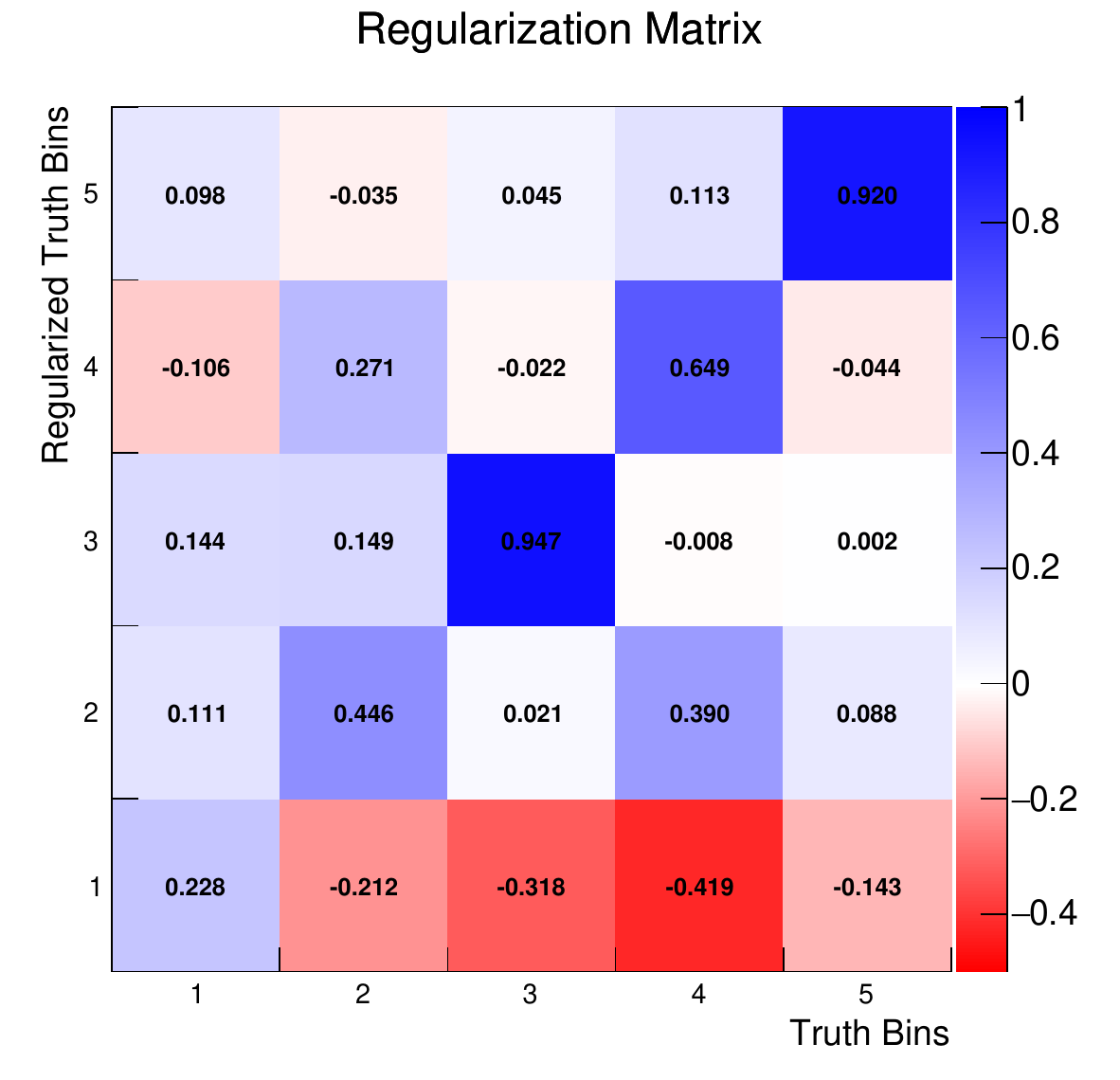}
         \put(3.5,10){\rotatebox{90}{\textbf{MicroBooNE}}}
    \end{overpic}
    \caption{Regularization matrix for the $p_{\mu}$ cross section. This needs to be applied to truth predictions when comparing them to the unfolded data results.}
    \label{fig:reg_mu_mom}
\end{figure}

\begin{figure}[!htb]
    \centering
     \begin{overpic}[width=0.6\textwidth]{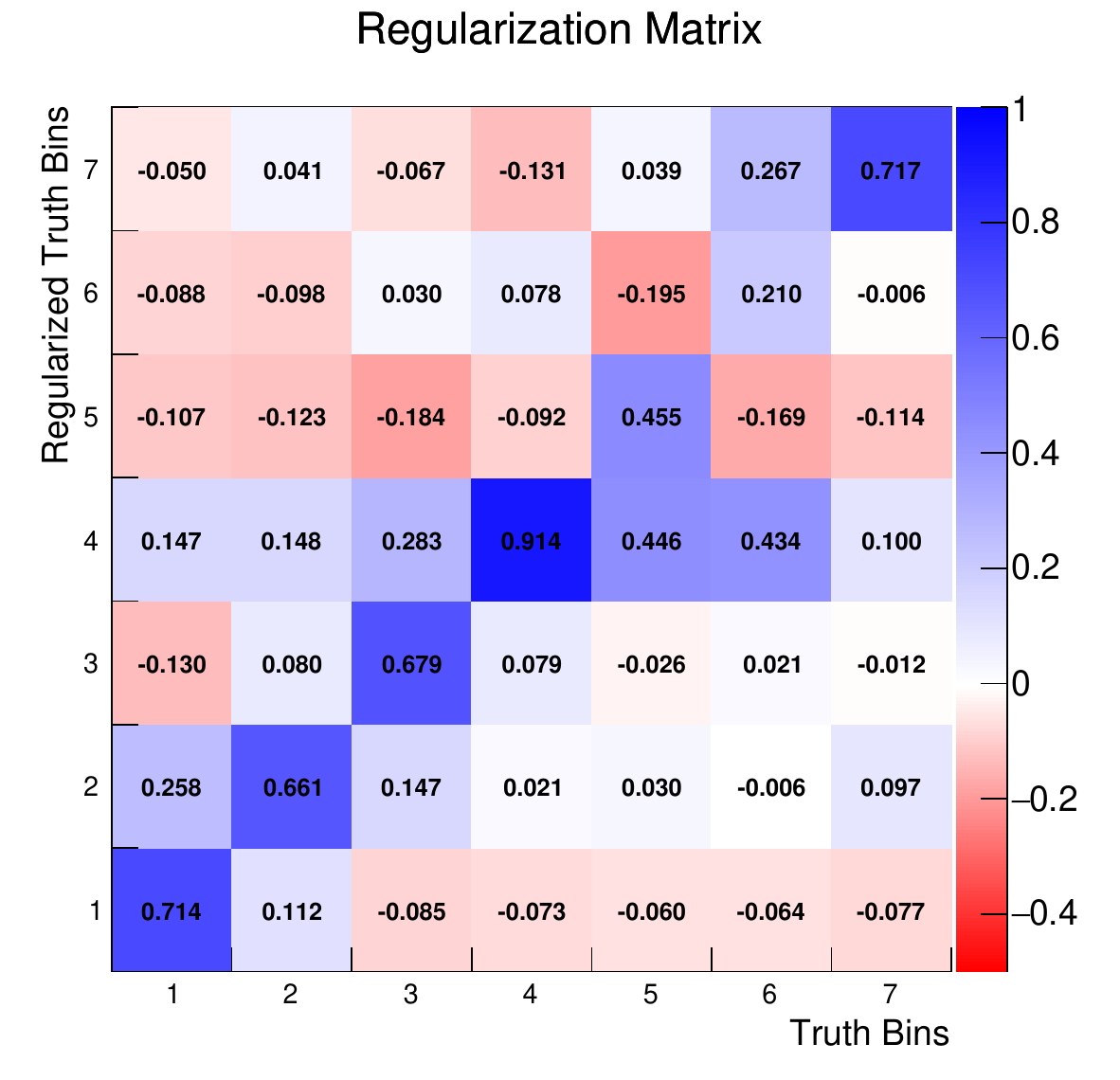}
         \put(3.5,10){\rotatebox{90}{\textbf{MicroBooNE}}}
    \end{overpic}
    \caption{Regularization matrix for the $\cos(\theta_{\pi})$ cross section. This needs to be applied to truth predictions when comparing them to the unfolded data results.}
    \label{fig:reg_pi_ang}
\end{figure}

\begin{figure}[!htb]
    \centering
     \begin{overpic}[width=0.6\textwidth]{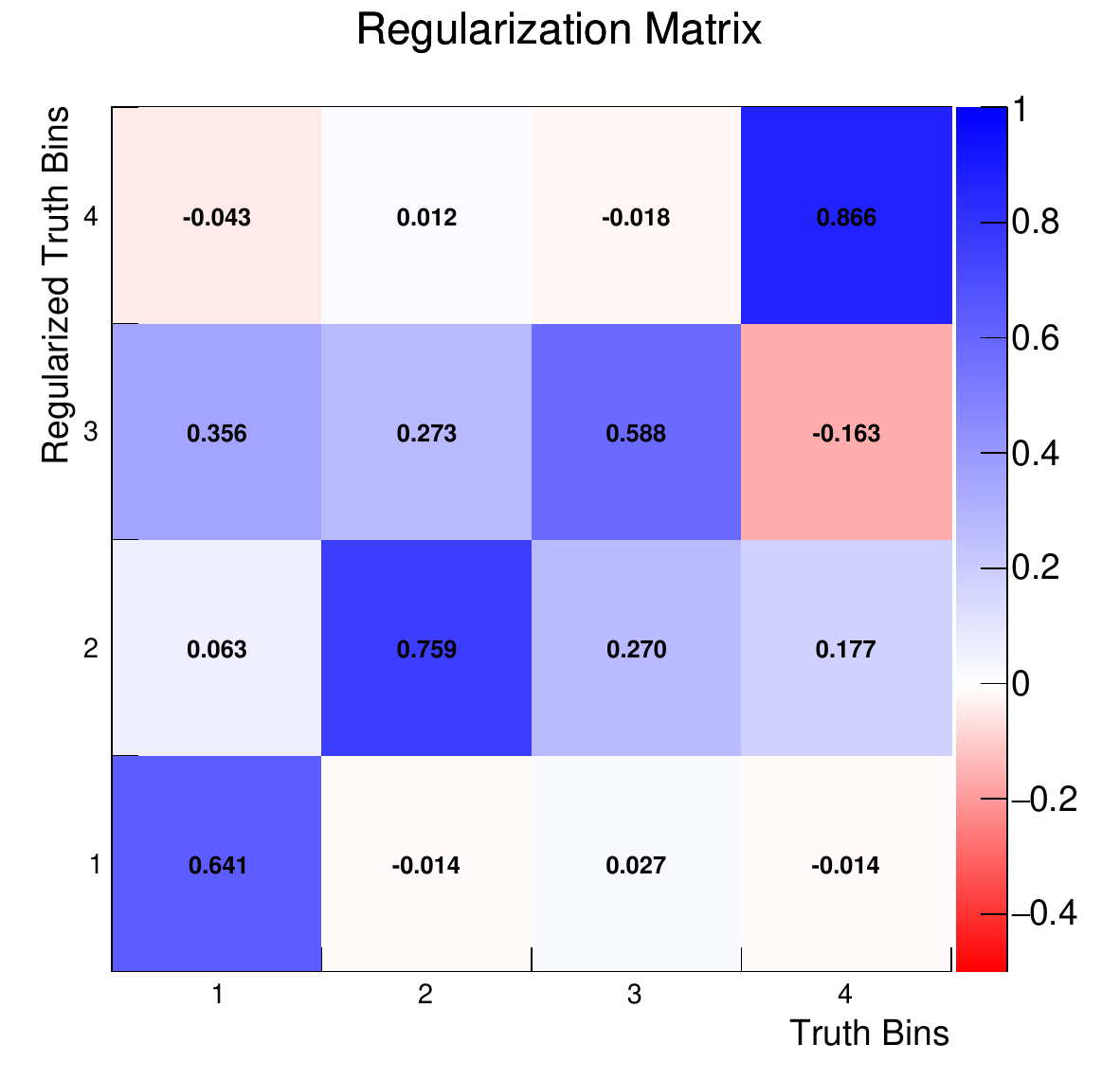}
         \put(3.5,10){\rotatebox{90}{\textbf{MicroBooNE}}}
    \end{overpic}
    \caption{Regularization matrix for the $p_{\pi}$ cross section. This needs to be applied to truth predictions when comparing them to the unfolded data results.}
    \label{fig:reg_pi_mom}
\end{figure}

\begin{figure}[!htb]
    \centering
     \begin{overpic}[width=0.6\textwidth]{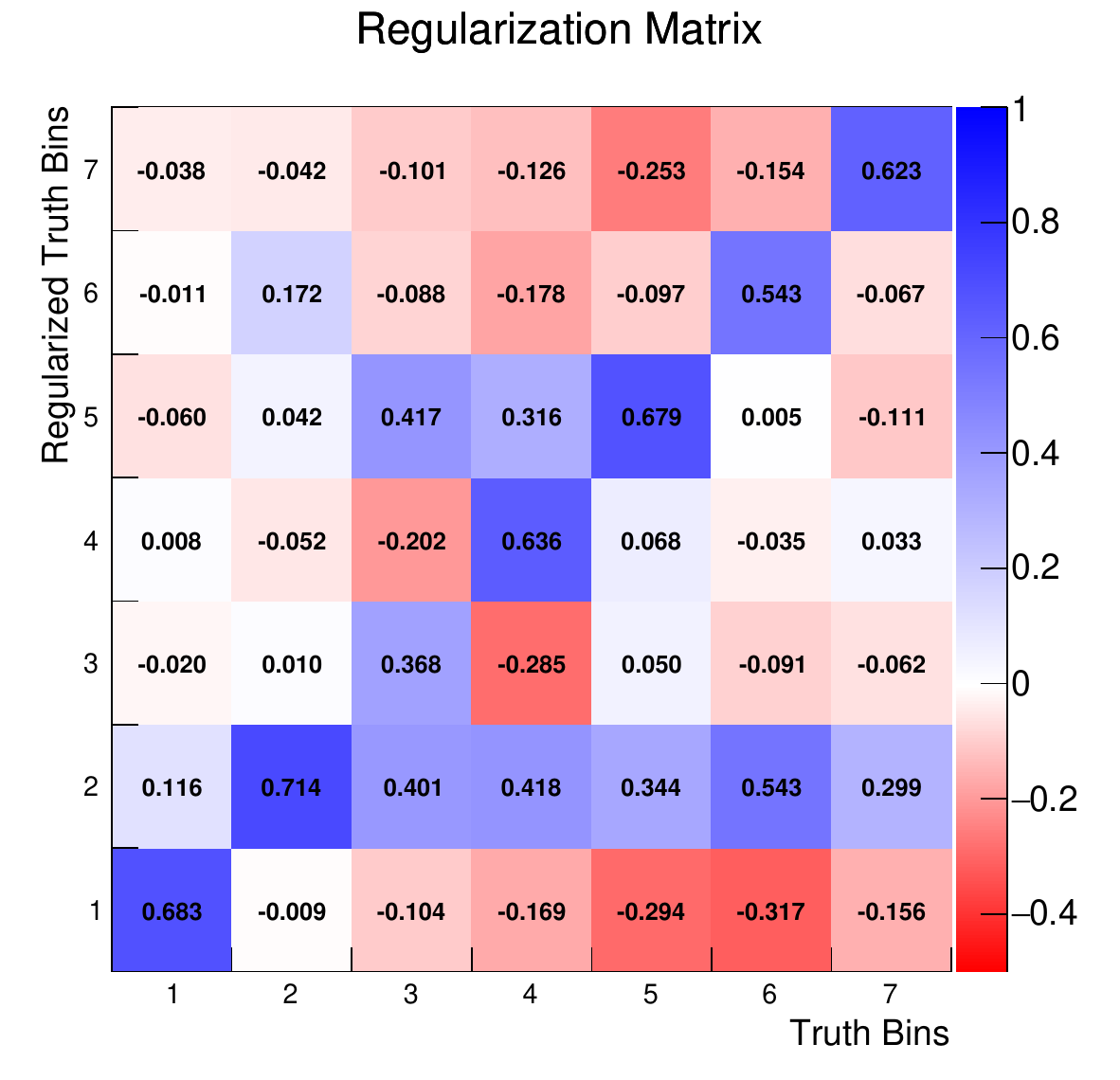}
         \put(3.5,10){\rotatebox{90}{\textbf{MicroBooNE}}}
    \end{overpic}
    \caption{Regularization matrix for the $\theta_{\mu \pi}$ cross section. Smearing effects into the second bin coming from the regularization are noticeable here. This needs to be applied to truth predictions when comparing them to the unfolded data results.}
    \label{fig:reg_mu_pi_ang}
\end{figure}

This section lists the bin values for the total and the differential cross sections with respect to muon angle, muon momentum, pion angle, pion momentum, and muon-pion opening angle in Tables \ref{tab:xsec_0}, \ref{tab:xsec_1}, \ref{tab:xsec_2}, \ref{tab:xsec_3}, \ref{tab:xsec_4}, and \ref{tab:xsec_5}. For the differential cross sections, Figs. \ref{fig:reg_mu_ang}, \ref{fig:reg_mu_mom}, \ref{fig:reg_pi_ang}, \ref{fig:reg_pi_mom}, and \ref{fig:reg_mu_pi_ang} list their regularization matrices.

\FloatBarrier
\subsection{Total}
\begin{table}[!htb]
\centering
\begin{tabular}{cc}
\toprule
\multicolumn{2}{c}{Total} \\
\midrule
Bin & Unfolded cross section \\
\midrule
1 & $3.745 \pm 0.801$ \\
\bottomrule
\end{tabular}
\caption{Total cross section in units of $10^{-38}$ cm$^2$/Ar.}
\label{tab:xsec_0}
\end{table}

\FloatBarrier
\subsection{Muon Angle}
\begin{table}[!htb]
\centering
\begin{tabular}{ccc}
\toprule
\multicolumn{3}{c}{$\cos(\theta_{\mu})$} \\
\midrule
Bin & Unfolded cross section & Bin range \\
\midrule
1 & $0.297 \pm 0.094$ & $[-1, -0.27)$ \\
2 & $0.591 \pm 0.163$ & $[-0.27, 0.29)$ \\
3 & $1.475 \pm 0.349$ & $[0.29, 0.46)$ \\
4 & $1.729 \pm 0.403$ & $[0.46, 0.58)$ \\
5 & $2.163 \pm 0.441$ & $[0.58, 0.67)$ \\
6 & $3.074 \pm 0.740$ & $[0.67, 0.77)$ \\
7 & $4.097 \pm 1.213$ & $[0.77, 0.82)$ \\
8 & $4.772 \pm 1.008$ & $[0.82, 0.88)$ \\
9 & $5.782 \pm 1.428$ & $[0.88, 0.93)$ \\
10 & $6.554 \pm 1.549$ & $[0.93, 0.97)$ \\
11 & $6.155 \pm 1.778$ & $[0.97, 1]$ \\
\bottomrule
\end{tabular}
\caption{Muon angle cross section in units of $10^{-38}$ cm$^2$/Ar.}
\label{tab:xsec_1}
\end{table}

\FloatBarrier
\subsection{Muon Momentum}
\begin{table}[!htb]
\centering
\begin{tabular}{ccc}
\toprule
\multicolumn{3}{c}{$p_{\mu}$} \\
\midrule
Bin & Unfolded cross section & Bin range \\
\midrule
1 & $4.058 \pm 0.736$ & $[0.15, 0.23)$ GeV \\
2 & $4.331 \pm 1.395$ & $[0.23, 0.32)$ GeV \\
3 & $4.863 \pm 0.921$ & $[0.32, 0.45)$ GeV \\
4 & $3.102 \pm 0.880$ & $[0.45, 0.66)$ GeV \\
5 & $1.531 \pm 0.302$ & $[0.66, \infty)$ GeV \\
\bottomrule
\end{tabular}
\caption{Muon momentum cross section in units of $10^{-38}$ cm$^2$/GeV/Ar.}
\label{tab:xsec_2}
\end{table}

\FloatBarrier
\subsection{Pion Angle}
\begin{table}[!htb]
\centering
\begin{tabular}{ccc}
\toprule
\multicolumn{3}{c}{$\cos(\theta_{\pi})$} \\
\midrule
Bin & Unfolded cross section & Bin range \\
\midrule
1 & $0.425 \pm 0.156$ & $[-1, -0.47)$ \\
2 & $0.800 \pm 0.201$ & $[-0.47, 0)$ \\
3 & $1.417 \pm 0.257$ & $[0, 0.39)$ \\
4 & $2.186 \pm 0.393$ & $[0.39, 0.65)$ \\
5 & $2.809 \pm 0.685$ & $[0.65, 0.84)$ \\
6 & $4.324 \pm 0.866$ & $[0.84, 0.93)$ \\
7 & $5.035 \pm 1.215$ & $[0.93, 1]$ \\
\bottomrule
\end{tabular}
\caption{Pion angle cross section in units of $10^{-38}$ cm$^2$/Ar.}
\label{tab:xsec_3}
\end{table}

\FloatBarrier
\subsection{Pion Momentum}
\begin{table}[!htb]
\centering
\begin{tabular}{ccc}
\toprule
\multicolumn{3}{c}{$p_{\pi}$} \\
\midrule
Bin & Unfolded cross section & Bin range \\
\midrule
1 & $7.519 \pm 6.253$ & $[0.1, 0.16)$ GeV \\
2 & $13.688 \pm 2.570$ & $[0.16, 0.19)$ GeV \\
3 & $9.174 \pm 2.055$ & $[0.19, 0.22)$ GeV \\
4 & $6.867 \pm 1.565$ & $[0.22, \infty)$ GeV \\
\bottomrule
\end{tabular}
\caption{Pion momentum cross section in units of $10^{-38}$ cm$^2$/GeV/Ar.}
\label{tab:xsec_4}
\end{table}

\FloatBarrier
\subsection{Muon-pion Opening Angle}
\begin{table}[!htb]
\centering
\begin{tabular}{ccc}
\toprule
\multicolumn{3}{c}{$\theta_{\mu \pi}$} \\
\midrule
Bin & Unfolded cross section & Bin range \\
\midrule
1 & $0.470 \pm 0.117$ & $[0, 0.49)$ rad \\
2 & $1.182 \pm 0.238$ & $[0.49, 0.93)$ rad \\
3 & $1.394 \pm 0.322$ & $[0.93, 1.26)$ rad \\
4 & $1.483 \pm 0.406$ & $[1.26, 1.57)$ rad \\
5 & $1.405 \pm 0.322$ & $[1.57, 1.88)$ rad \\
6 & $1.261 \pm 0.247$ & $[1.88, 2.21)$ rad \\
7 & $0.655 \pm 0.203$ & $[2.21, 2.65]$ rad \\
\bottomrule
\end{tabular}
\caption{Muon-pion opening angle cross section in units of $10^{-38}$ cm$^2$/rad/Ar.}
\label{tab:xsec_5}
\end{table}

\clearpage
\section{Sideband Study}
\label{sec:sideband}
\begin{figure}[!htb]
    \centering
    \includegraphics[width=0.6\textwidth]{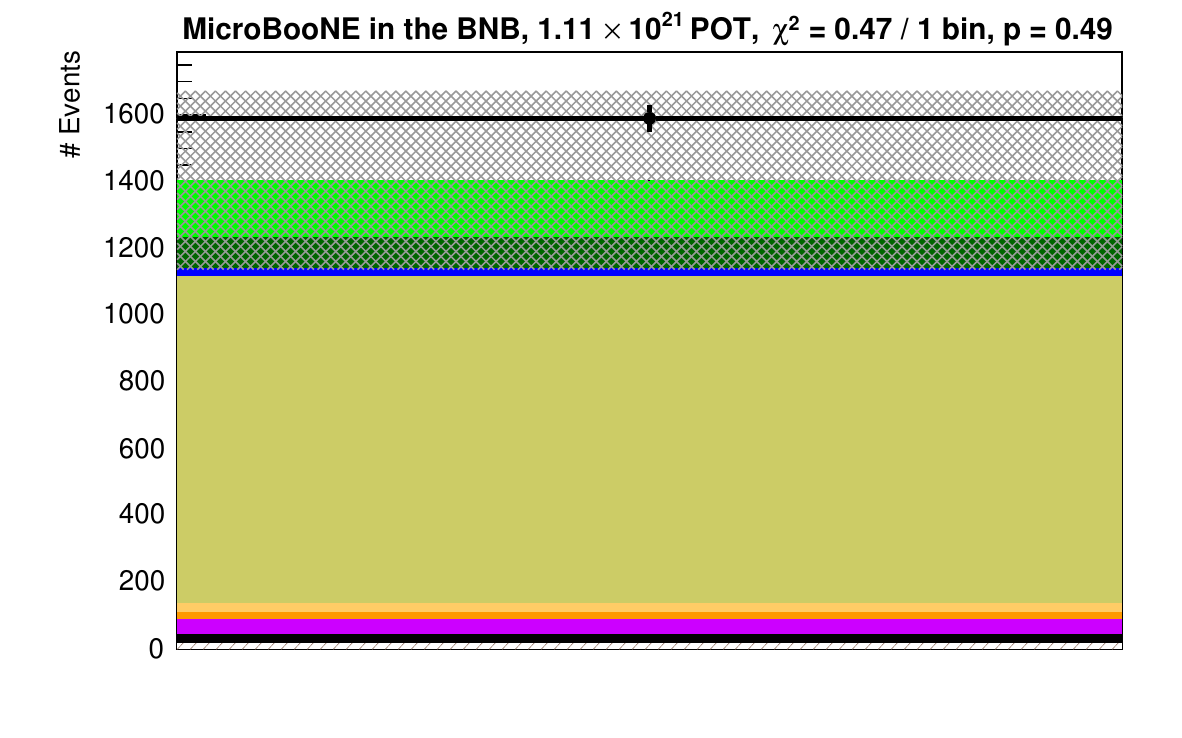}
    \caption{Total event rate of the sideband selection.}
    \label{fig:sideband_total}
\end{figure}

\begin{figure}[!htb]
    \centering
    \includegraphics[width=0.6\textwidth]{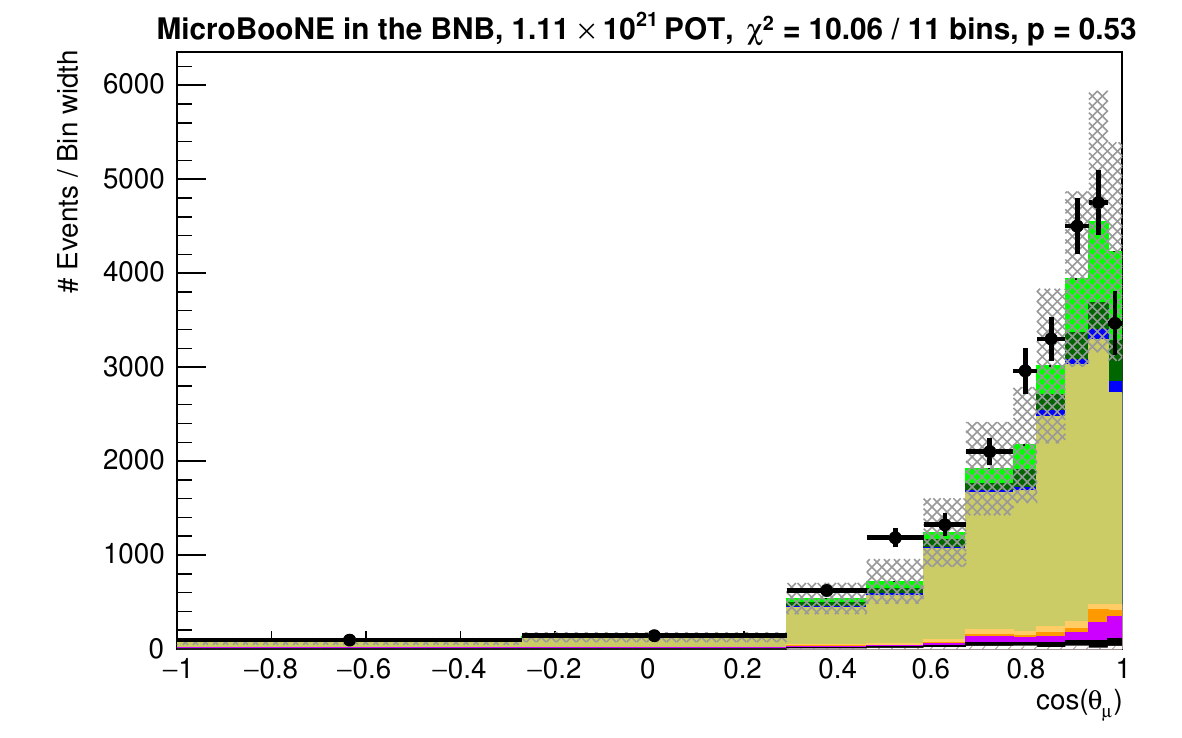}
    \caption{Event rate of the sideband selection for the muon angle}
    \label{fig:sideband_mu_angle}
\end{figure}

\begin{figure}[!htb]
    \centering
    \includegraphics[width=0.6\textwidth]{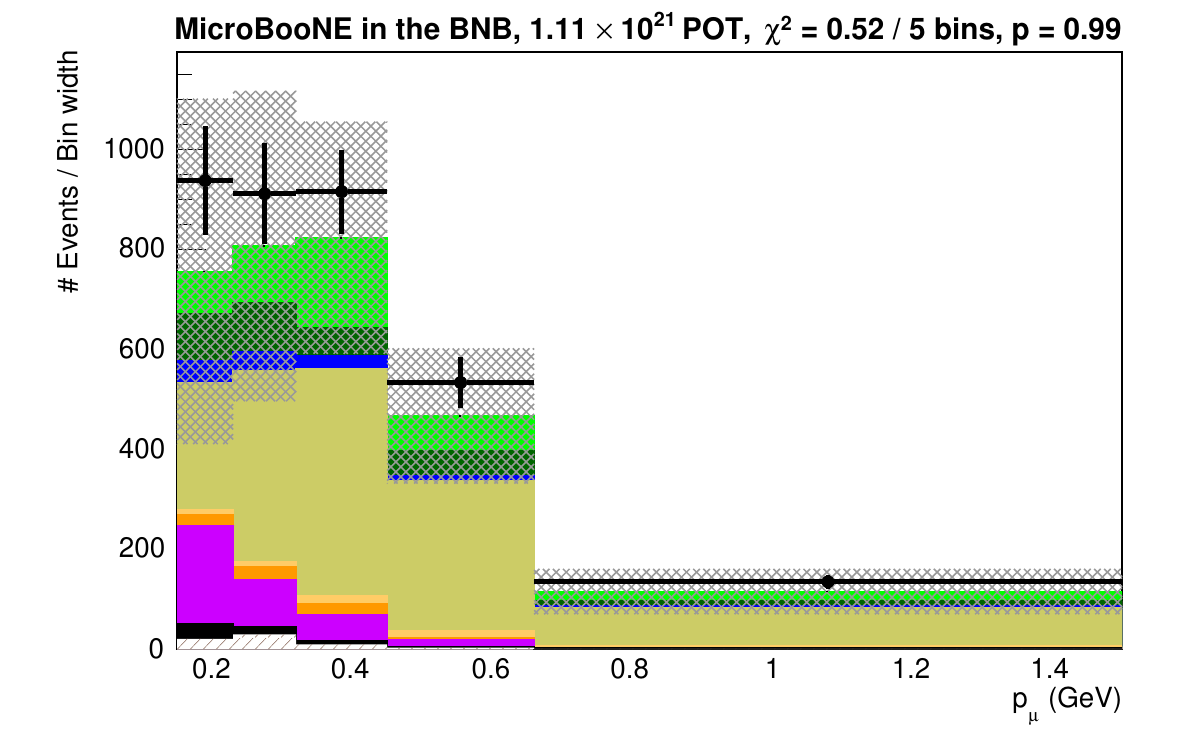}
    \caption{Stacked histogram of the event rate of the sideband selection for the muon momentum. Like the main selection, this only includes events fully contained inside the detector.}
    \label{fig:sideband_mu_momentum}
\end{figure}

\begin{figure}[!htb]
    \centering
    \includegraphics[width=0.6\textwidth]{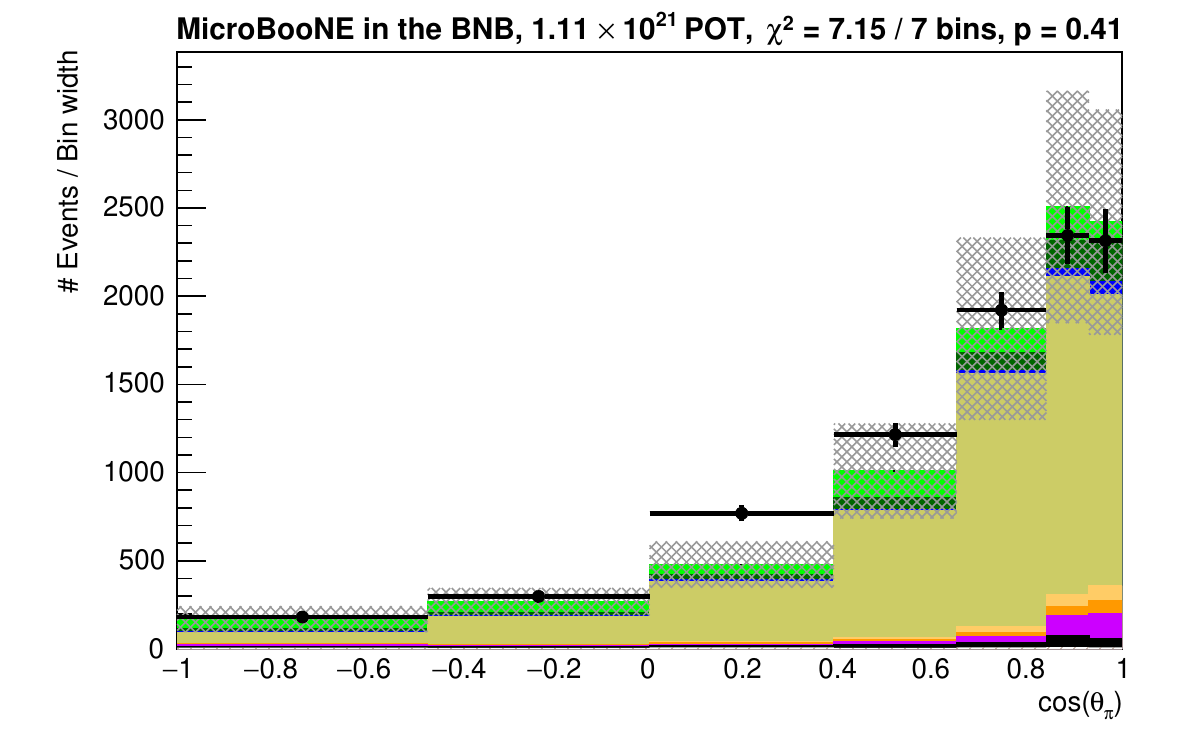}
    \caption{Event rate of the sideband selection for the pion angle kinematic variable using the leading proton angle.}
    \label{fig:sideband_pi_angle}
\end{figure}

\begin{figure}[!htb]
    \centering
    \includegraphics[width=0.6\textwidth]{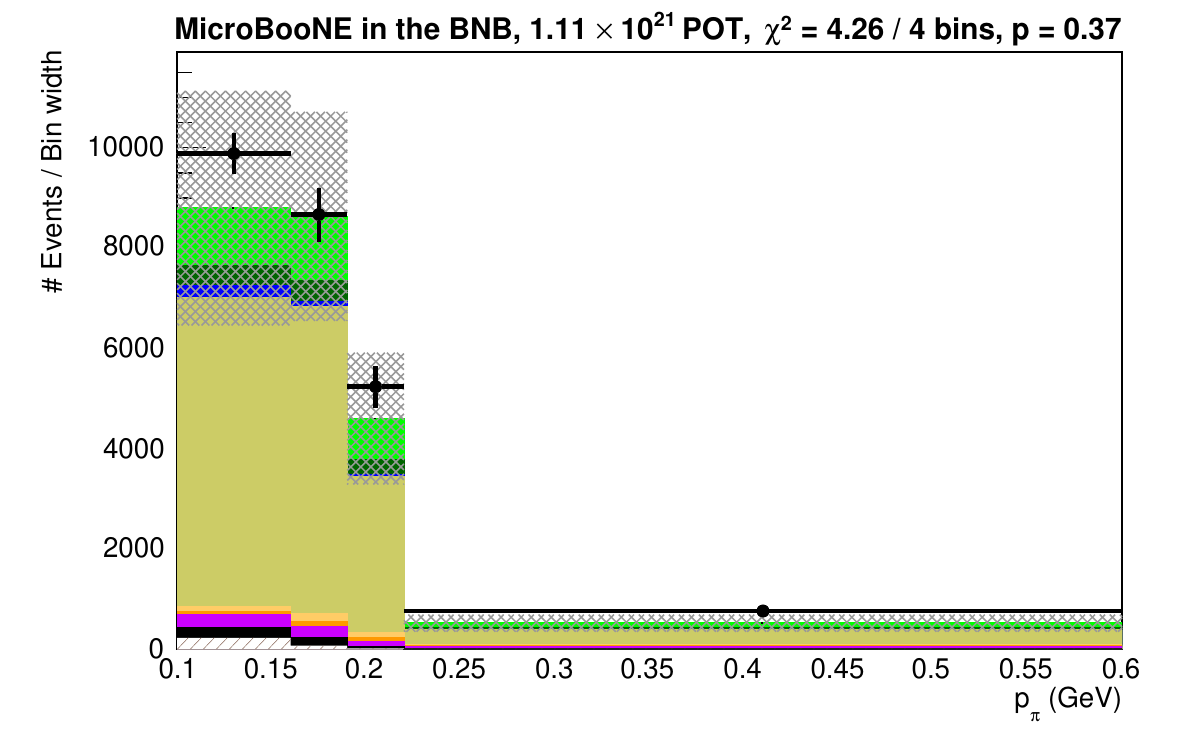}
    \caption{Event rate of the sideband selection for the pion momentum kinematic variable computed from the leading proton candidate. This includes the additional unscattered pion BDT cut but applied to the leading proton.}
    \label{fig:sideband_pi_momentum}
\end{figure}

\begin{figure}[!htb]
    \centering
    \includegraphics[width=0.6\textwidth]{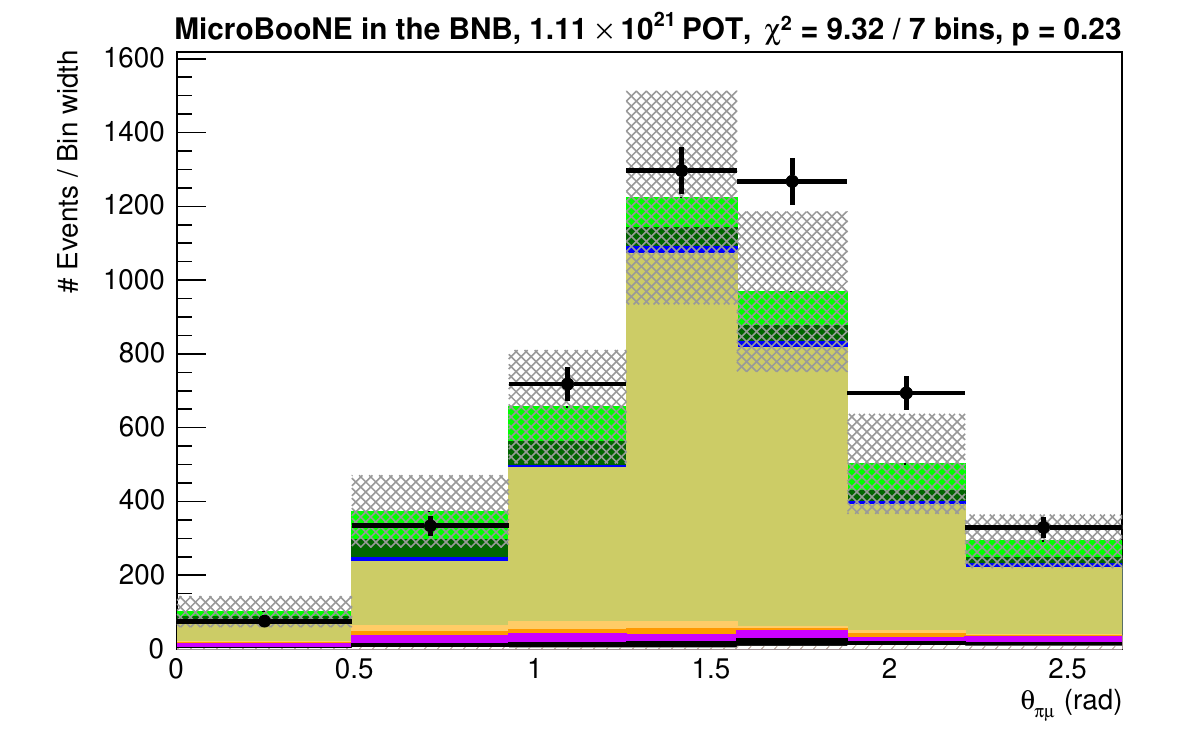}
    \caption{Event rate of the sideband selection for the muon-pion opening angle kinematic variable using the muon and leading proton candidates.}
    \label{fig:sideband_opening_angle}
\end{figure}

The dominant background in the analysis arises from muon neutrino charged-current events with zero pions, where proton tracks are misidentified as charged pions. To validate the background modeling, an orthogonal sideband is constructed by selecting events with one muon, at least one proton, and no other particles. Additional cuts are applied to mimic the kinematics of misidentified protons that end up as background in the main selection. Figures \ref{fig:sideband_total}, \ref{fig:sideband_mu_angle}, \ref{fig:sideband_mu_momentum}, \ref{fig:sideband_pi_angle}, \ref{fig:sideband_pi_momentum} and \ref{fig:sideband_opening_angle} show comparisons between data and simulation for the sideband using the same binning as the main selection. Good agreement between data and MC is seen across all distributions within statistical and systematic uncertainties.

\begin{figure*}[!htb]
    \centering
    \includegraphics[width=0.8\linewidth]{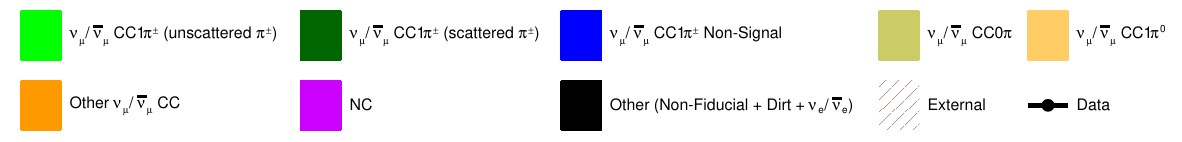}
\end{figure*}

\clearpage
\section{Fake Data Study}
\label{sec:fake_data}
\begin{figure}[!htb]
    \centering
    \includegraphics[width=0.8\textwidth]{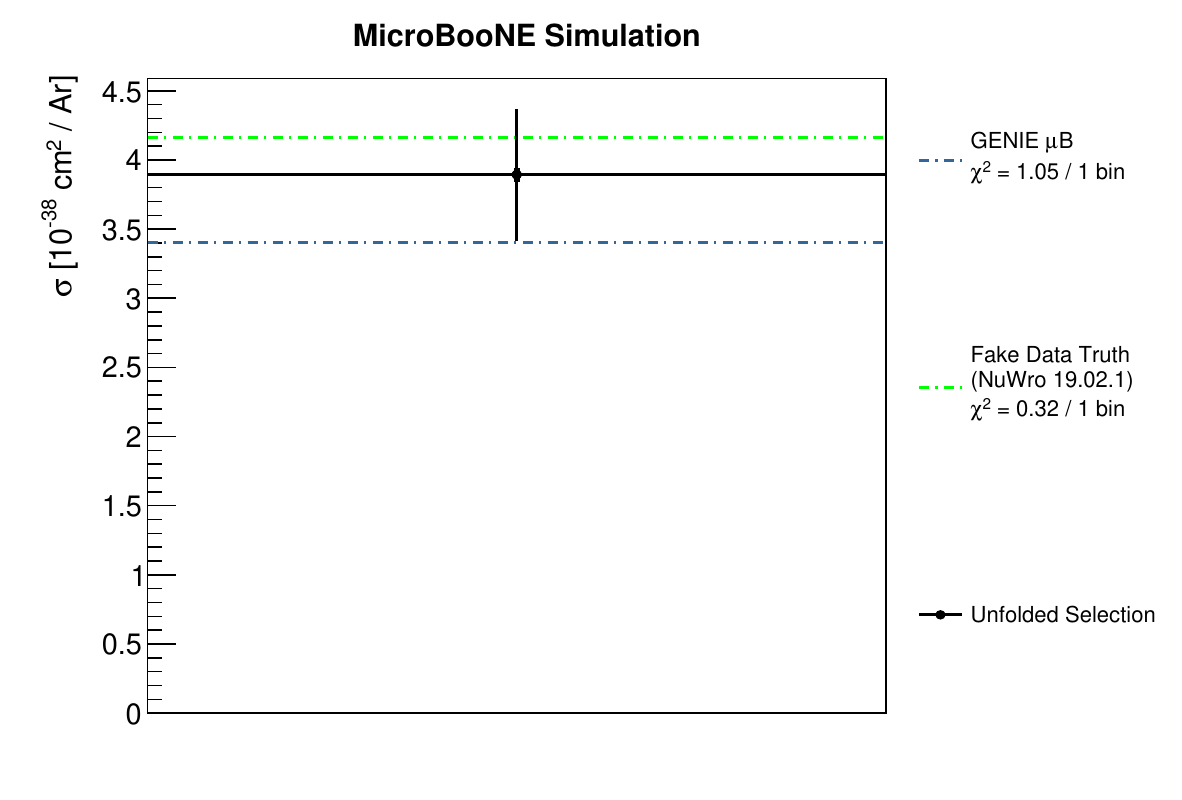}
    \caption{Unfolded total cross section using NuWro fake data.}
    \label{fig:unfolded_total}
\end{figure}

\begin{figure}[!htb]
    \centering
    \includegraphics[width=0.8\textwidth]{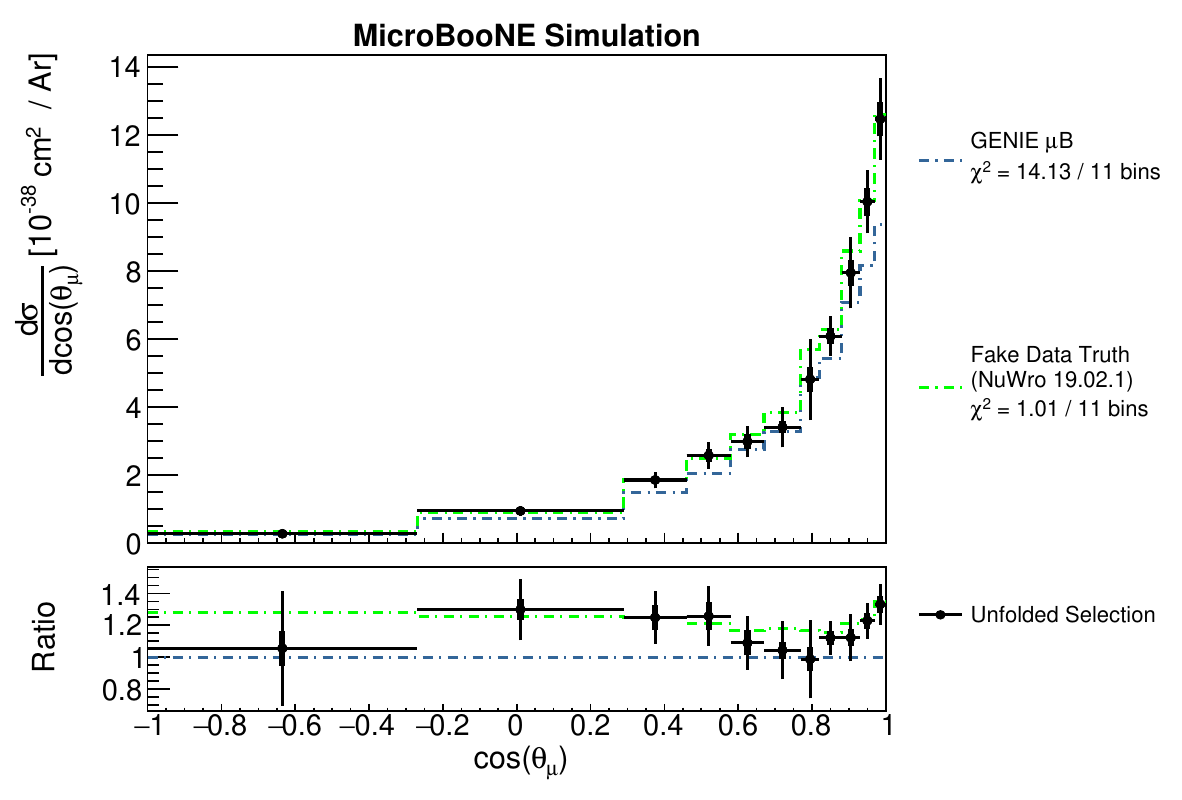}
    \caption{Unfolded $\cos(\theta_{\mu})$ cross section using NuWro fake data.}
    \label{fig:unfolded_mu_angle}
\end{figure}

\begin{figure}[!htb]
    \centering
    \includegraphics[width=0.8\textwidth]{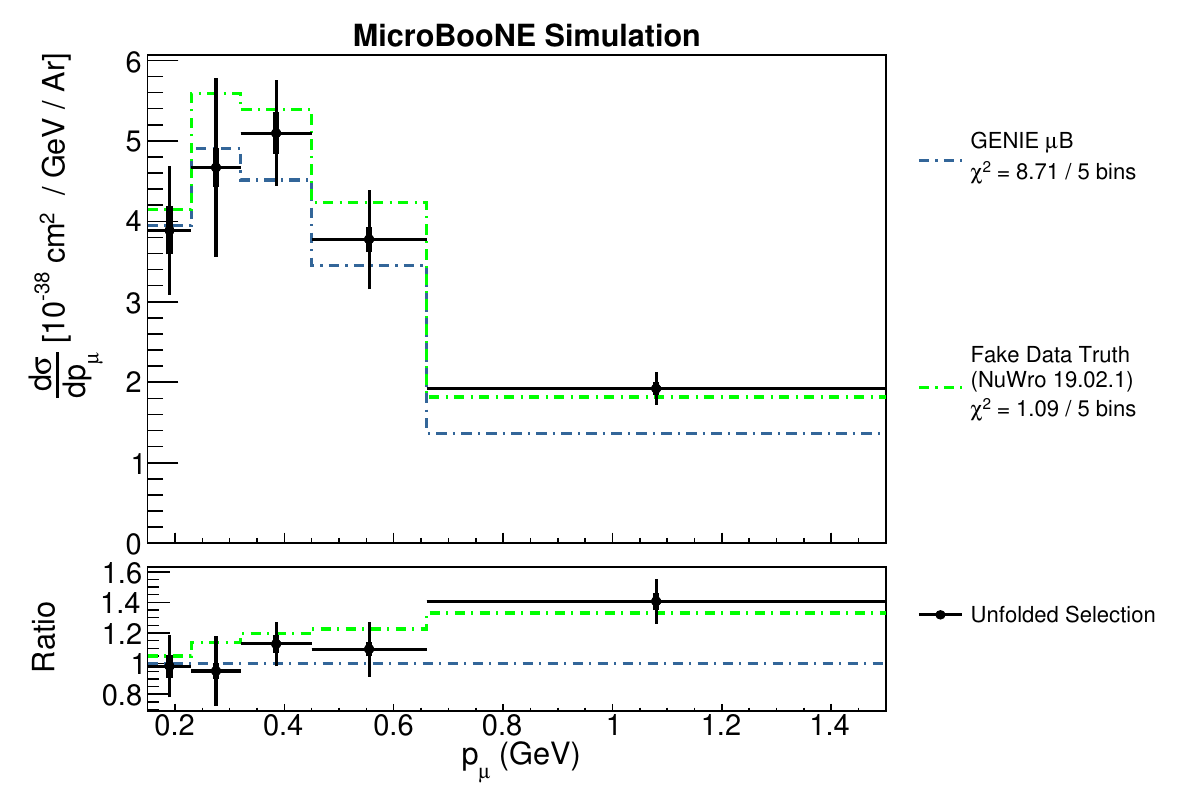}
    \caption{Unfolded $p_{\mu}$ cross section using NuWro fake data.}
    \label{fig:unfolded_mu_momentum}
\end{figure}

\begin{figure}[!htb]
    \centering
    \includegraphics[width=0.8\textwidth]{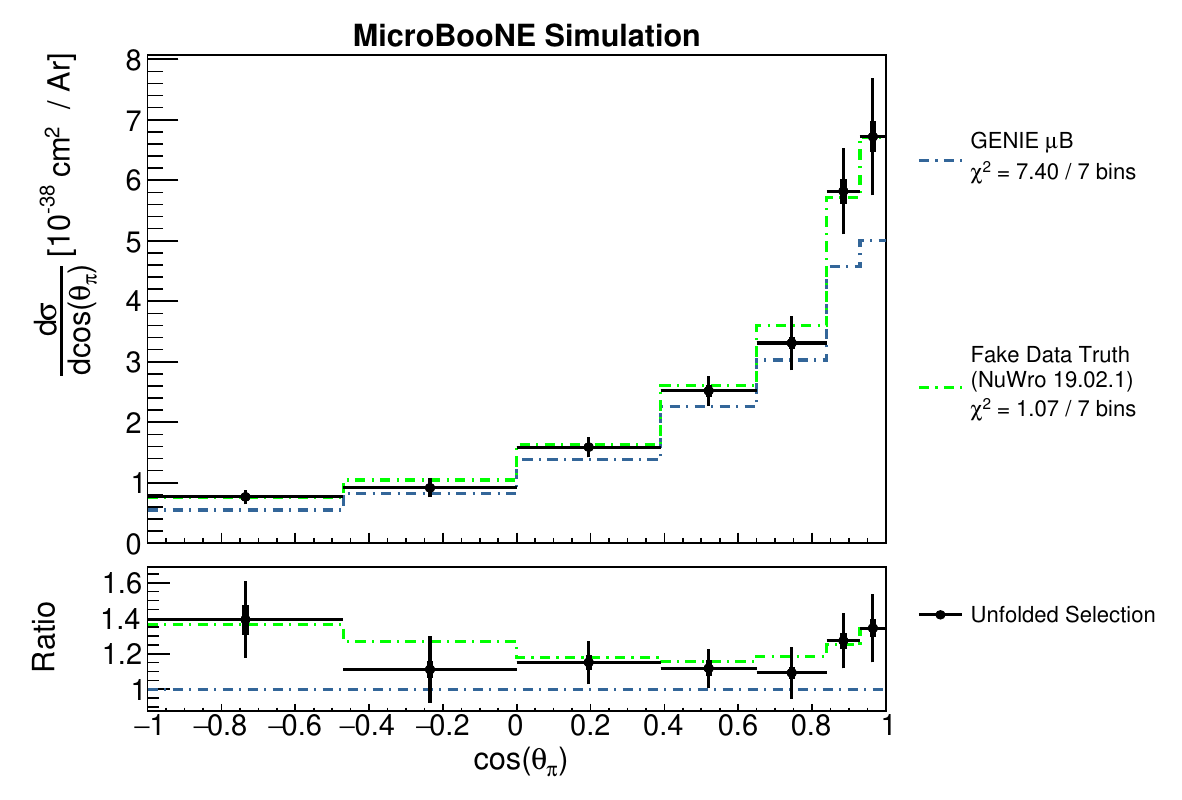}
    \caption{Unfolded $\cos(\theta_{\pi})$ cross section using NuWro fake data.}
    \label{fig:unfolded_pi_angle}
\end{figure}

\begin{figure}[!htb]
    \centering
    \includegraphics[width=0.8\textwidth]{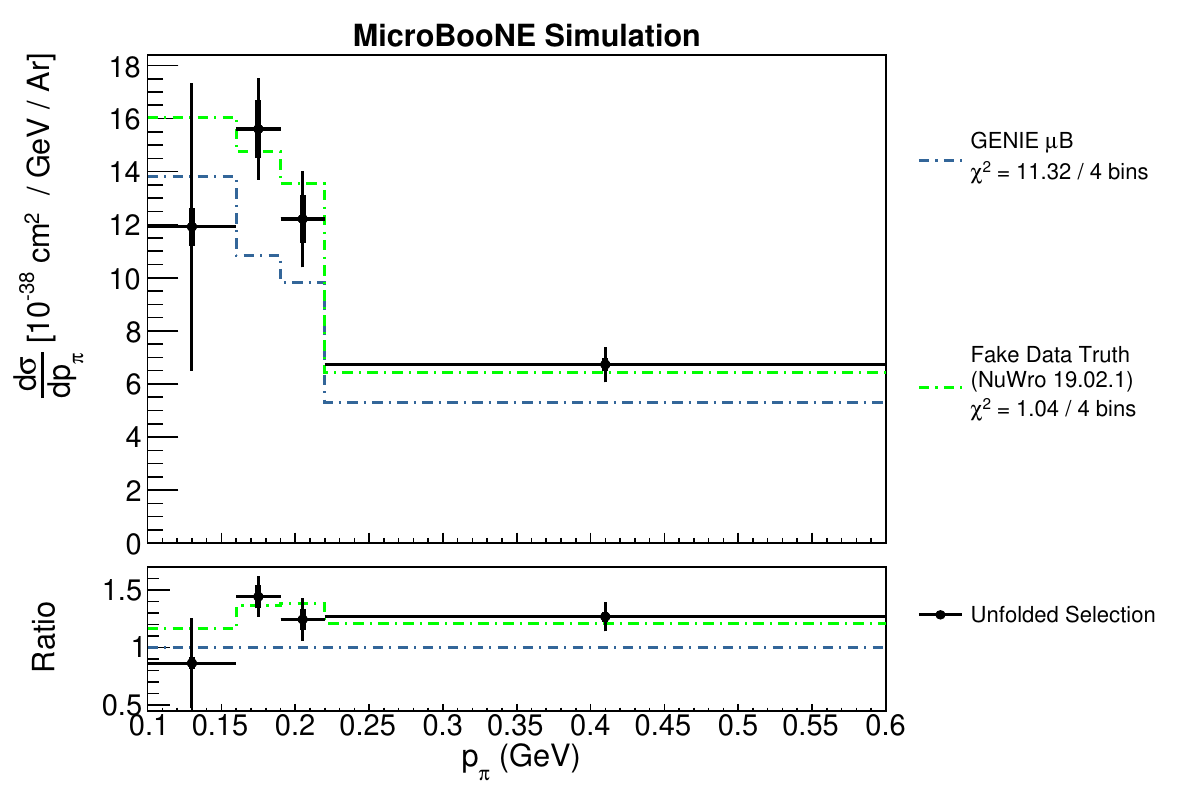}
    \caption{Unfolded $p_{\pi}$ cross section using NuWro fake data.}
    \label{fig:unfolded_pi_momentum}
\end{figure}

\begin{figure}[!htb]
    \centering
    \includegraphics[width=0.8\textwidth]{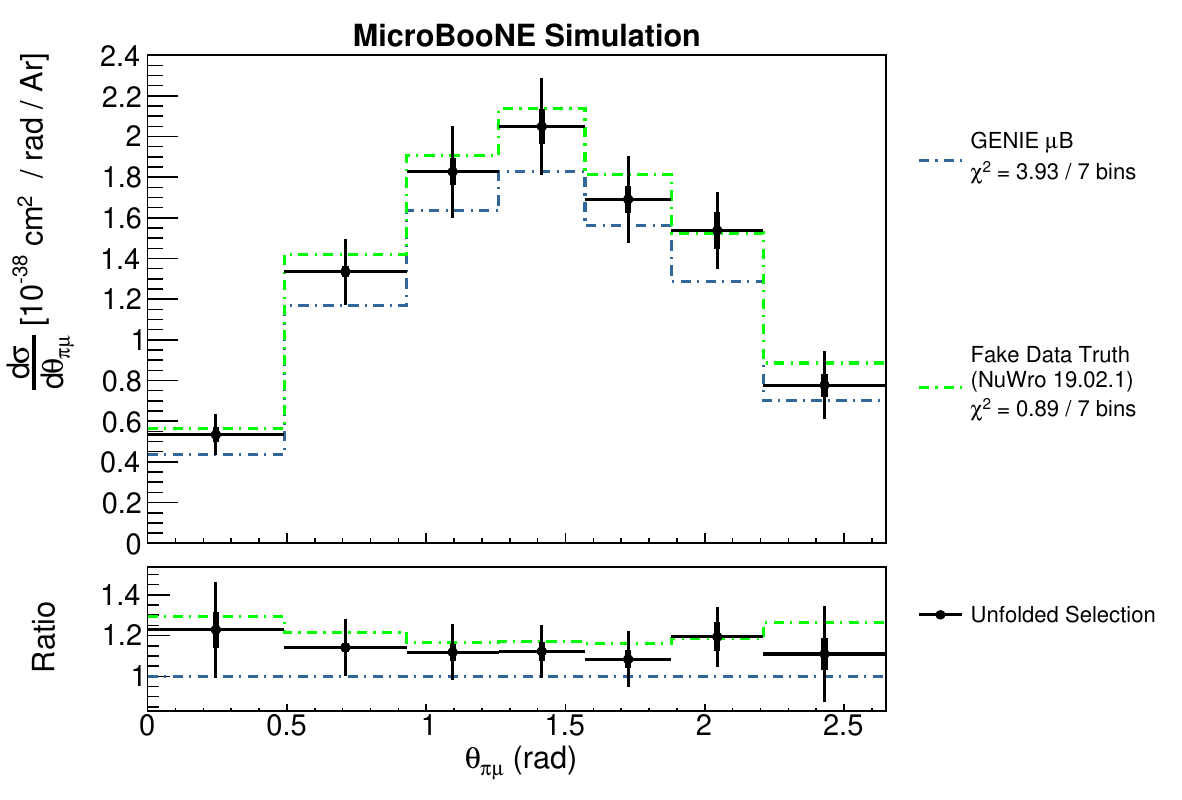}
    \caption{Unfolded $\theta_{\mu \pi}$ cross section using NuWro fake data.}
    \label{fig:unfolded_opening_angle}
\end{figure}

The unfolding procedure was validated using simulated events generated with NuWro version \texttt{19.02.1} treated as fake data. Only interaction model and statistical uncertainties are considered, as everything else of the simulation, such as flux and detector modeling, is identical to the simulated GENIE central value prediction. The unfolded fake data points are then compared to the corresponding true NuWro cross sections and show consistently good agreement as shown in Figs. \ref{fig:unfolded_total}, \ref{fig:unfolded_mu_angle}, \ref{fig:unfolded_mu_momentum}, \ref{fig:unfolded_pi_angle}, \ref{fig:unfolded_pi_momentum}, and \ref{fig:unfolded_opening_angle}.

\clearpage
\section{\texorpdfstring{$\chi^2$}{Chi-squared} Distributions for PID}
\label{sec:chi2_score_dist}
\begin{figure}[!htb] %
    \centering
    \begin{overpic}[width=0.6\textwidth]{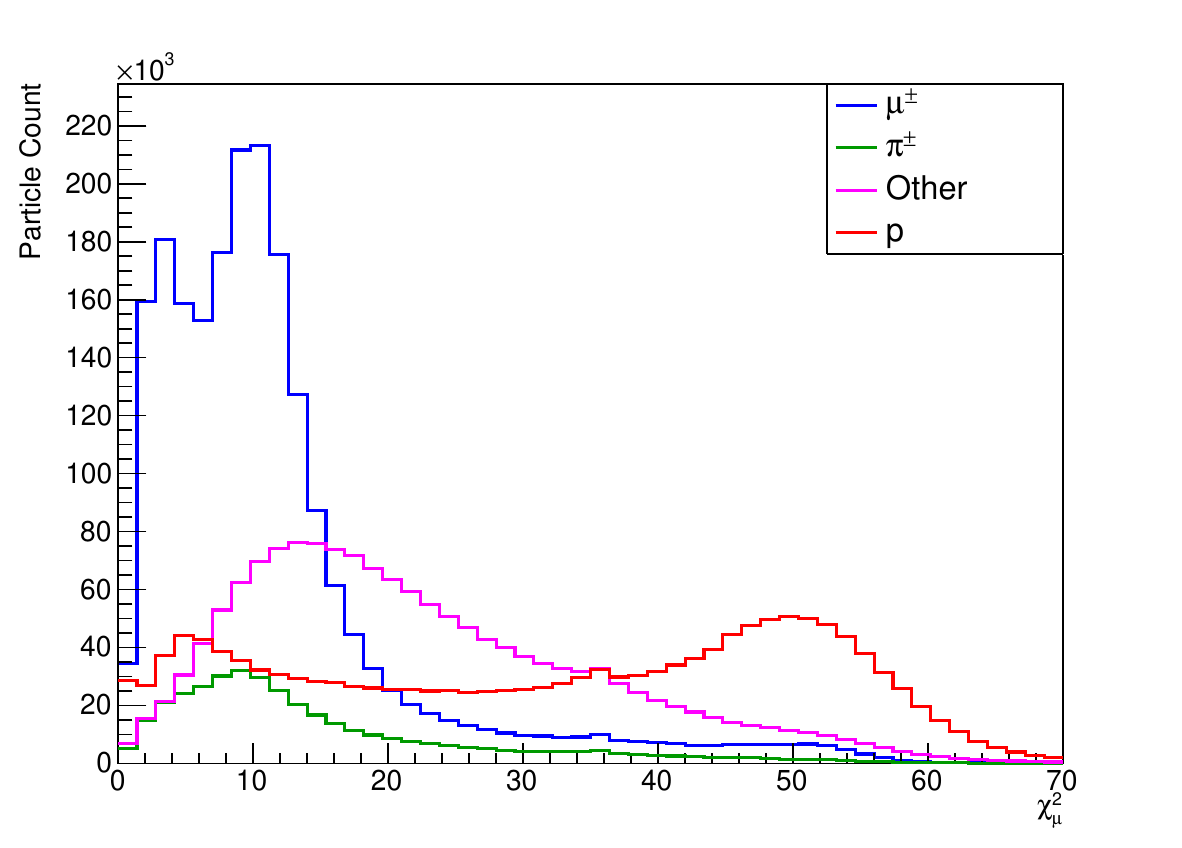}
         \put(46,66){\textbf{MicroBooNE Simulation}}
    \end{overpic}
    \caption{Particle identification metric \(\chi_\mu^2\) for simulated GENIE particles. The ends of many muon tracks lie outside the detector, resulting in score differences between contained and uncontained muons, and producing a double-peaked distribution.}
    \label{fig:chi2_mu}
\end{figure}

\begin{figure}[!htb] %
    \centering
    \begin{overpic}[width=0.6\textwidth]{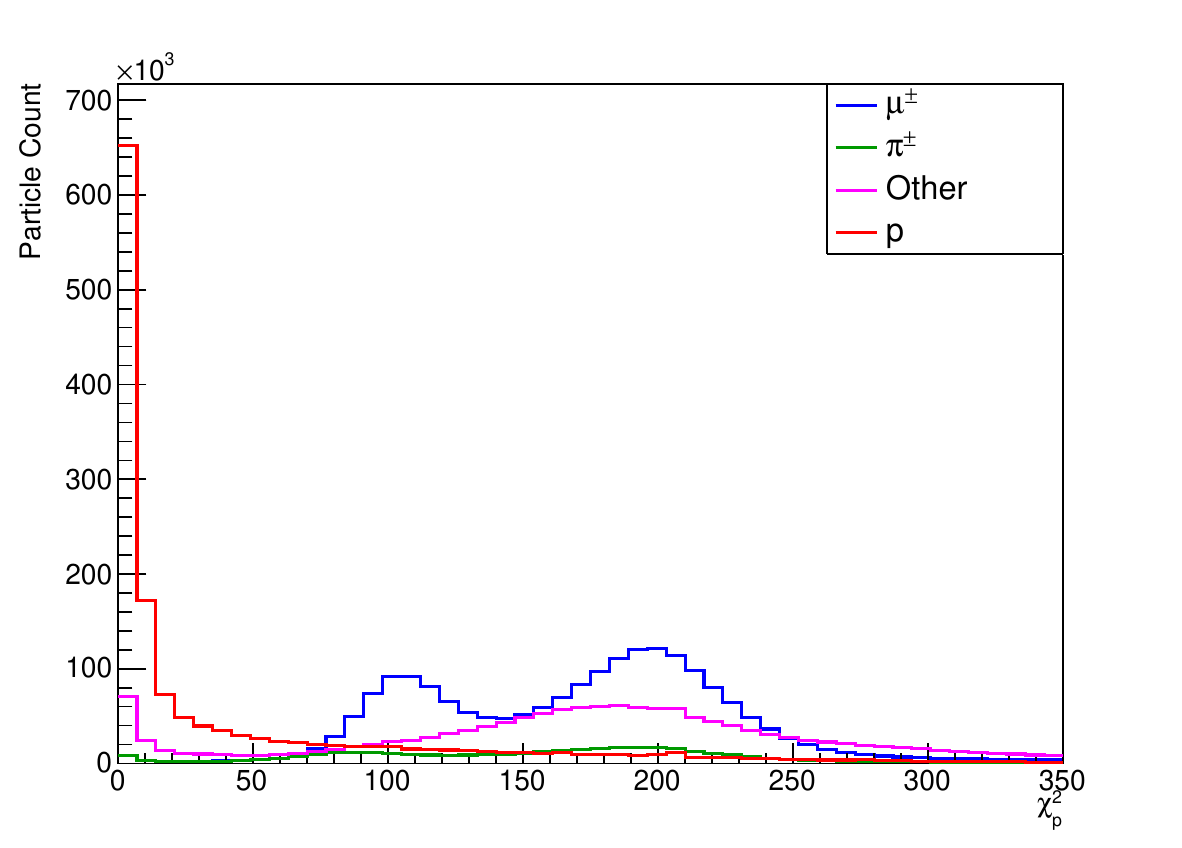}
         \put(46,66){\textbf{MicroBooNE Simulation}}
    \end{overpic}
    \caption{Particle identification metric \(\chi_p^2\) for simulated GENIE particles.}
    \label{fig:chi2_p}
\end{figure}

The agreement of the energy loss curve, $dE/dx$, for reconstructed calorimetric hits at the end of a track is compared to simulated distributions for muons and protons via a \(\chi^2\) metric. Figures \ref{fig:chi2_mu} and \ref{fig:chi2_p} show the distributions of the \(\chi^2\) metric for simulated particles under the muon and proton hypotheses.

\clearpage
\section{Phase Space Restrictions}
\label{sec:phase_space}
\begin{figure}[!htb]
    \centering
    \begin{overpic}[width=0.8\textwidth]{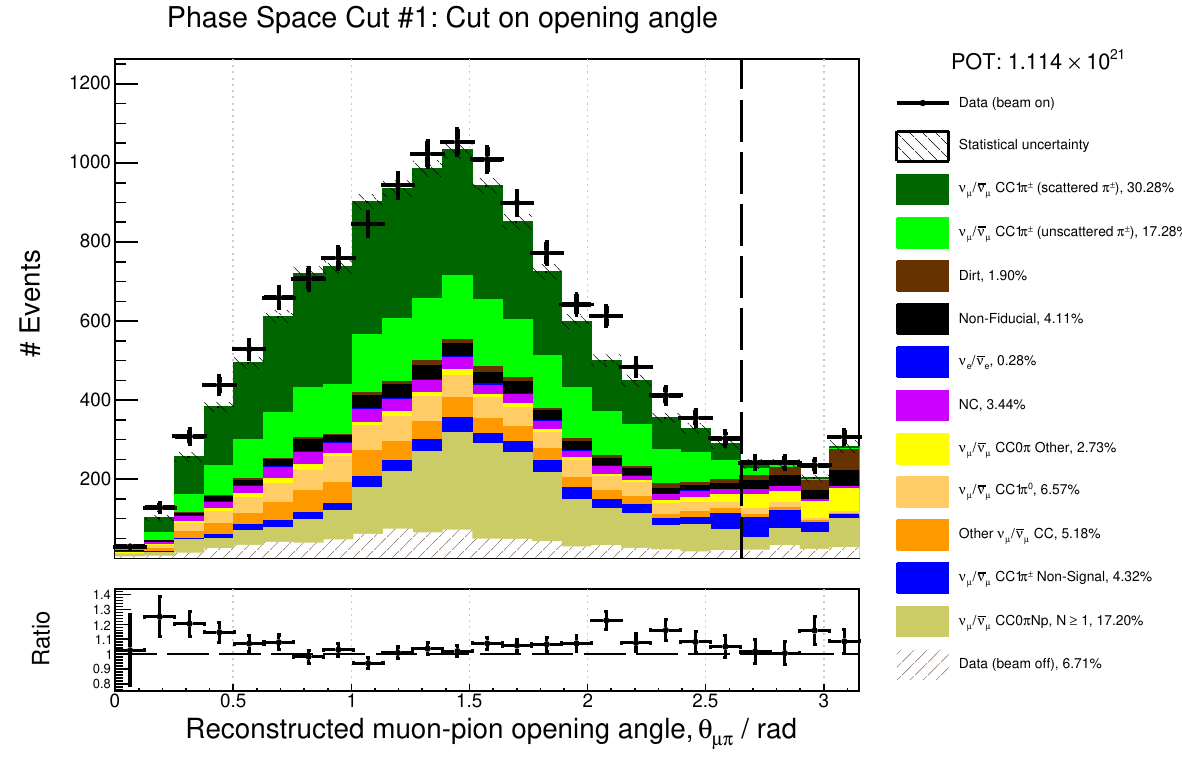}
         \put(76,62){\textbf{MicroBooNE}}
    \end{overpic}
    \caption{Phase space cut on the reconstructed opening angle. Events to the right of the line are rejected. Only statistical uncertainties are shown. The ratio shows the relative difference between data points and the prediction.}
    \label{fig:phase_space_1}
\end{figure}

\begin{figure}[!htb]
    \centering
    \begin{overpic}[width=0.8\textwidth]{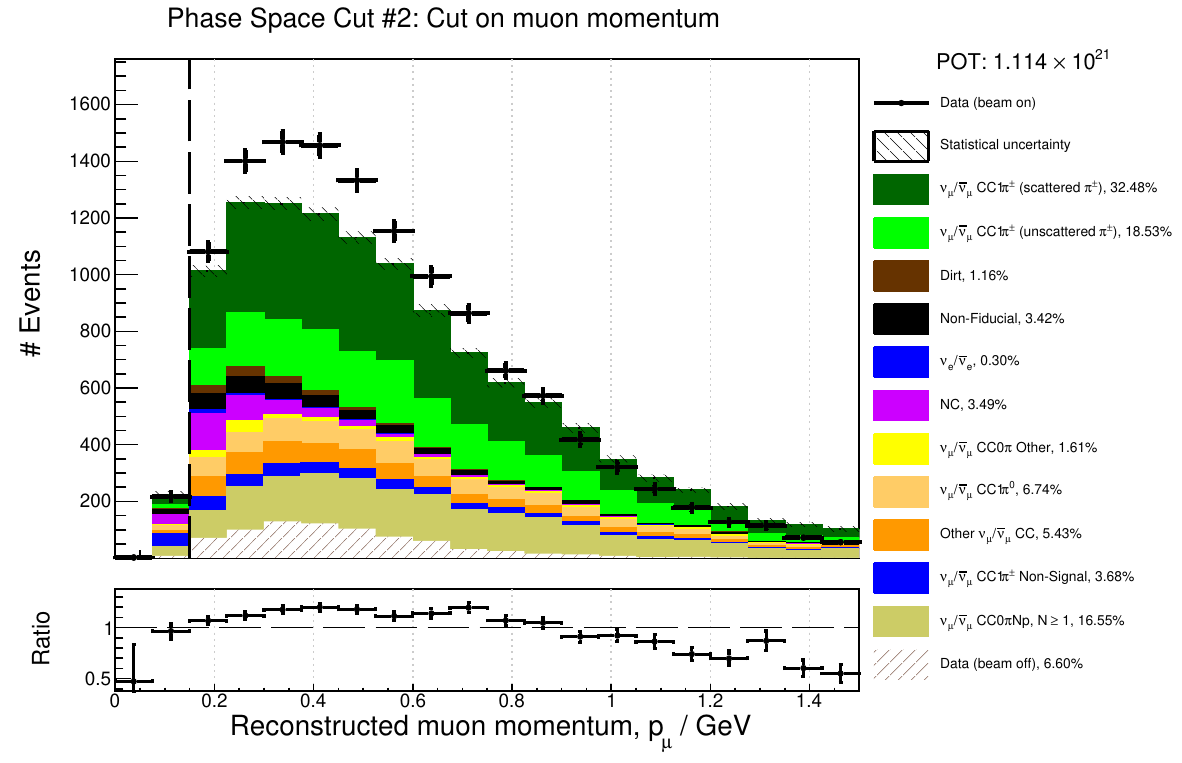}
         \put(76,62){\textbf{MicroBooNE}}
    \end{overpic}    
    \caption{Phase space cut on the reconstructed muon momentum (including contained and uncontained muons). Events to the left of the line are rejected. Only statistical uncertainties are shown. The ratio shows the relative difference between data points and the prediction.}
    \label{fig:phase_space_2}
\end{figure}

\begin{figure}[!htb]
    \centering
    \begin{overpic}[width=0.8\textwidth]{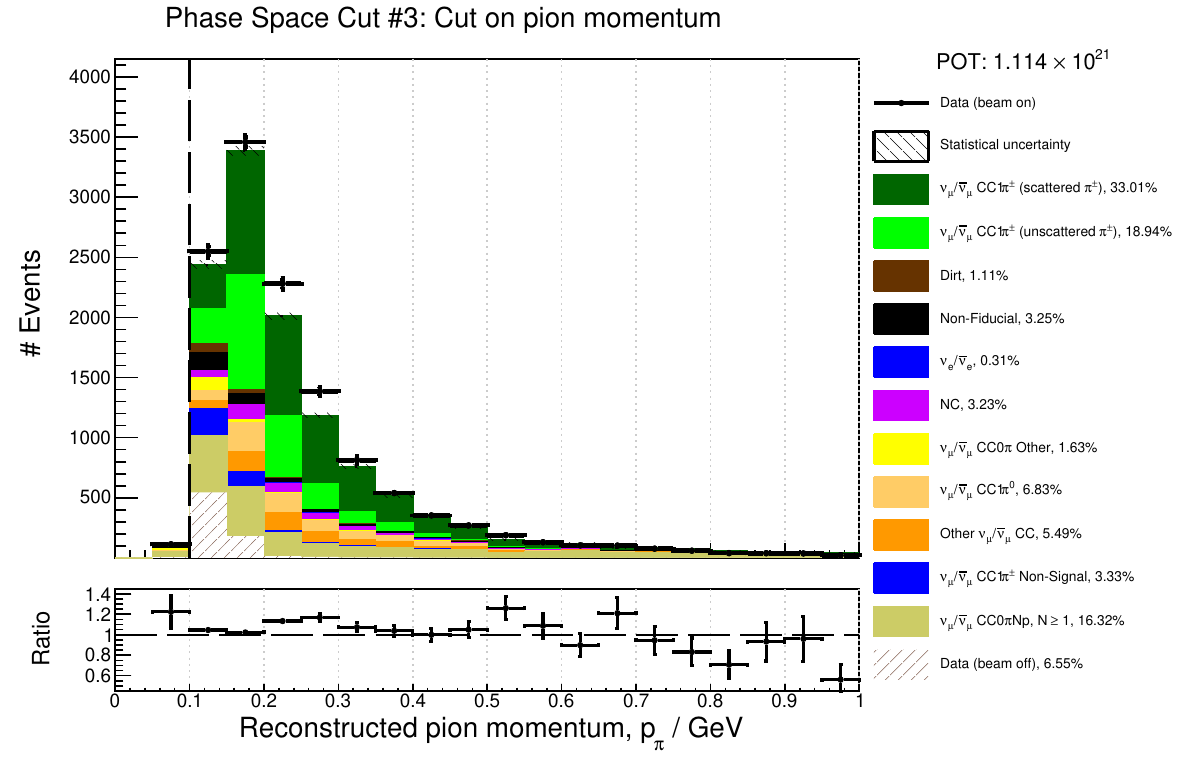}
         \put(76,62){\textbf{MicroBooNE}}
    \end{overpic}
    \caption{Phase space cut on the reconstructed pion momentum (including scattered and unscattered pions). Events to the left of the line are rejected. Only statistical uncertainties are shown. The ratio shows the relative difference between data points and the prediction.}
    \label{fig:phase_space_3}
\end{figure}

The three Figs. \ref{fig:phase_space_1}, \ref{fig:phase_space_2}, and \ref{fig:phase_space_3} show the phase space restrictions for the reconstructed variables. The restrictions are not applied in truth in these plots. The high background contamination at high muon-pion opening angles comes from long tracks being incorrectly split up during reconstruction and appearing as two back-to-back tracks.

\clearpage
\section{Fractional uncertainties}
\label{sec:fracUncert}
The Figs. \ref{fig:plot_frac_slice_07}, \ref{fig:plot_frac_slice_00}, \ref{fig:plot_frac_slice_02}, \ref{fig:plot_frac_slice_03}, \ref{fig:plot_frac_slice_05}, and \ref{fig:plot_frac_slice_06} show the fractional uncertainties for the unfolded cross-section distributions.

\newcommand{\legendX}{94.7}    %
\newcommand{\legendY}{2.3}    %
\newcommand{\legendW}{0.41\textwidth} %

\newcommand{\simlabelThree}{\put(50,62){\normalsize\bfseries MicroBooNE Simulation}}
\newcommand{\yLabelThree}{\put(-4,24){\rotatebox{90}{\small Fractional Uncertainties}}}

\begin{figure}[htbp]
    \centering
    \hspace{-0.35\textwidth}
    \begin{overpic}[width=0.6\textwidth,trim={1cm 0.88cm 0cm 0cm},clip]
      {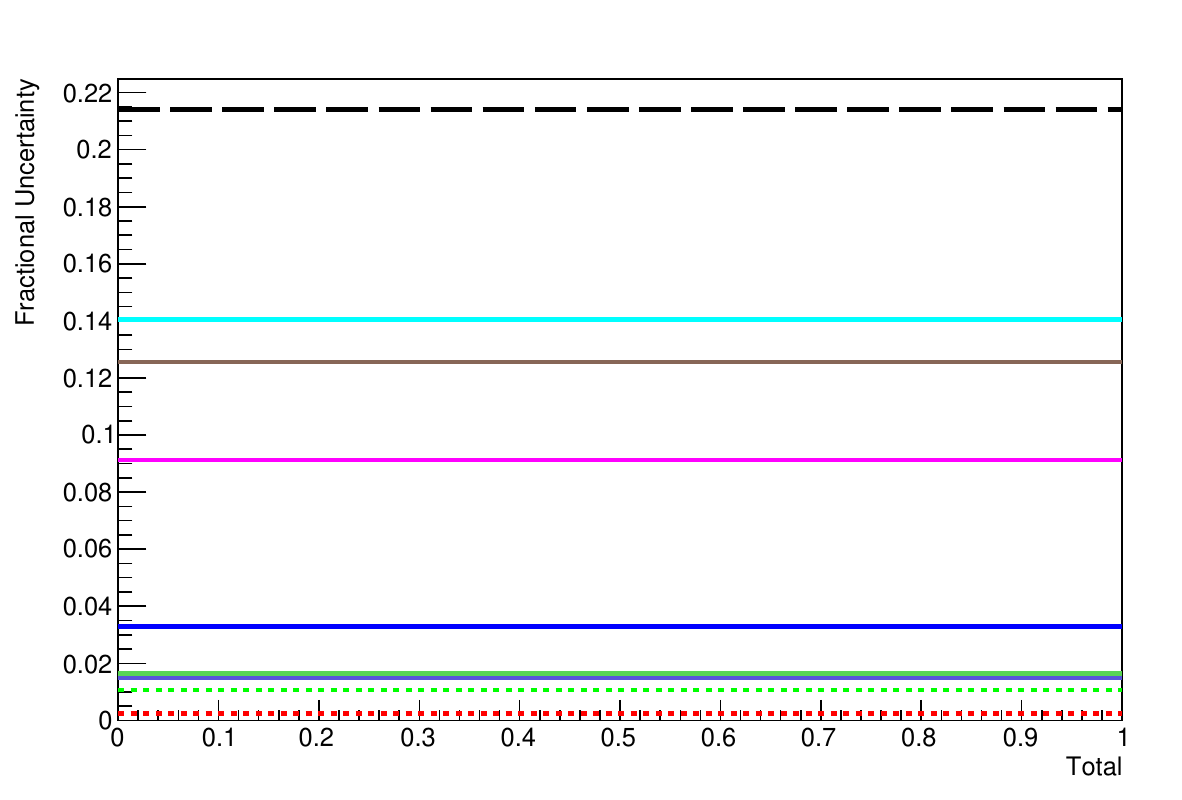}
        \simlabelThree
        \yLabelThree
        \put(50,-4){\small Total}
        \put(\legendX,\legendY){%
            \raisebox{0pt}{\includegraphics[width=\legendW,trim={2cm 1.8cm 0cm 0cm},clip]
              {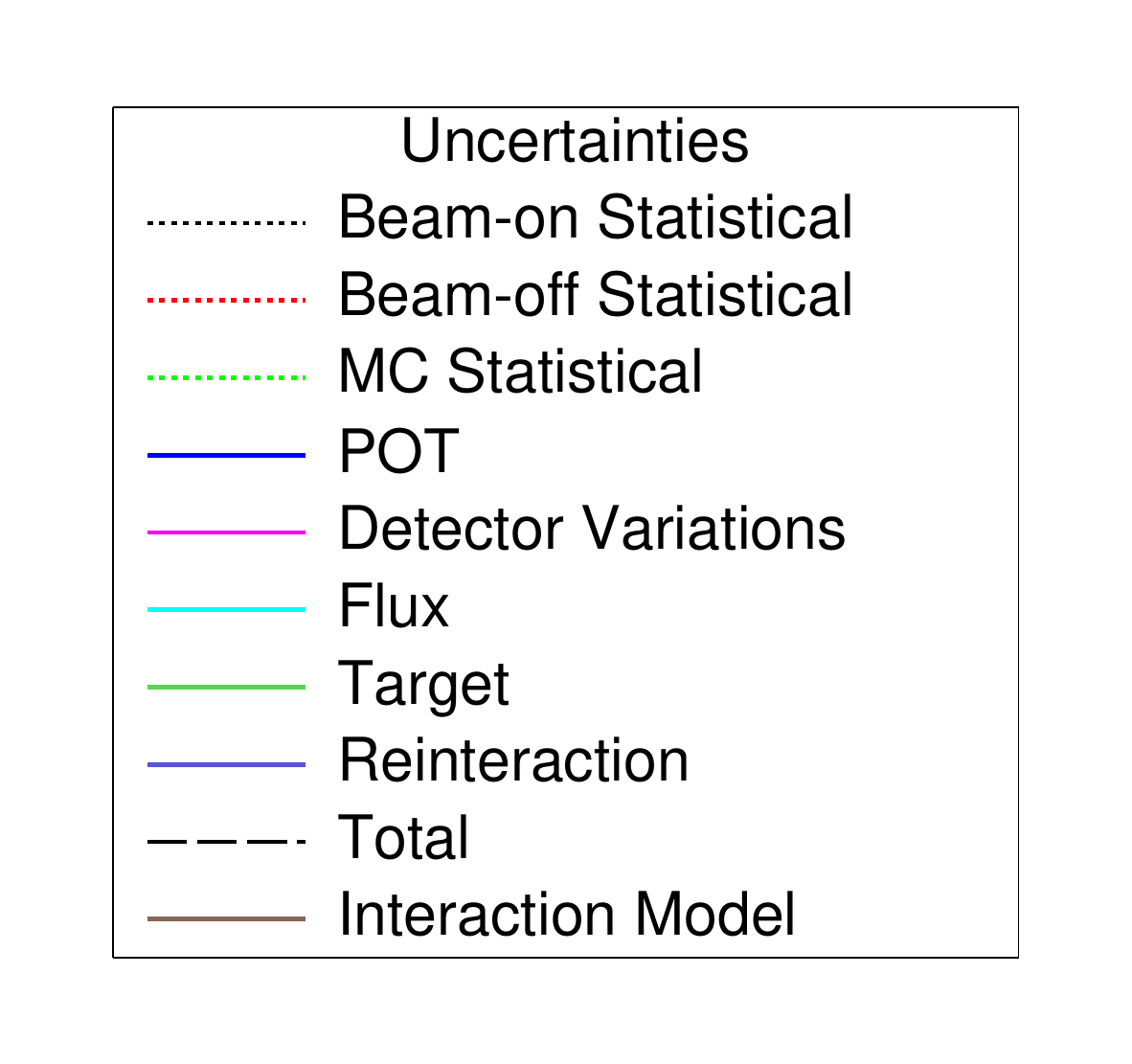}}%
        }
    \end{overpic}\vspace{0.3cm}
    \caption[Fractional uncertainties]{Fractional uncertainties for the reported total cross section.}
    \label{fig:plot_frac_slice_07}
\end{figure}

\begin{figure}[htbp]
    \centering
    \hspace{-0.35\textwidth}
    \begin{overpic}[width=0.6\textwidth,trim={1cm 0.88cm 0cm 0cm},clip]
      {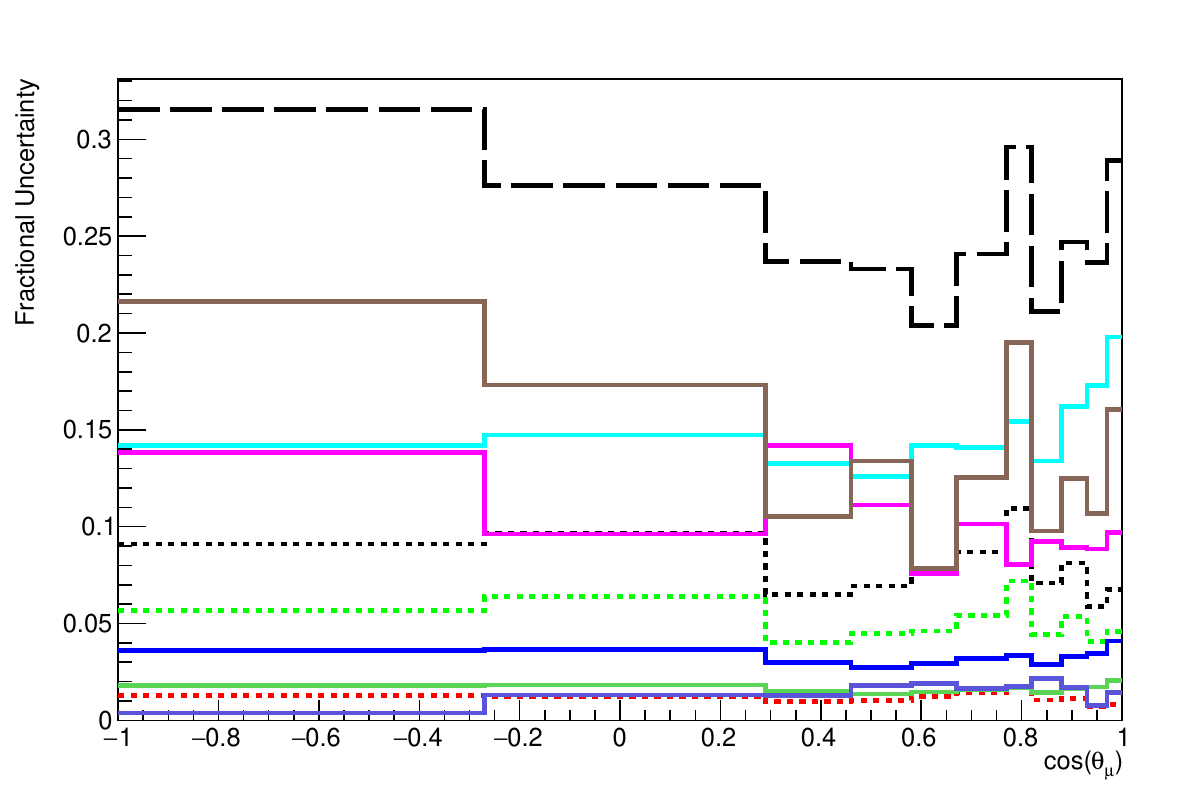}
        \simlabelThree
        \yLabelThree
        \put(48,-4){\small True \(\cos(\theta_\mu)\)}
        \put(\legendX,\legendY){%
            \raisebox{0pt}{\includegraphics[width=\legendW,trim={2cm 1.8cm 0cm 0cm},clip]
              {supplemental_plots/plot_frac_uncertainties_legend__bnb_fixedBackground_mergedOverflow_containedMuXSec_reco_total.pdf}}%
        }
    \end{overpic}\vspace{0.3cm}
    \caption[Fractional uncertainties]{Fractional uncertainties for the unfolded \(\cos\theta_\mu\) differential cross-section bins.}
    \label{fig:plot_frac_slice_00}
\end{figure}

\begin{figure}[htbp]
    \centering
    \hspace{-0.35\textwidth}
    \begin{overpic}[width=0.6\textwidth,trim={1cm 0.88cm 0cm 0cm},clip]
      {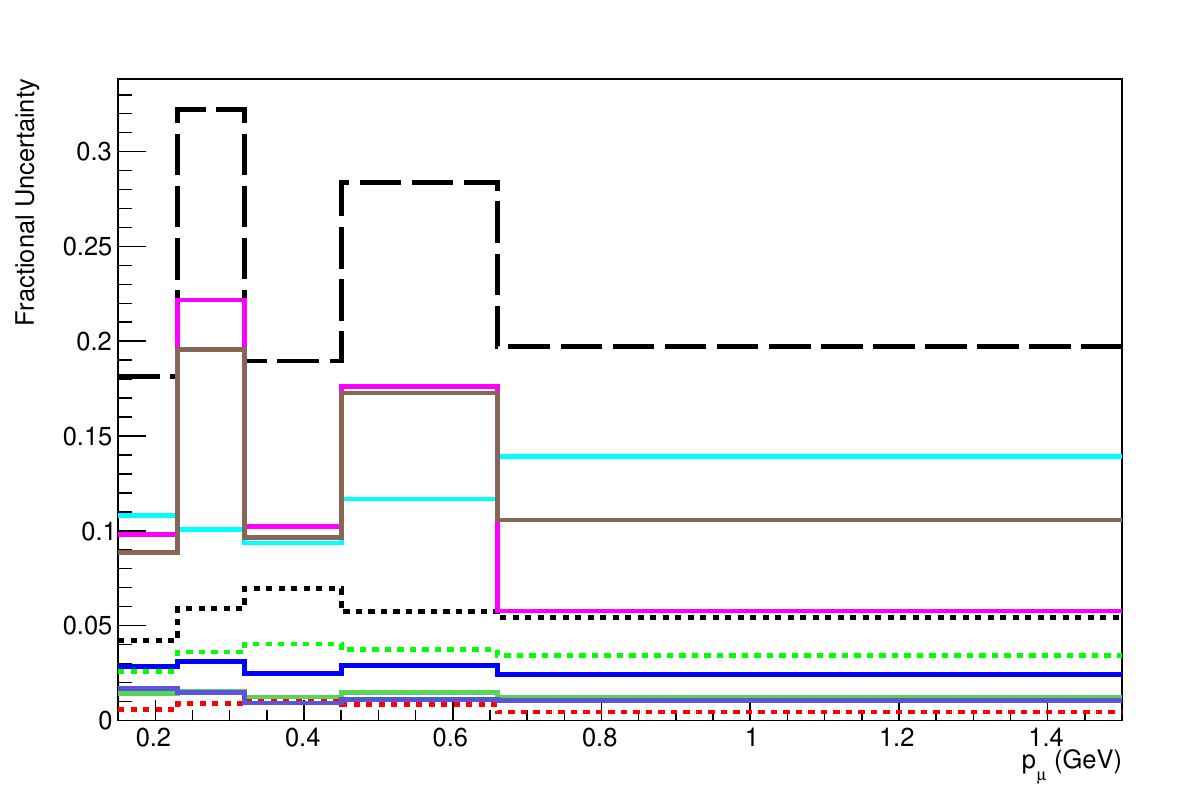}
        \simlabelThree
        \yLabelThree
        \put(45,-4){\small True \(p_\mu\) (GeV)}
        \put(\legendX,\legendY){%
            \raisebox{0pt}{\includegraphics[width=\legendW,trim={2cm 1.8cm 0cm 0cm},clip]
              {supplemental_plots/plot_frac_uncertainties_legend__bnb_fixedBackground_mergedOverflow_containedMuXSec_reco_total.pdf}}%
        }
    \end{overpic}\vspace{0.3cm}
    \caption[Fractional uncertainties]{Fractional uncertainties for the unfolded \(p_\mu\) differential cross-section bins.}
    \label{fig:plot_frac_slice_02}
\end{figure}

\begin{figure}[htbp]
    \centering
    \hspace{-0.35\textwidth}
    \begin{overpic}[width=0.6\textwidth,trim={1cm 0.88cm 0cm 0cm},clip]
      {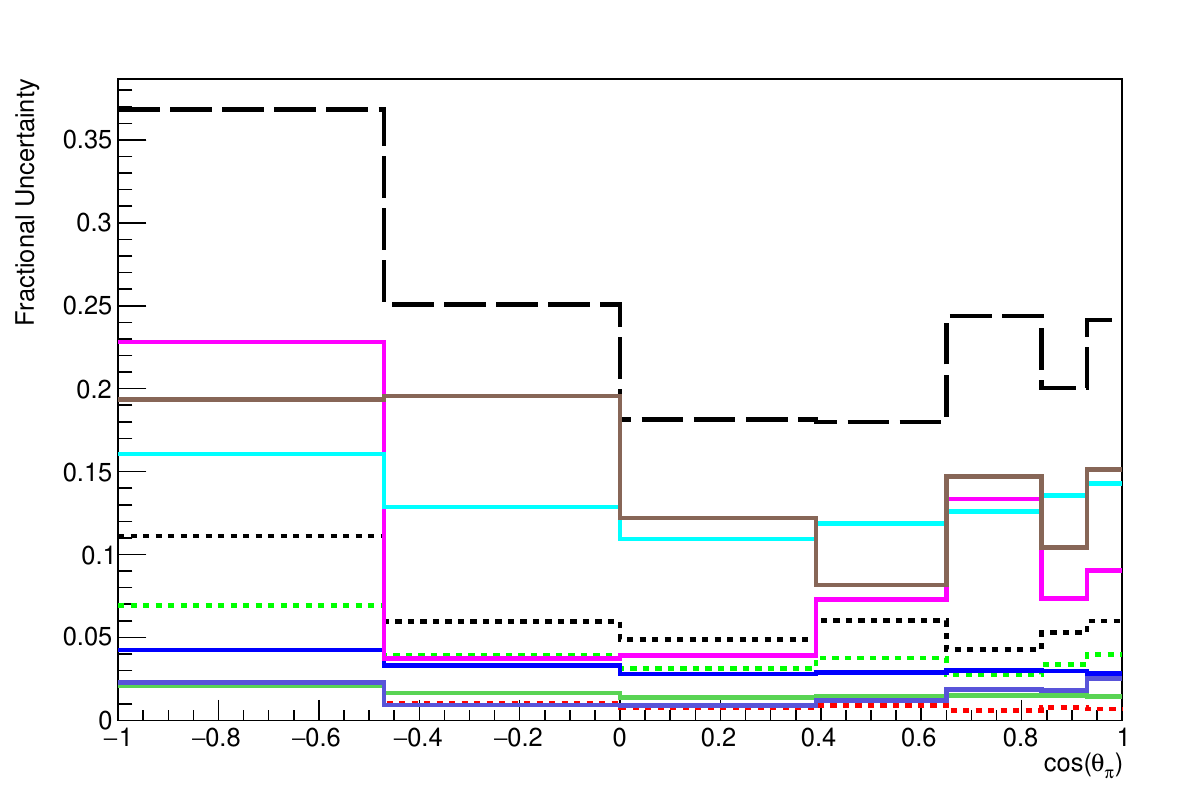}
        \simlabelThree
        \yLabelThree
        \put(45,-4){\small True \(\cos(\theta_\pi)\)}
        \put(\legendX,\legendY){%
            \raisebox{0pt}{\includegraphics[width=\legendW,trim={2cm 1.8cm 0cm 0cm},clip]
              {supplemental_plots/plot_frac_uncertainties_legend__bnb_fixedBackground_mergedOverflow_containedMuXSec_reco_total.pdf}}%
        }
    \end{overpic}\vspace{0.3cm}
    \caption[Fractional uncertainties]{Fractional uncertainties for the unfolded \(\cos\theta_\pi\) differential cross-section bins.}
    \label{fig:plot_frac_slice_03}
\end{figure}

\begin{figure}[htbp]
    \centering
    \hspace{-0.35\textwidth}
    \begin{overpic}[width=0.6\textwidth,trim={1cm 0.88cm 0cm 0cm},clip]
      {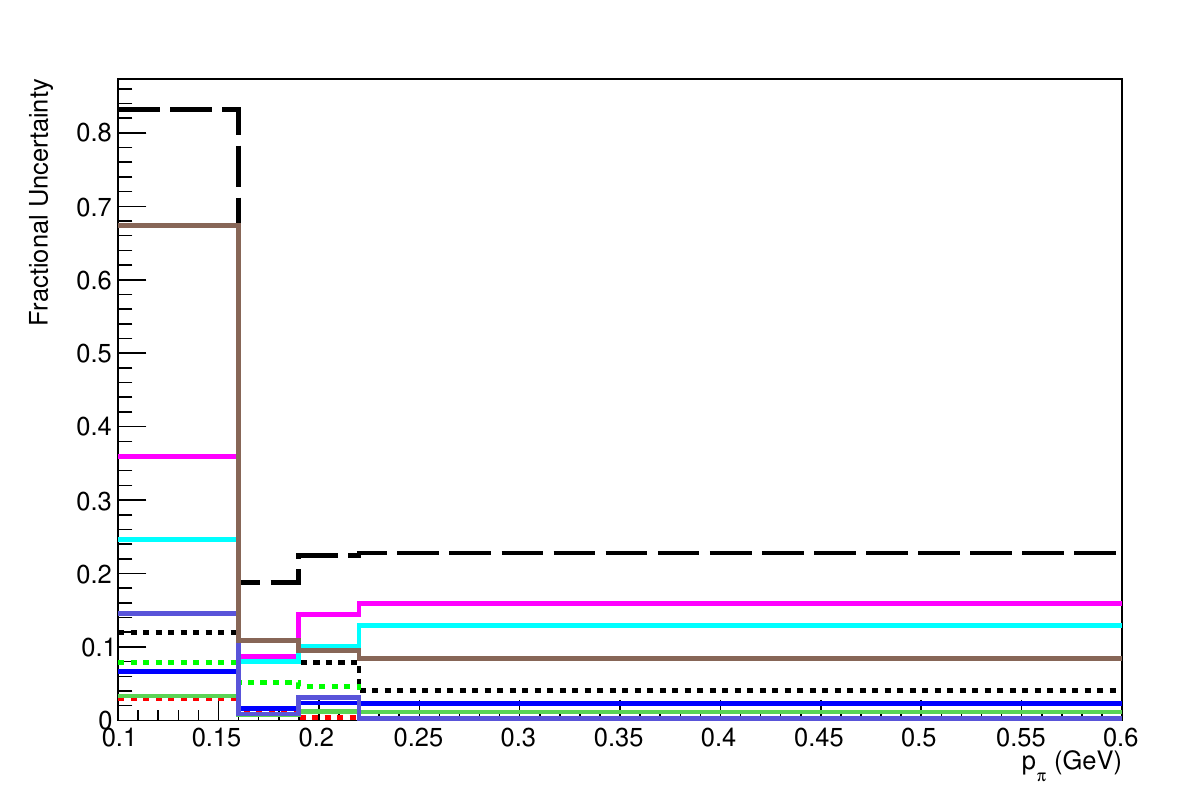}
        \simlabelThree
        \yLabelThree
        \put(45,-4){\small True \(p_\pi\) (GeV)}
        \put(\legendX,\legendY){%
            \raisebox{0pt}{\includegraphics[width=\legendW,trim={2cm 1.8cm 0cm 0cm},clip]
              {supplemental_plots/plot_frac_uncertainties_legend__bnb_fixedBackground_mergedOverflow_containedMuXSec_reco_total.pdf}}%
        }
    \end{overpic}\vspace{0.3cm}
    \caption[Fractional uncertainties]{Fractional uncertainties for the unfolded \(p_\pi\) differential cross-section bins.}
    \label{fig:plot_frac_slice_05}
\end{figure}

\begin{figure}[htbp]
    \centering
    \hspace{-0.35\textwidth}
    \begin{overpic}[width=0.6\textwidth,trim={1cm 0.88cm 0cm 0cm},clip]
      {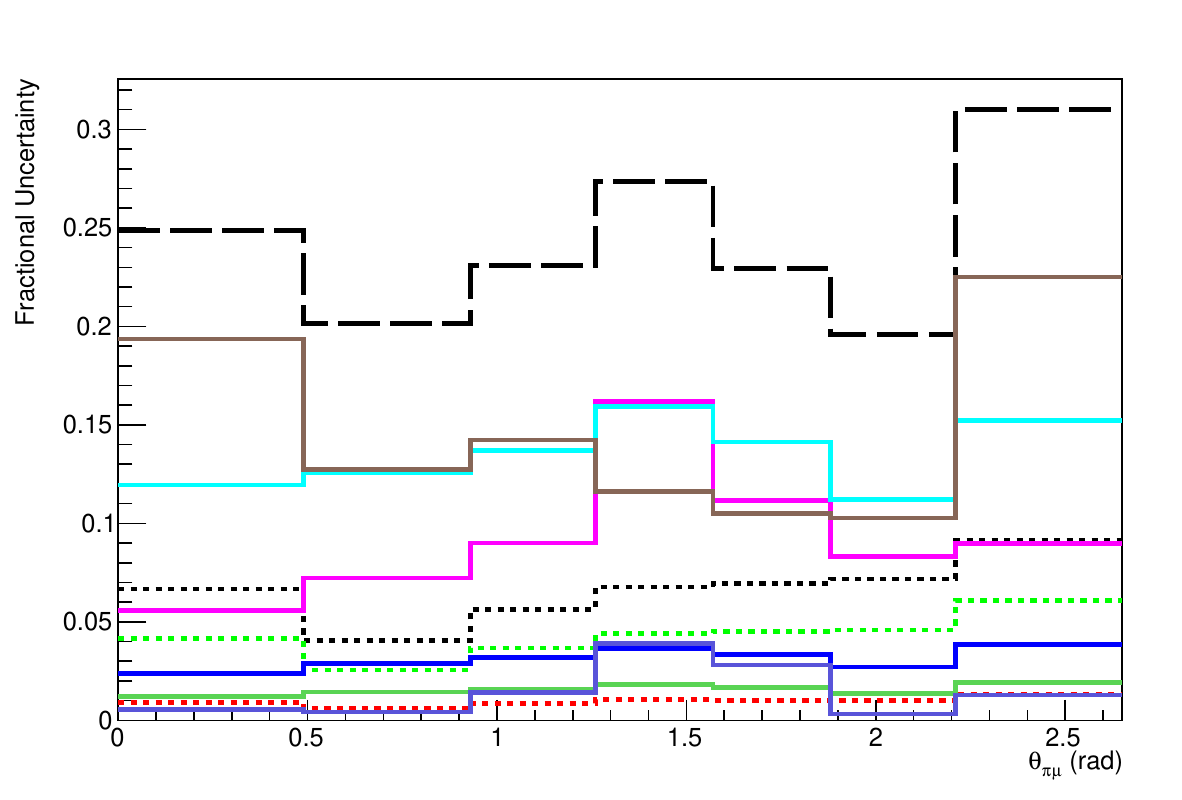}
        \simlabelThree
        \yLabelThree
        \put(45,-4){\small True \(\theta_{\mu\pi}\) (rad)}
        \put(\legendX,\legendY){%
            \raisebox{0pt}{\includegraphics[width=\legendW,trim={2cm 1.8cm 0cm 0cm},clip]
              {supplemental_plots/plot_frac_uncertainties_legend__bnb_fixedBackground_mergedOverflow_containedMuXSec_reco_total.pdf}}%
        }
    \end{overpic}\vspace{0.3cm}
    \caption[Fractional uncertainties]{Fractional uncertainties for the unfolded \(\theta_{\mu\pi}\) differential cross-section bins.}
    \label{fig:plot_frac_slice_06}
\end{figure}

\clearpage
\section{Generator Comparisons Without FSI}
\label{sec:noFSI}

\begin{figure}[!htb]
    \centering
    \includegraphics[width=0.7\textwidth]{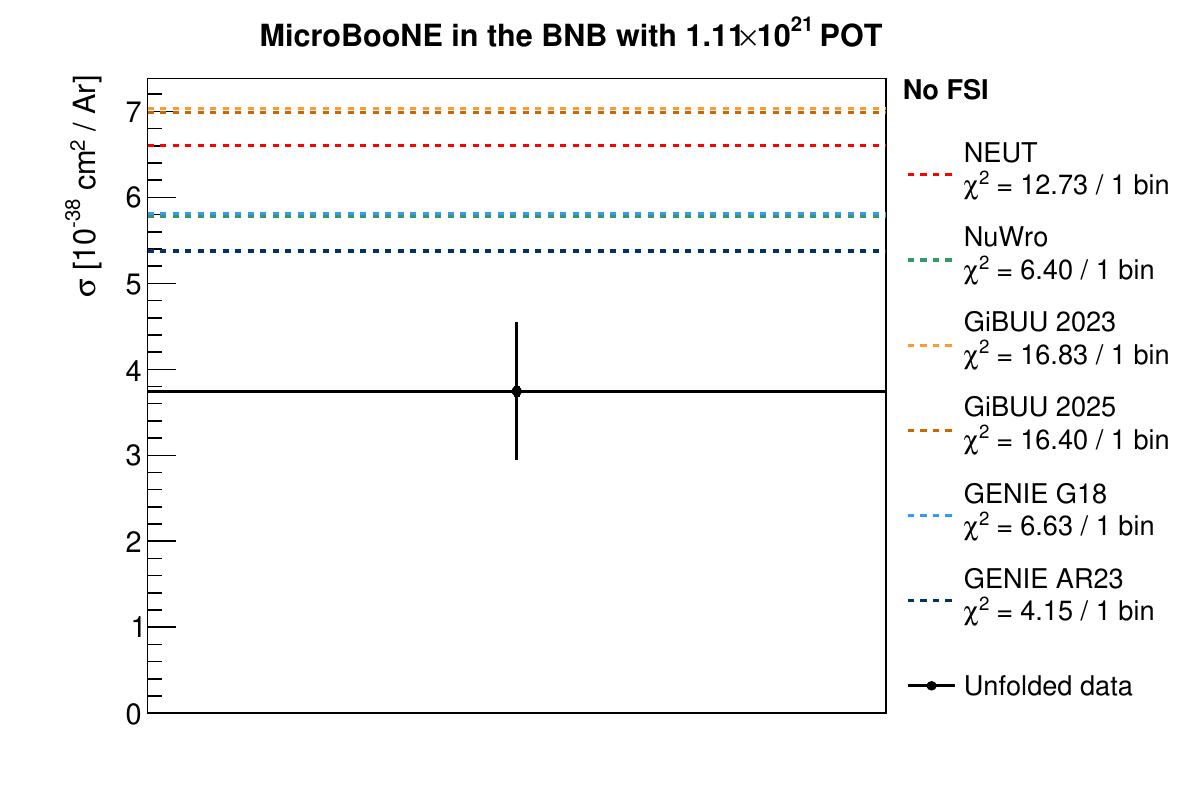}
    \caption{Unfolded total cross section in regularized truth space. The black dots represent the unfolded selection with the associated total uncertainty. A thicker inner error bar, that is very short here, shows the statistical uncertainty. The lines are generator predictions with FSI disabled.}
    \label{fig:total}
\end{figure}

\begin{figure}[!htb]
    \centering
    \includegraphics[width=0.7\textwidth]{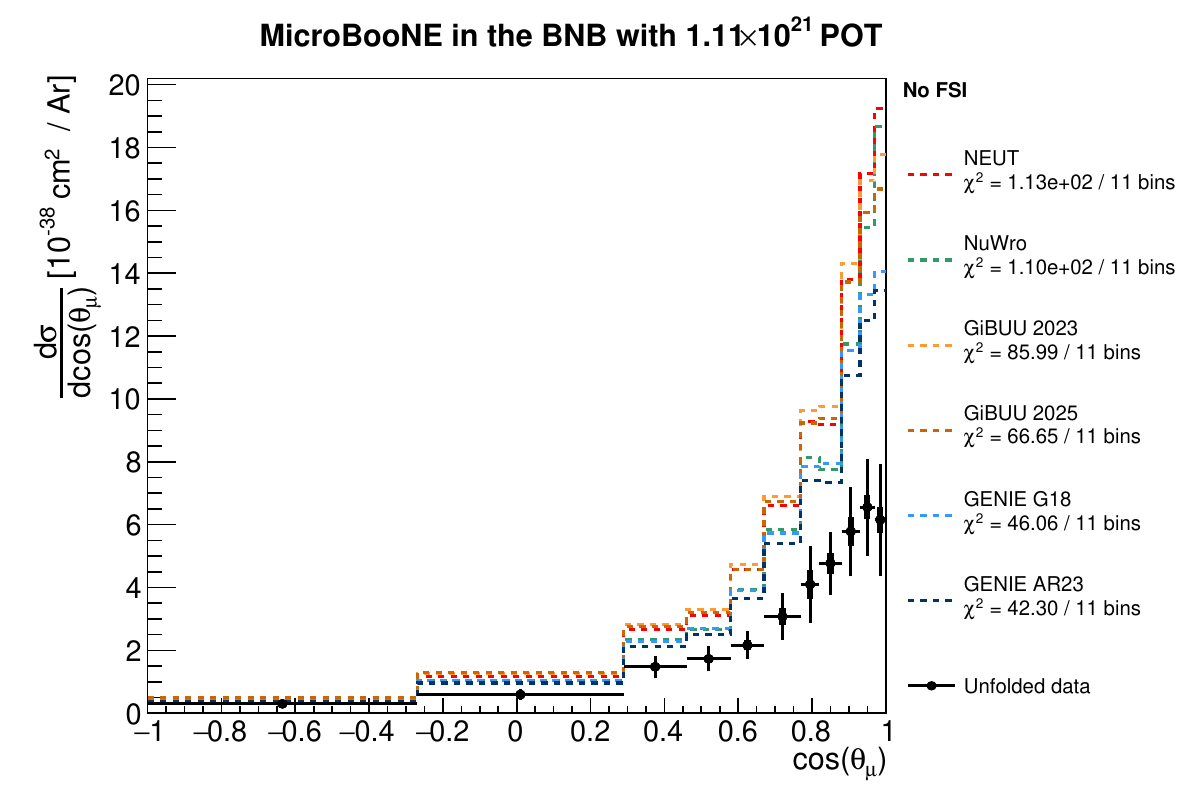}
    \caption{Unfolded muon angle cross section in regularized truth space. The black dots represent the unfolded selection with the associated total uncertainty. The thicker inner error bars are the statistical uncertainty only. The lines are generator predictions with FSI disabled.}
    \label{fig:xsec_00}
\end{figure}

\begin{figure}[!htb]
    \centering
    \includegraphics[width=0.7\textwidth]{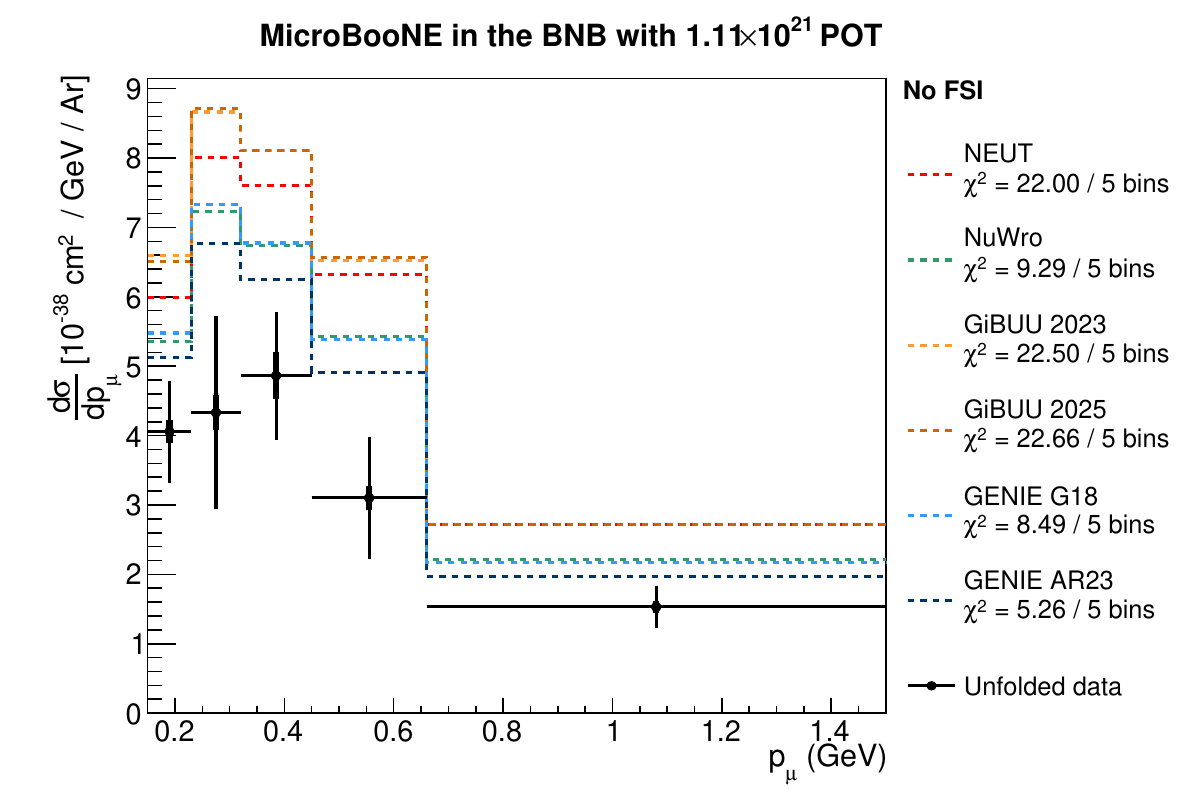}
    \caption{Unfolded muon momentum cross section in regularized truth space. The black dots represent the unfolded selection with the associated total uncertainty. The thicker inner error bars are the statistical uncertainty only. The lines are generator predictions with FSI disabled.}
    \label{fig:xsec_02}
\end{figure}

\begin{figure}[!htb]
    \centering
    \includegraphics[width=0.7\textwidth]{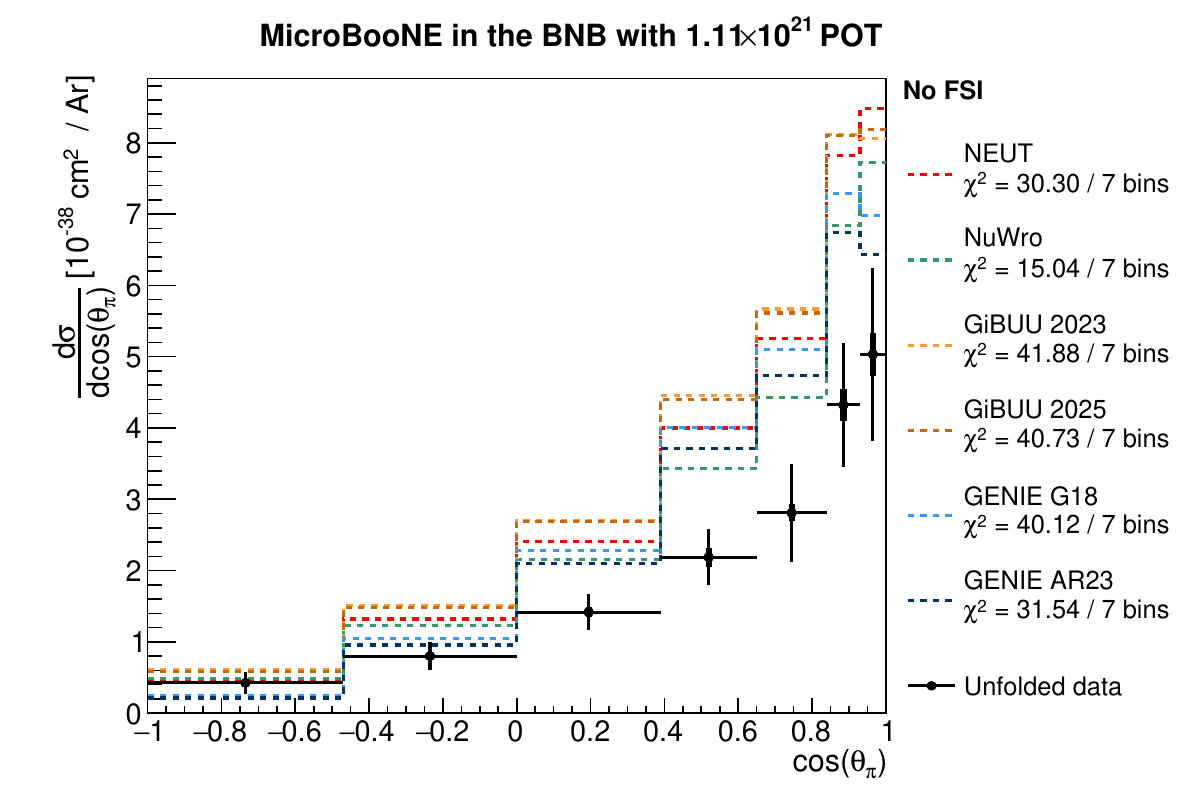}
    \caption{Unfolded pion angle cross section in regularized truth space. The black dots represent the unfolded selection with the associated total uncertainty. The thicker inner error bars are the statistical uncertainty only. The lines are generator predictions with FSI disabled.}
    \label{fig:xsec_03}
\end{figure}

\begin{figure}[!htb]
    \centering
    \includegraphics[width=0.7\textwidth]{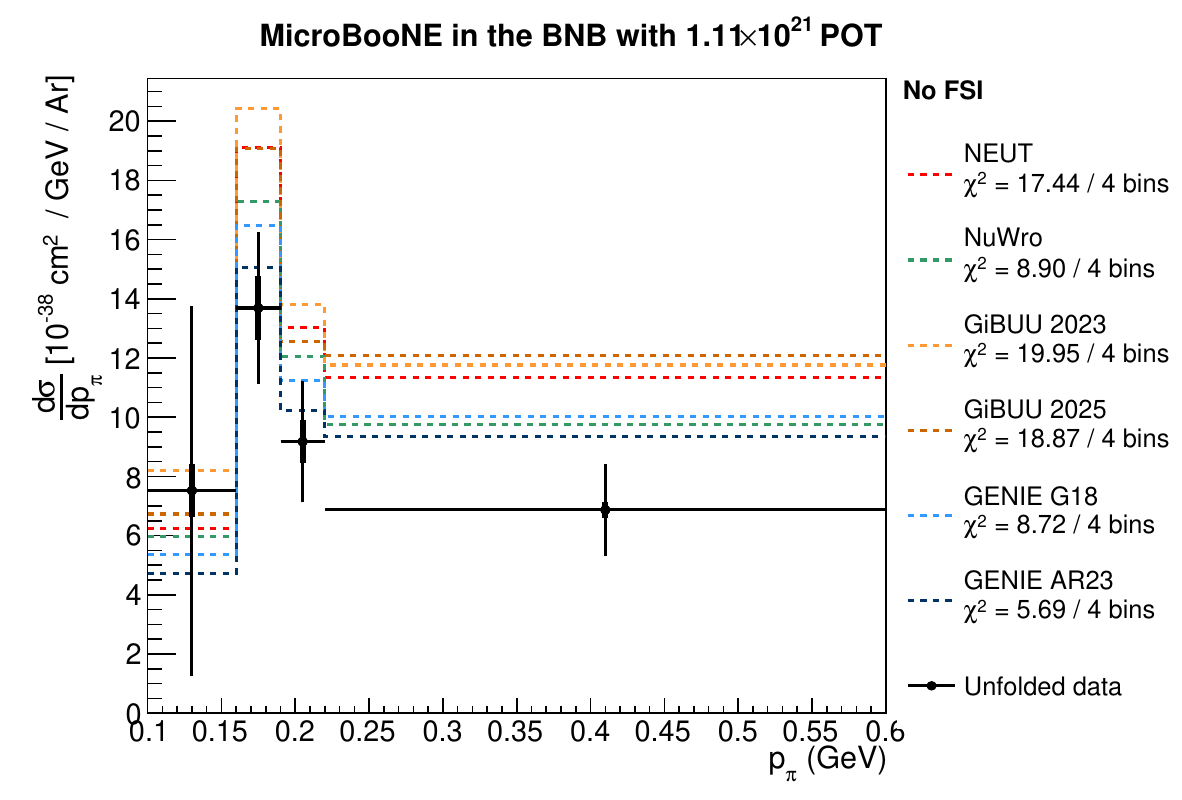}
    \caption{Unfolded pion momentum cross section in regularized truth space. The black dots represent the unfolded selection with the associated total uncertainty. The thicker inner error bars are the statistical uncertainty only. The lines are generator predictions with FSI disabled.}
    \label{fig:xsec_05}
\end{figure}

\begin{figure}[!htb]
    \centering
    \includegraphics[width=0.7\textwidth]{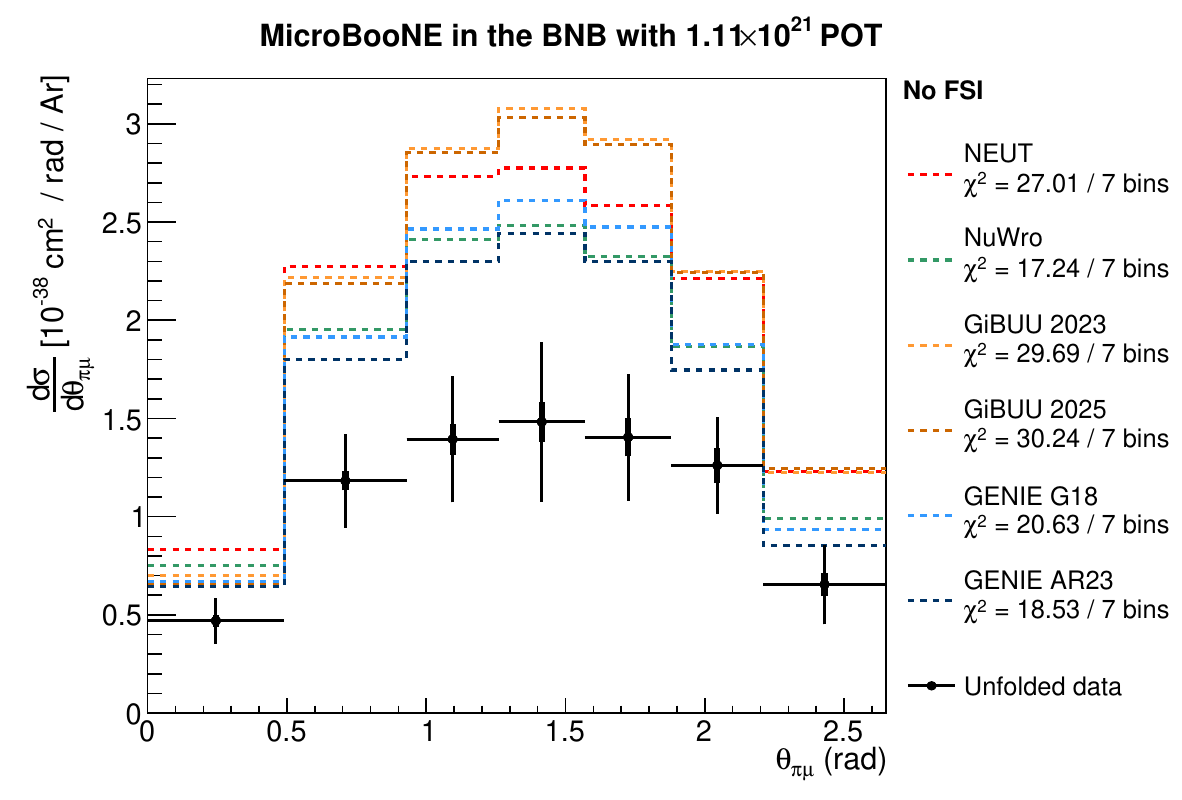}
    \caption{Unfolded muon-pion opening angle cross section in regularized truth space. The black dots represent the unfolded selection with the associated total uncertainty. The thicker inner error bars are the statistical uncertainty only. The lines are generator predictions with FSI disabled.}
    \label{fig:xsec_06}
\end{figure}

The generator comparisons shown in Figs. \ref{fig:total}, \ref{fig:xsec_00}, \ref{fig:xsec_02}, \ref{fig:xsec_03}, \ref{fig:xsec_05}, and \ref{fig:xsec_06} use the initial state particles produced in the generators without FSI.

